\documentclass[aps,prd,showpacs,nofootinbib]{revtex4}
\usepackage{graphicx}
\begin{document}


\def\theequation{\arabic{section}.\arabic{equation}}

\title{Nonlinear Landau damping of a plasmino in\\
the quark-gluon plasma}

\author{Yu.A. Markov}
\email{markov@icc.ru}
\author{M.A. Markova}
\email{markov@icc.ru}
\affiliation{Institute of System Dynamics\\
and Control Theory Siberian Branch\\
of Academy of Sciences of Russia,\\
P.O. Box 1233, 664033 Irkutsk, Russia}

\date{\today}

\begin{abstract}
On the basis of the Blaizot-Iancu equations, which are a local formulation
of the hard thermal loop (HTL) equations of motion for soft fluctuating
quark and gluon fields and their induced sources, the coupled kinetic equations
for plasminos and plasmons are obtained. The equality of matrix elements
for nonlinear scattering of a plasmino by hard particles in covariant and
temporal gauges is established by using effective Ward identities. The model
problem of the interaction of two infinitely narrow packets with fermion and boson
quantum numbers is considered. The kinematical relations between wave vectors
of the plasmino and plasmon are derived, when the effective pumping over of the plasma
excitation energy from the fermion branch of plasma excitations to the boson branch
and vice versa occur. The expression for the nonlinear
Landau damping rate of a plasmino at rest is found, and a comparison with a plasmino
damping constant obtained within the framework of the hard thermal loop
approximation is made. The nonlinear Landau damping rate for normal quark
excitations is shown to diverge like $1/\sqrt{q^2}$ near the light cone
where $q$ is a four-momentum of excitations, and the improved Blaizot-Iancu equations
removing this divergence are proposed.
\end{abstract}


\maketitle{}

\section{Introduction}\setcounter{equation}{0}
This work deals with the study of the interactions between soft-fermion
and gluon excitations in an off-equilibrium quark-gluon plasma (QGP) within
the hard thermal loop effective theory. The specific problem of
the damping of soft, collective, fermion excitations is ultimately considered.
This problem has already been addressed in the literature
(see the review below) in the formalism of thermal perturbation theory,
but nevertheless it would be interesting to see how it emerges in a real-time
formalism based on kinetic equations for soft excitations. In addition,
the calculation of the damping rate for {\it moving} soft excitations is
still incomplete.

The theoretical investigation of properties of plasma excitations bearing
fermion quantum numbers in a quark-gluon plasma originated in the pioneering
works by Klimov and Weldon \cite{klim}.
The gauge-independent dispersion law of fermion excitations was first
derived in a high-temperature region within the framework
of the imaginary time formalism (Matsubara technique) by Klimov and within the
framework of the real-time formalism (Keldysh-Swinger technique) by Weldon
at one-loop order. The quark propagator in the case of nonzero temperature
and massless fermions was shown to have
two poles corresponding to two different dispersion
relations, both with a positive energy. The first branch describes 
normal-particle excitations with the relation between chirality and helicity at
zero temperature. The second branch is collective excitations, where the
usual relation between chirality and helicity is flipped \cite{pis1}. It has been
called a plasmino \cite{bra} to emphasize that, like the plasmon mode of the gluons,
it is a purely collective branch of plasma excitations that have no analog at
zero temperature.

The damping rate of plasma modes is one of the most important characteristics
of high-temperature plasma. The calculation of the damping rate of fermion
excitations in the rigid one-loop approximation was made by
Kapusta and Toimela \cite{kap}. They showed that the standard one-loop
calculations were incomplete and resulted in an explicit dependence of the
damping rate on the gauge-fixing condition and were even negative in some gauges.
Overcoming this problem both in the case of fermion and boson modes
is known to lead to the development of an effective perturbation theory 
put forth by Braaten and Pisarski \cite{bra1}
and by Frenkel and Taylor \cite{fre} that enables one to resum in a systematic
way high-loop diagrams contributing to damping rates at the
leading order in the coupling constant $g$ and gives formal proof that
resummation produces gauge-invariant results for the damping rates of
both quarks and gluons. The calculation of the damping rate of a heavy
fermion \cite{pis2} by Pisarski is one of the first examples of the 
application of effective perturbation theory.

The damping rates depend in a nontrivial way on how fast the
quasiparticle is moving through the thermal medium. Within the framework of
the imaginary-time formalism using a dispersion relation method, Braaten
and Pisarski \cite{bra2} and independently Kobes {\it et al.} \cite{kob} present
by somewhat distinct approaches a complete calculation of the quark damping rate
at zero momentum at leading order in $g$ by evaluating
one-loop diagrams constructed out of the effective propagators and vertices.
It is shown that the damping rate of quark modes at rest is some pure number
times $g^2 T$, where $T$ is a temperature.

Much interest has been taken in the so-called damping rate of moving
(hard or soft) quarks in hot QCD plasma\footnote{Here, we have not discussed
the problems associated with calculation of the damping rate of moving fermions
in a hot QED plasma, which possesses proper specific features
(see, e.g. Ref.\,\cite{leb}).}.
The hard thermal loop (HTL) resummation
renders the infrared finite damping rate of plasma excitations at rest.
However, the damping rate of fast moving collective excitations is infrared
divergent, arising from the exchange of quasistatic, magnetic gluons,
reflecting the breakdown of the Braaten-Pisarski resummation scheme
in deriving the physical quantities that are sensitive to $O(g^2T)$
corrections. In hot QCD plasma this problem is avoided by the introduction of the magnetic
screening mass of order $g^2 T$, which is expected to come from soft-gluon
mutual interactions \cite{lin}. An accurate calculation of the damping rate for soft
moving quarks and gluons has been done by Pisarski in Ref.\,\cite{pis3}. He found that the
infrared-singular contributions were proportional to the group velocity of the
respective quark and gluon modes times $g^2 T \ln (1/g)$. In Ref.\,\cite{fle1} Flechsig
{\it et al.} presented a unified treatment of all next-to-leading-order
corrections to fermion and gluon dispersion laws in QGP's, which are infrared
singular due to mass-shell singularities.
They reproduced the result of Pisarski \cite{pis3} for moving
quasiparticle excitations.

The need for further resummation beyond the scheme of Braaten and Pisarski
also arises in an attempt to calculate the production rate of soft
real photons from equilibrium QGP's \cite{bai}. This rate was shown to be
logarithmically divergent owing to the so-called collinear (mass) singularities
of a HTL diagram. The reason for the collinear singularities is known to be
the use of bare
(massless) hard-quark propagators that are on the mass shell. In a somewhat
different context a similar difficulty was encountered in the research of the
behavior of the next-to-leading order longitudinal component of the
polarization function of the hot gluon system near the light cone \cite{fle2}. Here,
the singularity is even more strong and generated by a premature
restriction to soft loop momenta (the detailed study of the light cone singularity
in the simpler gauge theory of scalar electrodynamics was made by Kraemmer
{\it et al.} \cite{kra}).

The way to solve the problems connected with collinear
singularities  in hot QCD plasma was proposed by Flechsig and Rebhan \cite{fle3}.
They showed that the inclusion of asymptotic thermal masses for hard transverse
gluons\footnote{Notice that the fact of acquiring the transverse gluons in
hot QCD plasmas of the thermal mass was first mentioned by Shuryak \cite{shu}.}
and hard quarks removed the collinear singularities of the HTL amplitudes
without spoiling gauge invariance. The examples of employment of the effective
hard-quark propagator and effective photon--(hard)-quark vertex for deriving
the finite soft-photon production rate from the equilibrium QGP can be found in
Refs.\,\cite{fle3,aur}.

Here we would like to represent a somewhat different view about the
study of the above-mentioned problems, based solely on the kinetic theory.
More comprehensively, our purpose is to construct kinetic equations
that describe the evolution of off-equilibrium deviations in the
{\it number densities} of the soft excitations (fermions and gluons), and
to obtain damping rates from the collision terms.
As we have shown in the case of soft-boson modes \cite{mar1,mar2},
there is no direct connection between the results
obtained within the framework of
kinetic theory of soft excitations and the results derived within
the framework of effective perturbation theory. Here, we would like
to extend the analysis, carried out in Refs.\,\cite{mar1,mar2} on the fermion
degree of freedom of plasma
excitations in hot QCD plasma. Our approach is based on the fundamental system of
equations derived by Blaizot and Iancu \cite{bla2}. These equations are obtained on the basis of
a truncation of the Schwinger-Dyson hierarchy. They isolate consistently the dominant
terms in the coupling constant in a set of equations that describe the
response of the plasma to weak and slowly varying disturbances, and encompass all
HTL's. This system of equations is determined on the space-time scale
$(gT)^{-1}$. We show how to obtain the information
connected  with the damping of soft excitation QGP's by using this system,
i.e., with the processes determined on the other space-time scale.

In order to allow for the collective excitations with arbitrary quantum numbers,
Blaizot and Iancu considered both fermion and boson soft fields. The pure
gauge sector (when the influence of the soft-fermion field is neglected) of
this system of equations has been studied in detail. The sector connected with
a soft-fermion degree of freedom of plasma excitations is not actually studied.
In the present paper we have shown that  this sector of plasma excitations contains
some extended information on the dynamics not only of soft-fermion modes,
which are especially interesting for us, but also soft-boson modes
interacting with them.
Deriving a self-consistent
system of the kinetic equations for plasminos and plasmons enables us to have
a new look at the known results obtained from the diagrammatic technique. It
also allows us to represent
them in a more visual form and to obtain some new results, in particular, connected
with the possibility to observe in more detail the pumping dynamics
of plasma excitations energy over the oscillation spectrum and energy pumping from the fermion
branch of excitation into the boson branch and vice versa. Our strategy
for deriving the relevant kinetic equations is similar to that already used
in connection with the purely (soft) gauge sector \cite{mar2}.

From most papers close to the subject of our research
the recent work by Ni\'egawa \cite{nieg} is to be mentioned. Within the framework of the
closed-time-path formalism the generalized Boltzmann equation
that describes the evolution of the number-density functions of fermion
quasiparticles was derived by him.
The transport equations for ``normal and abnormal modes'' emerge
here from the requirement of the absence of large contributions (which is
the result of pinch singularity) of perturbative scheme proposed in Ref.\,\cite{nieg}.
In spite of the generality of a suggested approach, particular expressions
for collision terms unfortunately were not given.
It makes it difficult to establish the connection with other approaches,
also dealing with the problem of relaxation of plasma excitations, in particular,
with hard thermal loop effective theory.

The paper is organized as follows. In Sec.\,II, after summarizing the conventions
and the notations used in this paper, we write a starting coupling set of the 
soft-field equations and dynamical equations describing the
motion of hard particles in the presence of the soft quark and gluon fields.
The approach to the solution of a given nonlinear system based on
the approximation scheme method--the weak-field expansion--is proposed.
In Sec.\,III, the first-order approximation of the induced source is
considered, and the correlation functions of fermion and boson excitations
are introduced. In Sec.\,IV, the second- and third-order approximations of the
induced source and the color current induced by the soft-fermion field are
studied. In Sec.\,V, we discuss the consistency with gauge symmetry of the
approximation scheme used. In Sec.\,VI we derive the generalized kinetic
equation for soft-fermion excitations and supplement it with a generalized
kinetic equation for soft-boson excitations. The right-hand side of the last equation
contains the terms responsible for the nonlinear interaction between soft bosonic and
fermionic modes. Sec.\,VII presents a detailed consideration of the terms on the
right-hand side of the generalized kinetic equation obtained in the previous
section, their identification with specific physical processes. Then, kinetic
equations describing the nonlinear interactions between plasminos and plasmons
are extracted. In Sec.\,VIII, by means of the effective Ward identities, we show
that the
function defining the plasmino--hard-particle scattering matrix element squared is
the same in the covariant gauge as in the temporal one. In Sec.\,IX, the transformation
of the nonlinear Landau  damping rate for a plasmino similar to the transformation
in a pure gauge sector of soft QGP excitations is performed. In Sec.\,X,
on the basis of decomposition of the nonlinear Landau damping rate into positive
and negative parts, the model problem of two interacting infinitely narrow
packets
with fermion and boson quantum numbers is considered, and the kinematic relations
between wave vectors of excitations are defined, so that one or another
process of pumping over of energy occurs. In Sec.\,XI, an explicit expression
for the nonlinear Landau damping rate of the plasmino at rest is derived,
and a comparison with a similar expression for the damping rate obtained within
the framework of the HTL approach is carried out.
In Sec.\,XII, the behavior of the nonlinear
Landau damping rate for normal-particle excitations near the light cone is
considered, and this damping rate is shown to diverge owing to the
mass singularities of quarks. In the Conclusion
we briefly discuss a more general statement of the research problem of
soft excitation dynamics in hot QCD plasma within the framework of kinetic approach
outlined in this paper.

\section{Blaizot-Iancu equations}\setcounter{equation}{0}

We adopt conventions of Blaizot and Iancu \cite{bla2}.
We use the metric $g^{\mu \nu} = {\rm diag}(1,-1,-1,-1)$, choose units such
that $c=k_{B}=1$ and note $X=(X_0,{\mathbf X}), \,p=(p_0,{\mathbf p}), \ldots \,.$
We consider an ${\rm SU}(N_c)$ gauge theory with $N_f$ flavors of massless quarks.
The color indices for the adjoint representation, $a,b, \ldots$ run from 1
to $N_c^2-1$, while those for the fundamental representation, $i,j, \ldots \, ,$
run from 1 to $N_c$. The Greek indices $\alpha, \beta, \ldots$ for the spinor
representation run from 1 to 4.

The quark field $\psi$ and the gauge field $A_{\mu}=A_{\mu}^{a}t^{a}$ with
$N_{c}^{2}-1$ Hermitian generators in the fundamental representation obey
the field equations
\begin{equation}
i\!\not\!\!D \psi(X)=\eta(X),
\label{eq:2q}
\end{equation}
\begin{equation}
[D^{\nu},F_{\mu \nu}(X)] -
\xi^{-1} \partial_\mu \partial^\nu A_{\nu}(X) = g\,\bar{\psi}(X)\gamma_{\mu}
t^a \psi(X)t^a + j_{\mu}(X),
\label{eq:2w}
\end{equation}
where $\not\!\!D = \gamma^{\mu}D_{\mu}$ is the covariant derivative in the
fundamental representation:
\[
D_{\mu} = \partial_{\mu} + igA_{\mu}(X),
\]
$F_{\mu \nu}=F_{\mu \nu}^{a}t^{a}$ is field strength tensor with
$F_{\mu \nu}^{a} = \partial_\mu A_{\nu}^{a} - \partial_\nu A_{\mu}^{a}-
gf^{abc}A_{\mu}^{b}A_{\nu}^{c}$; $[\,,\,]$ denotes the commutator and
$\xi$ is a gauge parameter fixing a covariant gauge.

The induced source $\eta(X)$ on the right-hand side of Eq.\,(\ref{eq:2q}) can
be written as an integral over hard momentum ${\mathbf k}$ of a density
$\not\!\!\Lambda({\mathbf k},X) \equiv t^a \gamma^{\mu} \Lambda^a_{\mu}({\mathbf k},X)$
\cite{bla2}
\begin{equation}
\eta(X) = g\!\int\!\!\frac{{\rm d}{\bf k}}{(2 \pi)^3} \,\frac{1}{\epsilon_k}
\not\!\!\Lambda({\mathbf k},X), \;\; \epsilon_k \equiv \vert {\mathbf k} \vert.
\label{eq:2e}
\end{equation}
A density $\Lambda^a_{\mu}({\mathbf k},X)$ is a generalized one-body density matrix mixing
fermion and boson degrees of freedom.

The total induced color current $j_{\mu}$ on the right-hand side of field
equation (\ref{eq:2w}) is a sum of two parts: the current $j^A_{\mu}(X)$
induced by the soft-gauge field
\begin{equation}
j_{\mu}^A(X) = g\!\int\!\!\frac{{\rm d}{\mathbf k}}{(2 \pi)^3} \, v_{\mu}
[N_f (\delta n_{+}^A({\mathbf k},X) - \delta n_{-}^A({\mathbf k},X)) + 2N_c
\delta N^A({\mathbf k},X)],
\label{eq:2r}
\end{equation}
and the current $j^{\Psi}_{\mu}(X)$ induced by the soft-quark field
\begin{equation}
j_{\mu}^{\Psi}(X) = g\!\int\!\!\frac{{\rm d}{\mathbf k}}{(2 \pi)^3} \, v_{\mu}
[ \delta n_{+}^{\Psi}({\bf k},X) - \delta n_{-}^{\Psi}({\mathbf k},X)], \;
v_{\mu} = (1,{\mathbf v}), \; {\bf v}={\mathbf k}/ \vert {\mathbf k} \vert,
\label{eq:2t}
\end{equation}
where $\delta n^{A, \Psi}_{\pm} ({\mathbf k},X) =
\delta n^{A, \Psi \, a}_{\pm}t^a$ are soft fluctuations in the quark
and antiquark color densities and $\delta N^A({\mathbf k},X) = \delta N^{A \,a}t^a$
is a soft fluctuation in the gluon color density. The superscripts indicate the
nature of the background field that induces fluctuations.
On the space-time scale $(gT)^{-1}$ these functions and density
$\not\!\!\Lambda ({\mathbf k}, X)$ satisfy the following system of equations:
\begin{equation}
[v \cdot D_X, \delta n_{\pm}^A ({\mathbf k}, X)] = \mp\,g\,{\mathbf v} \cdot {\bf E}(X)
\, \frac{d n (\epsilon_k)}{d \epsilon_k} ,
\label{eq:2y}
\end{equation}
\begin{equation}
[v \cdot D_X, \delta N^A ({\mathbf k}, X)] = -\,g\,{\mathbf v}\cdot {\bf E}(X)
\, \frac{d N (\epsilon_k)}{d \epsilon_k} ,
\label{eq:2yy}
\end{equation}
\begin{equation}
[v \cdot D_X, \delta n_{\pm}^{\Psi} ({\mathbf k}, X)] = \pm\,\frac{ig}{2 \epsilon_k}
\, t^a\, ( \bar{\psi}(X) t^a\!\not\!\!\Lambda ({\mathbf k}, X)
- \bar{\not\!\!\Lambda}({\mathbf k},X) t^a \psi (X) ) ,
\label{eq:2yyy}
\end{equation}
\begin{equation}
(v \cdot D_X)\!\not\!\!\Lambda ({\mathbf k},X) = -\,ig \, C_F \, [N (\epsilon_k)
+ n(\epsilon_k)]\!\not\!v\,\psi (X) .
\label{eq:2yyyy}
\end{equation}
Here, ${\mathbf E} (X) = {\mathbf E}^a(X) t^a$ is the chromoelectric field,
$E^i = F^{i0}; n(\epsilon_k) = 1/(\exp(\epsilon_k/T) + 1)$ and
$N(\epsilon_k) = 1/(\exp(\epsilon_k/T) - 1)$ are the fermion and boson
occupation factors, $T$ is the temperature of the plasma, and $C_F$ is the quadratic
Casimir invariant of the fundamental representation. The function
$\bar{\not\!\!\Lambda} ({\mathbf k}, X)$ is related to
$\not\!\!\Lambda ({\mathbf k}, X)$ by $\bar{\not\!\!\Lambda} ({\mathbf k}, X) =
\not\!\!\Lambda^{\dagger} ({\mathbf k}, X)
\gamma^0$, where the Hermitian conjugation refers to color and spinor indices.

The self-consistent system of Eqs.\,(\ref{eq:2q})\,--\,(\ref{eq:2yyyy}) 
for the soft fluctuating fields
$\psi$ and $A_{\mu}$ and their induced sources was first derived in
Ref.\,\cite{bla2}. In the subsequent discussion it will be called the Blaizot-Iancu
equations. However, as distinct from the original paper \cite{bla2}, we
take Eqs.\,(\ref{eq:2y})\,--\,(\ref{eq:2yyyy}) not as kinetic equations,
i.e., time-irreversible ones, but as exact "microscopic"
dynamical equations coming from the hard thermal loop effective action
and describing the evolution of the collisionless plasma with zero expectation
values of the soft fields (or the associated induced sources).

The Blaizot-Iancu equations are solved by the approximation scheme
method -- {\it the weak-field expansion}. For this purpose first we
expand soft fluctuations of the (anti)quark and gluon color densities as
power series in the oscillation amplitudes of the functions $A_{\mu}$ and $\psi$,
\begin{equation}
\delta n^{A}_{\pm} = \sum_{n=1}^{\infty} \delta n^{A\,(n)}_{\pm}, \;
\delta N^{A} = \sum_{n=1}^{\infty} \delta N^{A\,(n)},
\label{eq:2u}
\end{equation}
\begin{equation}
\delta n^{\Psi}_{\pm} = \sum_{n=0}^{\infty} \delta n^{\Psi\,(n,2)}_{\pm},
\label{eq:2i}
\end{equation}
\begin{equation}
\not\!\!\Lambda = \sum_{n=0}^{\infty} \not\!\!\Lambda^{(n,1)},
\label{eq:2o}
\end{equation}
where the index $n$ shows that
$\delta n^{A\,(n)}_{\pm}, \, \delta N^{A\,(n)},\, \delta n^{\Psi\,(n,2)}_{\pm}$
and $\not\!\!\Lambda^{(n,1)}$
are proportional to the $n$th power of $A_{\mu}$. Since the fermion
fields appear in explicit form on the right-hand side of Eqs.\,(\ref{eq:2yyy})
and (\ref{eq:2yyyy}), the functions $\delta n^{\Psi\,(n,2)}_{\pm}$ and
$\not\!\!\Lambda^{(n,1)}$ also depend on amplitudes $\bar{\psi}$ and $\psi$.
By virtue of the structure of the right-hand side of Eqs.\,(\ref{eq:2yyy})
and (\ref{eq:2yyyy}), we have
\[
\delta n^{\Psi (n,2)}_{\pm} \sim \bar{\psi} \psi, \;
\not\!\!\Lambda^{(n,1)} \sim \psi
\]
for arbitrary values of $n$.

The induced color currents (\ref{eq:2r}) and (\ref{eq:2t}) and the induced source
(\ref{eq:2e}) are expressed as
\begin{equation}
j^{A}_{\mu} = \sum_{n=1}^{\infty} j^{A\,(n)}_{\mu},
\label{eq:2p}
\end{equation}
\begin{equation}
j^{\Psi}_{\mu} = \sum_{n=0}^{\infty} j^{\Psi \,(n,2)}_{\mu},
\label{eq:2a}
\end{equation}
where
\begin{equation}
j_{\mu}^{A\,(n)} = g\!\int\!\frac{{\rm d}{\mathbf k}}{(2 \pi)^3} \, v_{\mu}
[N_f (\delta n_{+}^{A\,(n)} - \delta n_{-}^{A\,(n)}) + 2N_c \delta N^{A\,(n)}],
\label{eq:2s}
\end{equation}
\begin{equation}
j_{\mu}^{\Psi\,(n,2)} = g\!\int\!\frac{{\rm d}{\mathbf k}}{(2 \pi)^3} \, v_{\mu}
[ \delta n_{+}^{\Psi\,(n,2)} - \delta n_{-}^{\Psi\,(n,2)}] ,
\label{eq:2d}
\end{equation}
and
\begin{equation}
\eta = \sum_{n=0}^{\infty} \eta^{(n,1)},
\label{eq:2f}
\end{equation}
where
\begin{equation}
\eta^{(n,1)} = g\!\int\!\frac{{\rm d}{\bf k}}{(2 \pi)^3} \frac{1}{\epsilon_k}
\not\!\!\Lambda^{(n,1)}.
\label{eq:2g}
\end{equation}

Now we turn to the field equations (\ref{eq:2q}) and
(\ref{eq:2w}), connecting the quark and gluon soft fields with induced source
$\eta$ and total color current $j_{\mu}$. Let us rewrite these equations,
explicitly separating the free parts in Eqs. (\ref{eq:2q}) and (\ref{eq:2w}) from
interaction terms. Taking into account the expansion of induced current
(\ref{eq:2p}), (\ref{eq:2a}) and induced source (\ref{eq:2f}), we have
\begin{equation}
i \not\!\partial \psi - \eta^{(0,1)} =
g\!\not\!\!A \psi + \eta_{NL},
\label{eq:2h}
\end{equation}
\begin{equation}
\partial_{\mu}(F^{\mu \nu})_{L} - \xi^{-1} \partial^{\nu}
\partial^{\mu}A_{\mu} - j^{A\,(1)\,\nu} =
j_{NL}^{\nu} + g\bar{\psi}\gamma^{\nu}t^a \psi t^a -
ig\,\partial_{\mu}[A^{\mu},A^{\nu}]
-\,ig[A_{\mu},(F^{\mu \nu})_{L}]
+\,g^{2}[A_{\mu},[A^{\mu},A^{\nu}]].
\label{eq:2j}
\end{equation}

Here, the indices $L$ and $NL$ denote the linear and nonlinear parts of
strength  tensor, the induced source
and the color-induced current with respect to $A_{\mu}$
and $\psi$. The approximation that will be made is the truncation of the
series expansion (\ref{eq:2p}), (\ref{eq:2a}) and (\ref{eq:2f}).
The accuracy of this approximation is controlled by the characteristic
amplitudes of the soft fields, which will be discussed in Sec. V. As will be
shown to account for the nonlinear interaction between  waves and hard
particles in QGP at leading order in $g$, it is sufficient to restrict
the consideration to the third order in total powers of $\psi$ and $A_{\mu}$
in expansions (\ref{eq:2p}), (\ref{eq:2a}), and (\ref{eq:2f}).

\section{Linear approximation of the induced source $\eta$.
Correlation functions of the fermion and boson excitations}
\setcounter{equation}{0}

We now come to the derivation of the kinetic equation for the soft-fermion 
modes. The starting point is Eq.\,(\ref{eq:2h}). Its
left-hand side contains the linear approximation of the induced source $\eta$, whose
explicit form can be easily found from Eq.\,(\ref{eq:2yyyy}). We prefer to work
in momentum space. The corresponding equations are obtained by using
\[
\psi (X) = \int\!dq \, \psi (q) \,{\rm e}^{-i q\cdot X}, \;
A_{\mu} (X) = \int\!dp \, A_{\mu} (p) \,{\rm e}^{-i p\cdot X}
\]
and similar transformations for $\delta n_{\pm}^{A},\,\delta N^{A}$, etc.
Here and in what follows
we denote  the momenta of the soft-quark fields by $q, \, q^{\prime}, \,
q_1, \ldots$ and the momenta  of the soft-gauge fields by $p, \, p^{\prime}, \,
p_1, \ldots\,$.

Let us linearize Eq.\,(\ref{eq:2yyyy}) with respect to $\psi$ and $A_{\mu}$.
The result of the Fourier transformation of the linearized equation is
\begin{equation}
\not\!\!\Lambda^{(0,1)}({\mathbf k},q) = gC_F \, \frac{\not\!v \, \psi(q)}
{v \cdot q + i \epsilon} \,[N(\epsilon_k) + n(\epsilon_k)], \;
\epsilon \rightarrow +0.
\label{eq:3q}
\end{equation}
By substituting Eq.\,(\ref{eq:3q}) into the Fourier transform of relation (\ref{eq:2g})
(for $n=0$) and performing the radial integration over $d \epsilon_k$, we obtain
the linear approximation with respect to the fermion field of the induced source,
\begin{equation}
\eta^{(0,1)}(q) = \delta \Sigma(q) \psi(q),
\label{eq:3w}
\end{equation}
where
\[
\delta \Sigma(q) = \omega_0^2\!\int\!\frac{d \Omega}{4\pi} \,
\frac{\not\!v}{v \cdot q + i \epsilon}
\]
is a well-known HTL expression for the soft-quark (retarded) self-energy,
$\omega_0^2 = g^2 C_F T^2/8$ is the plasma frequency of the quark sector
of plasma excitations, and $d \Omega$ is an angular measure.

Furthermore, we rewrite Eq.\,(\ref{eq:2h}) in momentum space. Taking into
account Eq.\,(\ref{eq:3w}) we obtain
\begin{equation}
\,^{\ast}\!S^{-1}(q^{\prime}) \psi^j(q^{\prime}) =
-\,g\!\int\!\!\not\!\!A(p_1) \psi^j (q_1)\,\delta(q^{\prime} - q_1 - p_1) dq_1 dp_1
-\,\eta^{(1,1)\,j}(q^{\prime}) - \eta^{(2,1)\,j}(q^{\prime}).
\label{eq:3e}
\end{equation}
Here $\,^{\ast}\!S(q) = [\,-\!\not\!q + \delta \Sigma(q)]^{-1}$ is an equilibrium
propagator for a soft quark, corrected to leading order in $g$.

Let us multiply Eq.\,(\ref{eq:3e}) by the Dirac conjugate amplitude
$\bar{\psi}^i(-q) = (\psi^i(q))^{\dagger}\gamma^0$ and
take an expectation value over the off-equilibrium ensemble:
\begin{equation}
\langle \bar{\psi}^i(-q)\,^{\ast}\!S^{-1}(q^{\prime}) \psi^j(q^{\prime}) \rangle =
-\,g\!\int\langle \bar{\psi}^i(-q) (\not\!\!A(p_1) \psi(q_1))^j \rangle\,
\delta(q^{\prime} - q_1 - p_1) dq_1 dp_1
\label{eq:3r}
\end{equation}
\[
-\,\langle \bar{\psi}^i(-q) \eta^{(1,1)\,j}(q^{\prime})\rangle
- \langle \bar{\psi}^i(-q) \eta^{(2,1)\,j}(q^{\prime})\rangle .
\]
We now introduce the correlation functions of soft-fermion and -boson
excitations
\begin{equation}
\Upsilon^{ji}_{\beta \alpha}(q^{\prime},q) =
\langle \bar{\psi}^i_{\alpha}(-q) \psi^j_{\beta}(q^{\prime}) \rangle, \;
I_{\mu \nu}^{ab}(p^{\prime},p)= \langle A_{\mu}^{\ast\,a}(p^{\prime})
A_{\nu}^{b}(p) \rangle,
\label{eq:3t}
\end{equation}
respectively.
The asterisk denotes a complex conjugate.
The considered soft excitations are necessarily {\it colorless} and have
{\it zero fermion number} by virtue of the fact that mean fields
$\langle A_{\mu}^a \rangle$ and $\langle \psi_{\alpha}^i \rangle$
or the associated mean induced color current and source are assumed to be vanishing.
Therefore, for the physical situation of interest,
the off-equilibrium two-point functions (\ref{eq:3t}) are diagonal in color space,
which will be implied in what follows.

For the conditions of
stationary and homogeneous QGP's (i.e., when correlation functions (\ref{eq:3t})
in the coordinate representation depend only on the difference of coordinates and
time $ \triangle X=X^{\prime} - X$), we have
\[
\Upsilon^{ji}_{\beta \alpha}(q^{\prime},q) =
\Upsilon^{ji}_{\beta \alpha}(q^{\prime}) \delta(q^{\prime} - q), \;
I_{\mu \nu}^{ab}(p^{\prime},p)=I_{\mu \nu}^{ab}(p^{\prime})
\delta (p^{\prime}-p).
\]
The QGP state becomes slightly heterogeneous and nonstationary
because of the effects of the nonlinear interaction between waves and particles.
This leads to
a $\delta$-function broadening, and $\Upsilon^{ji}_{\beta \alpha}$ and 
$I^{ab}_{\mu \nu}$ depend on both arguments.

Let us introduce
$\Upsilon_{\beta \alpha}^{ji}(q^{\prime},q)=\Upsilon_{\beta \alpha}^{ji}
(q, \triangle q)$, $\triangle q = q^{\prime} - q$ with
$\vert\,\triangle q / q\,\vert\ll\!1$
and $I_{\mu \nu}^{ab}(p^{\prime},p)=I_{\mu \nu}^{ab}
(p, \triangle p)$, $\triangle p = p^{\prime} - p$ with
$\vert\,\triangle p / p\,\vert\ll\!1$,
and insert the correlation functions in the Wigner form
\begin{equation}
\Upsilon_{\beta \alpha}^{ji}(q,x)= \int \Upsilon_{\beta \alpha}^{ji}
(q, \triangle q)\,{\rm e}^{- i \triangle q\cdot x} d \triangle q,\;
I_{\mu \nu}^{ab}(p,x)= \int I_{\mu \nu}^{ab}(p, \triangle p)
\,{\rm e}^{- i \triangle p\cdot x} d \triangle p,
\label{eq:3y}
\end{equation}
slowly depending on $x$. In Eq.\,(\ref{eq:3r}), we replace
$q \leftrightarrow q^{\prime}, \, i \leftrightarrow j$, take a complex
conjugation, and then subtract the resulting equation from Eq.\,(\ref{eq:3r}),
expanding beforehand the quark self-energy into ``Hermitian'' and
``anti-Hermitian'' parts:
\[
\delta \Sigma(q) = \delta \Sigma^{\rm H}(q) + \delta \Sigma^{\rm A}(q), \;
\gamma^0 ( \delta \Sigma^{\rm H}(q))^{\dagger}\gamma^0 =
\delta \Sigma^{\rm H}(q), \;
\gamma^0 ( \delta \Sigma^{\rm A}(q))^{\dagger}\gamma^0 =
-\,\delta \Sigma^{\rm A}(q).
\]
We assume that the anti-Hermitian part $\delta \Sigma^{\rm A}$ is small 
relative to the Hermitian part $\delta \Sigma^{\rm H}$ and is of the same
order as the nonlinear terms on the right-hand side. We can therefore set
$\delta \Sigma^{\rm A}(q) \simeq \delta \Sigma^{\rm A}(q^{\prime})$ and
move the term with $\delta \Sigma^{\rm A}$ into the right-hand side of
Eq.\,(\ref{eq:3r}).
We expand the remaining terms in the left-hand side with respect to
$\triangle q$ up to the first order. This corresponds to a {\it gradient
expansion} procedure usually used in the derivation of kinetic
equations. Multiplying the resulting equation by
${\rm e}^{-i \triangle q\cdot x}$ and integrating over $\triangle q$, we obtain
\begin{equation}
{\rm tr} \bigg( \frac{\partial}{\partial q^{\mu}} [\,-\!\not\!q +
\delta \Sigma^{\rm H}(q)]
\frac{\partial \Upsilon^{ji}(q,x)}{\partial x_{\mu}} \bigg) =
2i \, {\rm tr}\,(\delta \Sigma^{\rm A}(q) \Upsilon^{ji}(q,x))
\label{eq:3u}
\end{equation}
\[
-ig\!\!\int\!dq^{\prime} dq_1 dp_1 \Big\{\!\langle
(\bar{\psi}(-q_1)\!\not\!\!A(-p_1))^i
\psi^j(q^{\prime}) \rangle \delta(q - q_1 - p_1) -
\langle \bar{\psi}^i(-q)(\not\!\!A(p_1)\psi(q_1))^j \rangle
\delta(q^{\prime} - q_1 - p_1)\!\Big\}
\]
\[
-i\!\!\int\!dq^{\prime}\Big\{\!(
\langle \bar{\eta}^{(1,1)\,i}(-q) \psi^j(q^{\prime})\rangle -
\langle \bar{\psi}^i(-q) \eta^{(1,1)\,j}(q^{\prime})\rangle )
+ ( \langle \bar{\eta}^{(2,1)\,i}(-q) \psi^{j}(q^{\prime}) \rangle
- \langle \bar{\psi}^i (-q) \eta^{(2,1)\,j}(q^{\prime})\rangle )\!\Big\}.
\]
Here, the Dirac trace is represented by ${\rm tr}$, $\bar{\eta}(-q) =
\eta^{\dagger}(q) \gamma^0$, and we take into account the reality of the
gauge field: $A^{\ast\,a}_{\mu}(p) = A^a_{\mu}(-p)$. The linear term with
$\delta\Sigma^{\rm A}$ on the right-hand side corresponds to the linear Landau
damping of soft-fermion excitations.

\section{Second and third approximations of the induced source
$\eta$}
\setcounter{equation}{0}

Now we are concerned with computation of the nonlinear corrections to the induced
source on the right-hand side of Eq.\,(\ref{eq:3u}).
At first, we define $\not\!\!\Lambda^{(1,1)}$. Substituting the series expansion
(\ref{eq:2o}) into (\ref{eq:2yyyy}) and keeping only the terms of the second order
in fields, we find
\[
(v \cdot \partial_X)\!\not\!\!\Lambda^{(1,1)}({\mathbf k},X) =
-\,ig (v\cdot A(X))\!\not\!\!\Lambda^{(0,1)}({\mathbf k},X).
\]
Performing the Fourier transformation of the last equation and taking into account
an explicit form of $\not\!\!\Lambda^{(0,1)}$ (\ref{eq:3q}), we derive
\begin{equation}
\not\!\!\Lambda^{(1,1)}({\mathbf k},q) =
g^2 C_F \, [N(\epsilon_k) + n(\epsilon_k)] \,
\frac{v^{\mu} \not\!v}{v\cdot q + i \epsilon}
\,t^a\!\!\int\!\frac{1}{v\cdot q_1 + i \epsilon} \,
A_{\mu}^a(p_1) \psi(q_1)\,\delta(q - q_1 - p_1) dq_1 dp_1.
\label{eq:4q}
\end{equation}
Substituting the obtained expression into Eq.\,(\ref{eq:2g}) (for $n=1$) and
performing the radial integration over $d \epsilon_k$, we find the required
induced source correction
\begin{equation}
\eta^{(1,1)}(q) =  g\,\omega_0^2\,t^a\!\!\int\!\frac{{\rm d}\Omega}{4 \pi}\,
\frac{v^{\mu} \not\!v}{(v\cdot q + i \epsilon) (v\cdot q_1 + i \epsilon)}
A_{\mu}^a(p_1) \psi(q_1)\,\delta(q - q_1 - p_1) dq_1 dp_1.
\label{eq:4w}
\end{equation}

The expression for the induced source of the third order in the fields is defined
by means of reasoning similar to the previous one. Here, we have
\begin{equation}
\eta^{(2,1)}(q) =  g^2\omega_0^2 \,t^a t^b\!\int\!\frac{{\rm d}\Omega}{4 \pi}\,
\frac{v^{\mu}v^{\nu} \not\!v}{(v\cdot q + i \epsilon)
(v\cdot (q - p_1) + i \epsilon)(v\cdot q_1 + i \epsilon)} \,
A_{\mu}^a(p_1) A^b_{\nu}(p_2) \psi(q_1)
\delta(q - q_1 - p_1 - p_2) dq_1 dp_1 dp_2.
\label{eq:4e}
\end{equation}

Now we return to the initial equation for the soft-fermion field (\ref{eq:3u}).
We substitute nonlinear corrections to the induced source $\eta$
(Eqs.\,(\ref{eq:4w}) and (\ref{eq:4e})) into the right-hand side of this equation.
After simple algebraic transformations, instead of Eq.\,(\ref{eq:3u}), we find
$$
{\rm tr} \bigg( \frac{\partial}{\partial q^{\mu}} [\,-\!\not\!q +
\delta \Sigma^{\rm H}(q)]
\frac{\partial \Upsilon^{ji}(q,x)}{\partial x_{\mu}} \bigg) =
2i \, {\rm tr}\,( \delta \Sigma^{\rm A}(q) \Upsilon^{ji}(q,x))
$$
$$
-\,ig\!\int\!dq^{\prime} dq_1 dp_1 \Big\{
\langle \bar{\psi}^i(-q)A^{a\mu}(p_1)(\,^{\ast}\Gamma^{(Q)\,a}_{\mu}(p_1;q_1,-q^{\prime})
\psi(q_1))^j \rangle
\delta(q^{\prime} - q_1 - p_1)
$$
\begin{equation}
-\,\langle (\bar{\psi}(-q_1)
\,^{\ast}\Gamma^{(Q)\,a}_{\mu}(-p_1;-q_1,q))^iA^{\ast\,a\mu}(p_1)
\psi^j(q^{\prime}) \rangle \delta(q - q_1 - p_1)
\label{eq:4r}
\end{equation}
$$
+\,ig^2 \int\!dq^{\prime}dq_1dp_1dp_2\Big\{
\langle A^{a\mu}(p_1) A^{b\nu}(p_2) \bar{\psi}^i(-q)
(\delta\Gamma^{(Q)\,ab}_{\mu\nu}(-p_1,-p_2;q^{\prime},-q_1)\psi(q_1))^j \rangle
\delta(q^{\prime} - q_1 - p_1 - p_2)
$$
$$
-\,\langle A^{\ast\,a\mu}(p_1) A^{\ast\,b\nu}(p_2)(\bar{\psi}(-q_1)
\delta\Gamma^{(Q)\,ba}_{\mu\nu}(p_2,p_1;-q,q_1))^i\psi^j(q^{\prime})\rangle
\delta(q - q_1 - p_1 - p_2)\Big\}.
$$
Here,
\begin{equation}
\,^{\ast}\Gamma^{(Q)\,a}_{\mu}(p;q_1,q_2) =
t^a\,^{\ast}\Gamma^{(Q)}_{\mu}(p;q_1,q_2) \equiv
t^a(\gamma_{\mu} + \delta \Gamma^{(Q)}_{\mu}(p;q_1,q_2))
\label{eq:4t}
\end{equation}
is an effective (i.e., HTL-resummed) vertex between a quark pair and a gluon,
that is, a sum of the bare vertex $\gamma_{\mu}$ and a corresponding HTL correction
\cite{bra1,tay,bla2}
\begin{equation}
\delta \Gamma^{(Q)}_{\mu}(p;q_1,q_2) =
- \, \omega^2_0\!\int\!\frac{{\rm d} \Omega}{4 \pi} \, \frac{v_{\mu} \not\!v}
{(v\cdot q_1 + i \epsilon)(v\cdot q_2 - i \epsilon)},
\label{eq:4y}
\end{equation}
and
\begin{equation}
\delta \Gamma^{(Q)\,ab}_{\mu\nu}(p_1,p_2;q_1,q_2)=
-\, \omega_0^2\!\int\!\frac{{\rm d} \Omega}{4 \pi}\,
\frac{v_{\mu}v_{\nu} \not\!v}{(v\cdot q_1 + i \epsilon)
(v\cdot q_2 - i \epsilon)}
\bigg(\frac{t^at^b}{v\cdot (q_1+p_1) + i \epsilon}
+ \,\frac{t^bt^a}{v\cdot (q_1+p_2) + i \epsilon} \bigg)
\label{eq:4yy}
\end{equation}
is an effective vertex between a quark pair and two gluons (this vertex does
not exist at tree level, and in the leading order it arises entirely from the
HTL \cite{bra1,tay}).
The superscript $(Q)$ denotes that
the vertex $\delta \Gamma_{\mu}(X;Y_1,Y_2)$ corresponds to the function
(\ref{eq:4y}) in the coordinate representation, where the time
arguments satisfy $Y_1^0 \geq X^0 \geq Y_2^0$ (boundary conditions),
the vertex function
(\ref{eq:4yy}) corresponds to $\delta \Gamma_{\mu}(X_1,X_2;Y_1,Y_2),$ where
the time arguments satisfy $Y_1^0 \geq X_1^0 \geq X_2^0 \geq Y_2^0$
for the first term in parentheses on the right-hand side of Eq.\,(\ref{eq:4yy})
and for the second
term we have $Y_1^0 \geq X_2^0 \geq X_1^0 \geq Y_2^0$, i.e., the time argument of
the external quark leg incoming in the vertex functions (\ref{eq:4y}) and
(\ref{eq:4yy}), is largest. In deriving Eq.\,(\ref{eq:4r}) we have dropped
the terms proportional to $\langle A_{\mu}\bar{\psi}\rangle\langle
A_{\nu} \psi\rangle$ not contributing to the right-hand side of kinetic
equation for the soft-fermion excitations at the leading order in $g$, in which
we are interested.

Because of nonlinear wave interactions, phase correlation effects occur.
By virtue of their smallness, the expectation value of a product of four
quantities $\langle A_{\mu}A_{\nu}\bar{\psi} \psi \rangle$ can be expressed
approximately as a product of the expectation values of two fields
$\langle A_{\mu}A_{\nu}\rangle$ and
$\langle \bar{\psi} \psi \rangle$. For a product of three fields this approach
yields zero; in this case, the weak correlation between the fields  is to be
taken into account. The third-order correlation functions on the right-hand
side of Eq.\,(\ref{eq:4r}) contain amplitudes of waves with different statistic,
nonlinear corrections to that are defined by corresponding field equations.
Let us consider first of all the nonlinear correction to a free quark field.
For this purpose, we use field equation (\ref{eq:3e}), keeping in the right-hand side
only the second-order terms (with respect to $A_{\mu}$ plus $\psi$)
\[
\,^{\ast}\!S^{-1}(q) \psi(q) = -\, g\!\int\!\,^{\ast}\Gamma^{(Q)\, \mu}(p_1;q_1,-q)
A_{\mu}(p_1) \psi(q_1)\,\delta(q - q_1 - p_1) dq_1 dp_1.
\]
The approximate solution of this equation has the form
\begin{equation}
\psi(q) = \psi^{(0)}(q) -
g \,^{\ast}\!S(q)\!\int\!\,^{\ast}\Gamma^{(Q)\, \mu}(p_1;q_1,-q)
A^{(0)}_{\mu}(p_1) \psi^{(0)}(q_1)\,\delta(q - q_1 - p_1) dq_1 dp_1,
\label{eq:4u}
\end{equation}
where $A^{(0)}_{\mu}$ and $\psi^{(0)}$ are solutions of the appropriate
homogeneous field equations corresponding to free fields.

Now we consider the nonlinear correction to a free gauge field. For this
purpose, we use field equation (\ref{eq:2j}) is rewritten in momentum
space, keeping on the right-hand side the terms of the second order in fields:
\[
\,^{\ast} {\cal D}^{-1 \, \mu \nu}(p) A_{\nu}^a(p) =
j^{A (2)\, a \mu}(p) + j^{\Psi (0,2)\, a \mu}(p)
+ g\!\!\int\!\bar{\psi}(-q_1) \gamma^{\mu} t^a \psi(q_2)\,
\delta(p + q_1 - q_2) dq_1 dq_2
\]
\begin{equation}
-\,\frac{i}{2}\,g f^{abc}\!\!\int\!\Gamma^{\mu \nu \lambda}(p,-p_1,-p_2) \,
A^b_{\nu}(p_1)A^c_{\lambda}(p_2)\,
\delta(p-p_1-p_2) dp_1 dp_2.
\label{eq:4i}
\end{equation}
Here, $\,^{\ast}{\mathcal D}^{\mu \nu}(p) = -\,[\,p^2 g^{\mu \nu} - (1 + \xi^{-1})
p^{\mu}p^{\nu} + \delta \Pi^{\mu \nu}(p)]^{-1}$ represents the medium-modified
(retarded) gluon propagator with the soft-gluon polarization tensor,
\[
\delta \Pi^{\mu \nu}(p) =
3\,\omega^2_{\rm pl} \, \Big( g^{\mu 0} g^{\nu 0} - p^0\!\int\!\frac{{\rm d} \Omega}
{4 \pi} \, \frac{v^{\mu}v^{\nu}}{v\cdot p + i \epsilon} \Big) ,
\]
corrected at the leading order in $g$, $\omega^2_{\rm pl} = g^2T^2(N_f + 2N_c)/18$ is
a plasma frequency of the gauge sector of plasma excitations, and
$\Gamma^{\mu \nu \lambda}(p,p_1,p_2)$
is a bare three-gluon vertex. The first and the last terms on the right-hand
side of Eq.\,(\ref{eq:4i}) give the contribution to a correlation function
$\langle A \bar{\psi} \psi \rangle$ proportional to
$\langle AA \bar{\psi} \psi \rangle - \langle AA \rangle \langle \bar{\psi}
\psi \rangle$. By replacing interacting fields by free ones and
dividing the fourth-order correlation into a product of the second-order
correlators, this contribution vanishes. Therefore these terms in this
approximation can be dropped.

Now we consider the remaining terms on the right-hand side of Eq.\,(\ref{eq:4i}).
For determining the explicit form of $j_{\mu}^{\Psi(0,2) \,a}$ it is necessary
to derive the expression $\delta n^{\Psi (0,2)\,a}_{\pm}$ according
to Eq.\,(\ref{eq:2d}).
For this purpose, we substitute the expansion (\ref{eq:2i}) into Eq.\,(\ref{eq:2yyy}),
keeping only the terms with quark fields $\bar{\psi},\,
\psi$ on the right-hand side. Performing the Fourier transformation and taking into account
Eq.\,(\ref{eq:3q}), we find
\[
\delta n^{\Psi (0,2)\, a}_{\pm}({\mathbf k}, p) =
\pm\,  g^2 C_F \, \frac{1}{2 \epsilon_k} \,[N(\epsilon_k) + n(\epsilon_k)]
\int\!\frac{1}{(v\cdot q_1 - i \epsilon)(v\cdot q_2 + i \epsilon)} \,
\bar{\psi}(-q_1)\!\not\!v t^a \psi(q_2)
\delta(p + q_1 - q_2) dq_1 dq_2.
\]
Substituting the obtained expression into Eq.\,(\ref{eq:2d}) and performing the radial
integration over $d \epsilon_k$, we find the required current correction
$j^{\Psi(0,2)\,a}_{\mu}$:
\begin{equation}
j^{\Psi(0,2)\,a}_{\mu}(p) = -\,g\,\omega_0^2\!\int\!\frac{d \Omega}{4 \pi}\,
\frac{v_{\mu}}{(v\cdot p + i \epsilon)} \bigg( \frac{1}
{v\cdot q_2 + i \epsilon} - \frac{1}{v\cdot q_1 - i \epsilon} \bigg)
\bar{\psi}(-q_1)\!\!\not\!v t^a \psi(q_2)
\delta(p + q_1 - q_2) dq_1 dq_2.
\label{eq:4o}
\end{equation}
Taking into account the discussion above, we obtain the following equation,
instead of Eq.\,(\ref{eq:4i}):
\begin{equation}
\,^{\ast} {\mathcal D}^{-1 \, \mu \nu}(p) A_{\nu}^a(p) =
g\!\!\int \bar{\psi}(-q_1)\!\,^{\ast}\Gamma^{(G)\mu}(p;q_1,-q_2)t^a
\psi(q_2)\,\delta(p + q_1 - q_2) dq_1 dq_2.
\label{eq:4p}
\end{equation}
Here,
\begin{equation}
\,^{\ast}\Gamma^{(G)}_{\mu}(p;q_1,q_2) = \gamma_{\mu} +
\delta \Gamma^{(G)}_{\mu}(p;q_1,q_2),
\label{eq:4a}
\end{equation}
\[
\delta \Gamma^{(G)}_{\mu}(p;q_1,q_2) =
- \, \omega^2_0\!\int\!\frac{d \Omega}{4 \pi} \, \frac{v_{\mu} \not\!v}
{(v\cdot q_1 - i \epsilon)(v\cdot q_2 - i \epsilon)}
\]
is the effective two-quarks--one-gluon vertex function, where now (in the coordinate
representation) the time arguments satisfy $X^0 \geq {\rm max}\,(Y_1^0,Y_2^0)$
and the chronological order of $Y_1$ and $Y_2$ is arbitrary \cite{bla2}. The time
argument of an external gluon leg incoming in the vertex is largest, as indicated
by superscript $(G)$. The approximate solution of Eq.\,(\ref{eq:4p}) (with
accuracy required for our further calculations) is of the form
\begin{equation}
A^a_{\mu}(p) = A^{(0)\,a}_{\mu}(p) +
g\,^{\ast} {\cal D}_{\mu \nu}(p)
\!\int \bar{\psi}^{(0)}(-q_1) \,^{\ast}\Gamma^{(G)\, \nu}(p;q_1,-q_2) \,
t^a \psi^{(0)}(q_2)\,\delta(p + q_1 - q_2) dq_1 dq_2.
\label{eq:4s}
\end{equation}

At the end of this section we present the expression for the next term in the
expansion of current $j^{\Psi}_{\mu}$, which is needed for deriving
general kinetic equation for soft-gluon modes in Sec.\,VI. Performing similar
calculations, we obtain
\[
j^{\Psi (1,2)\,a}_{\mu}(p) =
- \, g^2 \omega^2_0\!\int\!\frac{{\rm d} \Omega}{4 \pi} \, \
\frac{v^{\mu} v^{\nu} (\not\!v)_{\alpha \beta}}{v\cdot p + i \epsilon}
\Big[ \,(T^a)^{bc} (t^c)^{ij} \frac{1}
{(v\cdot q_1 - i \epsilon)(v\cdot q_2 + i \epsilon)}
\]
\begin{equation}
+ (t^at^b)^{ij} \frac{1}{(v\cdot (q_2 + p_1) + i \epsilon)(v\cdot q_2 + i \epsilon)}
- (t^bt^a)^{ij} \frac{1}
{(v\cdot (q_1 -p_1) - i \epsilon)(v\cdot q_1 - i \epsilon)} \, \Big]
\label{eq:4d}
\end{equation}
\[
\times A^b_{\nu}(p_1) \bar{\psi}^i_{\alpha}(-q_1) \psi^j_{\beta}(q_2)\,
\delta(p - p_1 + q_1 - q_2) dp_1 dq_1 dq_2,
\]
where $(T^a)^{bc} = -if^{abc}$. The appearance of the term with the production
of generators of the ${\rm SU}(N_c)$ group in other representations (the adjoint and
the fundamental ones) is a special feature of the last expression.
This term is not vanishing only in the processes of the
higherorder in field powers, the processes of Boltzmann type, i.e.,
the scattering processes between the soft-fermion and -gluon excitations
(see Conclusion).

\section{Characteristic amplitudes of the soft fields}
\setcounter{equation}{0}

In this section we will estimate the typical amplitudes of the soft fields,
both in coordinate and in momentum space, such that the truncation of the
series expansions (\ref{eq:2p}), (\ref{eq:2a}), and (\ref{eq:2f}) can be made
and in the long run one can derive a closed system of gauge-invariant kinetic
equations for soft plasma excitations.

For this purpose, let us discuss in more detail the approximation scheme
used in this paper. In fact, the value $\delta\equiv gRA_{\mu}$ is a
(dimensionless) parameter of the expansion in powers of nonlinearities in
Eqs.\,(\ref{eq:2q})\,--\,(\ref{eq:2yyyy}), where $R\sim\partial_{X}^{-1}$ is
a typical length for off-equilibrium deviations. It follows, for example,
from rewriting the equations of motion for the soft fields (\ref{eq:2q}) and
(\ref{eq:2w}) in the form (\ref{eq:2h}) and (\ref{eq:2j}), where the gauge fields
$A_{\mu}(X)$ are expanded from the covariant derivatives
$D_{\mu}=\partial_{\mu} + igA_{\mu}(X)$ and the field strength tensor
$F_{\mu\nu}(X)$. The disadvantage of the proposed iterated scheme for deriving
the terms in expansions (\ref{eq:2p}), (\ref{eq:2a}), and (\ref{eq:2f}) is
explicit breaking of the non-Abelian gauge symmetry of the dynamical equations
for the soft fluctuations (\ref{eq:2y})\,--\,(\ref{eq:2yyyy}) at each step
of the approximate calculations. Here, the answer to the nontrivial question on
the recovery of a gauge symmetry depends directly on the order of
the magnitudes of the soft fields, the dynamics of which we would like to describe.
Below, we discuss two relevant cases.

Let us assume that the amplitudes of soft-boson and -fermion excitations are
of orders $g$ coinciding with the orders of $A_{\mu}$ and $\psi$ presented
in Ref.\,\cite{bla2}, i.e.,
\[
\vert A_{\mu}(X) \vert \sim T \, \; (\vert A_{\mu}(p) \vert \sim 1/g(gT)^3),
\]
\[
\vert \bar{\psi}(X) \vert \vert \psi(X) \vert \sim g T^3 \; \,
(\vert \bar{\psi}(q) \vert \vert \psi(q) \vert \sim 1/g^2(g T)^5).
\]
In this case, by using the obtained expressions for the terms in the expansion of
the induced source $\eta$ (\ref{eq:3w}), (\ref{eq:4w}), and (\ref{eq:4e}) and
the color currents $j^{A}_{\mu}$ \cite{mar2} and
$j_{\mu}^{\Psi}$ (\ref{eq:4o}) and (\ref{eq:4d}), we have the following
estimates:
$$
\hspace{0.9cm} \eta^{(0,1)}(q) \sim \eta^{(1,1)}(q) \sim \eta^{(2,1)}(q) \sim
\ldots \sim gT\, \vert \psi(q) \vert,
$$
$$
j^{A\,(1)}_{\mu}(p) \sim j^{A\,(2)}_{\mu}(p) \sim j^{A\,(3)}_{\mu}(p) \sim
\ldots \sim \frac{1}{g^2T},
$$
$$
j^{\Psi\,(0,2)}_{\mu}(p) \sim j^{\Psi\,(1,2)}_{\mu}(p) \sim
\ldots \sim g(gT)^4 \vert \bar{\psi}(q) \vert \vert \psi(q) \vert\,
\bigg( \sim \frac{1}{g^2T} \bigg).
$$
Thus, all terms in each expansion (\ref{eq:2p}), (\ref{eq:2a}), and
(\ref{eq:2f}) are of the same order of magnitude and the problem of resummation
of all the relevant contributions arises. Thus, gauge symmetry is recovered.
In a pure-gauge case such an approach is represented, for instance, in the work of
Jackiw and Nair \cite{jac} dealing with the derivation of the non-Abelian 
version of the Kubo formula.
Here, the explicit expression for the induced current $j_{\mu}^A(p)$,
including the contribution of all higher point functions, that actually coincide with
the iterations $j_{\mu}^{A(1)},\;j_{\mu}^{A(2)},\;\ldots$ \cite{mar2}, is defined.
The requirement of gauge invariance leads to the completely nonlinear theory and
principal impossibility of deriving closed kinetic equations for
number densities (i.e. two-point functions) of the soft-fermion and
-gluon excitations.

In this work, we restrict consideration only to a finite number of
terms in expansions of the induced source and color currents. This
imposes more rigid restrictions on the order of the magnitudes of soft fields.
Blaizot and Iancu \cite{ian} (Sec.\,1.1) showed in the special case
when the soft fields were thermal fluctuations at the soft scale $gT$
(namely, this situation takes place in our case), their typical
amplitudes would be of the orders
\[
\vert A_{\mu}(X) \vert \sim g^{1/2}T \;\,
\bigg(\vert A_{\mu}(p) \vert \sim \frac{1}{g^{1/2}(gT)^3}\bigg),
\]
\[
\vert \bar{\psi}(X) \vert \vert \psi(X) \vert \sim g^2T^3 \;\,
\bigg(\vert \bar{\psi}(q) \vert \vert \psi(q) \vert
\sim \frac{1}{g(g T)^5}\bigg).
\]
In this case the parameter of expansions $\delta\sim g^{1/2}\ll 1$ and
we have the estimates
$$
\hspace{2cm} \eta^{(0,1)}(q) \sim gT\, \vert \psi(q) \vert,\;
\eta^{(1,1)}(q) \sim g^{3/2}T\, \vert \psi(q) \vert,\;
\eta^{(2,1)}(q) \sim g^{5/2}T\, \vert \psi(q) \vert, \ldots \, ,
$$
$$
j^{A\,(1)}_{\mu}(p) \sim \frac{1}{g^{3/2}T},\;
j^{A\,(2)}_{\mu}(p) \sim \frac{1}{gT},\;
j^{A\,(3)}_{\mu}(p) \sim \frac{1}{g^{1/2}T}, \ldots \,,
$$
$$
j^{\Psi\,(0,2)}_{\mu}(p) \sim g(gT)^4 \vert \bar{\psi}(q) \vert \vert
\psi(q) \vert\, \bigg(\sim\frac{1}{gT}\bigg), \;
j^{\Psi\,(1,2)}_{\mu}(p) \sim g^{3/2}(gT)^4 \vert \bar{\psi}(q) \vert
\vert \psi(q) \vert\, \bigg(\sim\frac{1}{g^{1/2}T}\bigg),
$$
$$
\ldots \,,
$$
i.e., every successive term in the source and
current expansions is suppressed by one power of $g^{1/2}$ in comparison
with the preceding ones, and the use of the perturbation theory is therefore
justified. By taking into account the
nonlinear interaction between soft fields and hard particles at leading order
in $g$, it is sufficient to keep only the first three terms in the expansions
(\ref{eq:2p}) and (\ref{eq:2f}), and first two terms in expansion (\ref{eq:2a}),
as mentioned at the end of Sec.\,II.
The recovery of a gauge symmetry here, occurs another way, by
the account of weak correlation in the calculation of the expectation value
of three soft fields in a slightly heterogeneous, slightly nonstationary
quark-gluon plasma (see the following section). In this case we derive a closed system
of the gauge-invariant kinetic equations that describe the evolution of the
off-equilibrium deviations in the number densities, and obtain the damping rates
from the collision terms, which are closely allied in form to the
corresponding damping rates in the HTL approximation \cite{bra2,kob}.

\section{Generalized kinetic equation for soft
fermion excitations}
\setcounter{equation}{0}

Now we turn to the equation for the soft-fermion excitations (\ref{eq:4r}). We
substitute Eqs.\,(\ref{eq:4u}) and (\ref{eq:4s}) into third-order correlation
functions entering into Eq.\,(\ref{eq:4r}).
Because of the fact that $\bar{\psi}^{(0)}, \, \psi^{(0)}$, and
$A_{\mu}^{(0)}$ represent the amplitudes of entirely uncorrelated waves,
the correlation function $\langle A_{\mu}^{(0)} \bar{\psi}^{(0)} \psi^{(0)} \rangle$
drops out. In this case, each term in
$\langle \bar{\psi}^i(-q)A^{a\mu}(p_1)(\,^{\ast}\Gamma^{(Q)\,a}_{\mu}
\psi(q_1))^j \rangle$ and
$\langle (\bar{\psi}(-q_1)
\,^{\ast}\Gamma^{(Q)\,a}_{\mu})^iA^{\ast\,a\mu}(p_1)\psi^j(q^{\prime})\rangle$
should be defined more exactly. In the fourth-order correlation functions, within the
accepted accuracy, we replace the fields $\bar{\psi},\, \psi$, and $A_{\mu}$ by
$\bar{\psi}^{(0)},\, \psi^{(0)}$, and $A_{\mu}^{(0)}$.

Furthermore, we make the correlation decoupling of the fourth-order 
correlation functions in terms of the pair correlation functions by the rules
$$
\langle \bar{\psi}^{(0)}(-q_1) \psi^{(0)}(q_2)
\bar{\psi}^{(0)}(-q_3) \psi^{(0)}(q_4) \rangle =
\langle \bar{\psi}^{(0)}(-q_1) \psi^{(0)}(q_2) \rangle
\langle \bar{\psi}^{(0)}(-q_3) \psi^{(0)}(q_4) \rangle
$$
$$
- \, \langle \bar{\psi}^{(0)}(-q_1) \psi^{(0)}(q_4) \rangle
\langle \bar{\psi}^{(0)}(-q_3) \psi^{(0)}(q_2) \rangle,
$$
$$
\langle A^{(0)}(p_1) A^{(0)}(p_2) \bar{\psi}^{(0)}(-q_1)
\psi^{(0)}(q_2) \rangle
= \langle A^{(0)}(p_1) A^{(0)}(p_2) \rangle
\langle \bar{\psi}^{(0)}(-q_1) \psi^{(0)}(q_2) \rangle.
$$
Here, we suppress the color, spinor, and Lorentz indices, and take into
consideration the
Grassmanian nature of the quark field. Setting $\Upsilon_{\beta \alpha}^{ij} =
\delta^{ji} \Upsilon_{\beta \alpha}$ and $I^{ab}_{\mu \nu} = \delta^{ab}
I_{\mu \nu}$, taking into account
$\delta^{ab}\delta\Gamma_{\mu\nu}^{ab}(p_1,p_2;q_1,q_2)\equiv
C_F\delta\Gamma_{\mu\nu}(p_1,p_2;q_1,q_2)$, after cumbersome calculations,
we come to the following generalized kinetic equation\footnote{The term
{\it generalized} here means that we have not yet restricted ourselves to any
mass-shell conditions.}  for the soft-fermionic excitation QGP, 
instead of Eq.(\ref{eq:4r})
\begin{equation}
{\rm tr} \bigg( \frac{\partial}{\partial q^{\mu}} [\,-\!\not\!q +
\delta \Sigma^{\rm H}(q)]
\frac{\partial \Upsilon(q,x)}{\partial x_{\mu}} \bigg) =
2i\,{\rm tr}\,( \delta \Sigma^{\rm A}(q) \Upsilon(q,x))
\label{eq:6q}
\end{equation}
$$
+\,2g^2 C_F \Bigg(\!\int\!\!dq_1\, {\rm Im}\Big[\,^{\ast}{\mathcal D}_{\mu \nu}(q - q_1)
\,{\rm tr}\Big\{
\Upsilon(q) \,^{\ast}\Gamma^{(Q)\,\mu}(q-q_1;q_1,-q)
\Upsilon(q_1) \,^{\ast}\Gamma^{(G)\,\nu}(q-q_1;q_1,-q)\Big\}\Big]
$$
$$
- \int\!dq_1 dp_1\,I_{\mu \nu}(p_1)\,{\rm Im} \Big[\, {\rm tr}
\Big\{\!\,^{\ast}\Gamma^{(Q)\,\mu}(-p_1;-q_1,q)\!\,^{\ast}\!S(q)\!
\,^{\ast}\Gamma^{(Q)\,\nu}(p_1;q_1,-q)\Upsilon(q_1) \Big\} \Big]
\delta(q-q_1-p_1)
$$
$$
-\!\!\int\!\!dp\,I_{\mu \nu}(p)\,{\rm Im}\Big[
\,{\rm tr} \Big\{ \Big( \delta \Gamma^{(Q)\, \mu \nu}(p,-p;q,-q) -\!
\,^{\ast}\Gamma^{(Q)\,\mu}(p;q-p,-q)\!\,^{\ast}\!S(q-p)\!
\,^{\ast}\Gamma^{(Q)\,\nu}(-p;q,-q+p)
$$
$$
- \,^{\ast}\Gamma^{(Q)\,\nu}(-p;q+p,-q) \,^{\ast}\!S(q+p)
\,^{\ast}\Gamma^{(Q)\,\mu}(p;q,-q-p) \Big) \Upsilon(q) \Big\}\Big]\Bigg).
$$
On the right-hand side of Eq.\,(\ref{eq:6q}) the $x$ dependence of
$\Upsilon_{\beta \alpha}(q)$ and
$I_{\mu \nu}(p)$ is understood, although not explicitly written.

In deriving Eq.\,(\ref{eq:6q})
we use the following properties of effective two-quark--one-gluon and
two-quark--two-gluon vertex
functions, which immediately follows from initial definitions (\ref{eq:4t}),
(\ref{eq:4y}), (\ref{eq:4yy}) and (\ref{eq:4a})
$$
\gamma^0 \,^{\ast}\Gamma^{(Q)\dagger}_{\mu}(p;q_1,q_2) \gamma^0
= \,^{\ast}\Gamma^{(Q)}_{\mu}(p;q_2,q_1)
= \,^{\ast}\Gamma^{(Q)}_{\mu}(-p;-q_1,-q_2),
$$
$$
\hspace{1cm}\gamma^0 \,^{\ast}\Gamma^{(G)\dagger}_{\mu}(p;q_1,q_2) \gamma^0
= \,^{\ast}\Gamma^{(G)}_{\mu}(-p;-q_1,-q_2)
= \,^{\ast}\Gamma^{(G)}_{\mu}(-p;-q_2,-q_1),
$$
$$
\gamma^0 \delta \Gamma^{(Q)\dagger}_{\mu \nu}(p_1,p_2;q_1,q_2) \gamma^0
= -\,\delta\Gamma^{(Q)}_{\mu \nu}(p_1,p_2;q_2,q_1)
= -\,\delta\Gamma^{(Q)}_{\mu \nu}(-p_1,-p_2;-q_1,-q_2).
$$
Moreover, we assume that under Hermitian conjugation the Wigner function
$\Upsilon(q,x)$ behaves like an ordinary $\gamma$ matrix:
\begin{equation}
\gamma^0 \Upsilon^{\dagger}(q,x) \gamma^0 = \Upsilon(q,x).
\label{eq:6w}
\end{equation}
Below this property is shown to ensure that the physical variable such as the
fermion number density is real.

Equation (\ref{eq:6q}) is incomplete, since the unknown function $I_{\mu \nu}(p,x)$
enters this equation. This function obeys the kinetic equation,
which is similar to Eq.\,(\ref{eq:6q}). The soft-field equation (\ref{eq:2j})
is initial for derivation of this
kinetic equation. We use the obtained expressions for the induced color current
$j^{\Psi}_{\mu}$ (\ref{eq:4o}) and (\ref{eq:4d}) and corresponding expressions
for the induced color current $j^{A}_{\mu}$ defined in Ref.\,\cite{mar2}. Performing
calculations similar to previous ones and keeping in the right-hand side only the terms
responsible for the nonlinear interaction between soft-fermion excitations
and soft-gluon ones (purely gauge sector was considered in detail
in Ref.\,\cite{mar2}), we come to the following generalized kinetic equation for 
soft-gluon excitation QGP's (again, the $x$ dependence of most quantities is suppressed)
\begin{equation}
\frac{\partial}{\partial p_{\lambda}}[\,p^{2}g^{\mu \nu} - (1 + \xi^{-1})
p^{\mu}p^{\nu} + \delta \Pi^{{\rm H}\,\mu\nu}(p)]
\frac{\partial I_{\mu \nu}(p,x)}{\partial x^{\lambda}} =
-\,2i\,\delta\Pi^{{\rm A}\,\mu\nu}(p)I_{\mu\nu}(p,x)
\label{eq:6e}
\end{equation}
$$
-\,2g^2 T_F \Bigg({\rm Im} \bigl( \,^{\ast}{\cal D}_{\mu \nu}(p)\bigr)\!
\int\!dq_1dq_2 \,{\rm tr}\Big\{
\Upsilon(q_1) \,^{\ast}\Gamma^{(G)\,\mu}(p;q_1,-q_2)
\Upsilon(q_2) \,^{\ast}\Gamma^{(G)\,\nu}(-p;-q_1,q_2)\Big\}
$$
$$
+\,I_{\mu \nu}(p)\!\!\int\!\!dq\,{\rm Im} \Big[
\,{\rm tr} \Big\{\!\Big( \delta \Gamma^{(G)\, \mu \nu}(p,-p;q,-q) +
\!\,^{\ast}\Gamma^{(Q)\,\nu}(-p;p-q,q)\!\,^{\ast}\!S(p-q)
\!\,^{\ast}\Gamma^{(G)\,\mu}(p;-p+q,\!-q)
$$
$$
- \,^{\ast}\Gamma^{(G)\,\mu}(p;q,-p-q) \,^{\ast}\!S(p+q)
\,^{\ast}\Gamma^{(Q)\,\nu}(-p;p+q,-q) \Big) \Upsilon(q) \Big\}\Big]\Bigg).
$$
Here, $T_F$ is the index of the fundamental representation,
$\delta\Pi^{{\rm H}\,\mu\nu}(p)$ and $\delta\Pi^{{\rm A}\,\mu\nu}(p)$
are the Hermitian and anti-Hermitian parts of the soft-gluon self-energy,
respectively, and $\delta\Gamma^{(G)\,\mu\nu}$ is the effective
two-quark--two-gluon vertex function defined by an expression
\begin{equation}
\delta \Gamma^{(G)\, \mu \nu}(p_1,p_2;q_1,q_2)=
-\, \omega_0^2\!\int\!\frac{{\rm d} \Omega}{4 \pi}\,
\frac{v^{\mu} v^{\nu} \not\!v}{(v\cdot q_1 - i \epsilon)
(v\cdot q_2 - i \epsilon)}
\bigg(\frac{1}{v\cdot (q_1+p_1) + i \epsilon}
+ \,\frac{1}{v\cdot (q_1+p_2) - i \epsilon} \bigg),
\label{eq:6r}
\end{equation}
where the first term in large parentheses satisfies
the chronological order: $X_1^0 \geq Y_2^0, \, X_1^0 \geq X_2^0 \geq Y_1^0$
in the coordinate representation,
and the second term satisfies the chronological order
$X_1^0 \geq Y_1^0, \, X_1^0 \geq X_2^0 \geq Y_2^0$; i.e., the time argument
of one of external gluon legs coming into the vertex is largest.
The mean of the superscript $(G)$ is evident here. Notice that in deriving
Eq.\,(\ref{eq:6e}) we use the properties of Hermitian conjugate of the vertex
function (\ref{eq:6r}):
$$
\gamma^0 \delta \Gamma^{(G)\dagger}_{\mu \nu}(p_1,p_2;q_1,q_2) \gamma^0
= \delta\Gamma^{(G)}_{\mu \nu}(p_2,p_1;q_1,q_2)
= -\,\delta\Gamma^{(G)}_{\mu \nu}(p_1,p_2;q_2,q_1).
$$

In closing this section we again call attention to the
chronological order of arguments in the vertex functions entering into the
right-hand side of generalized equations (\ref{eq:6q}) and (\ref{eq:6e}),
which appear here in nonevident manner. Also, we call attention to the
fact that the right-hand side of Eq.\,(\ref{eq:6e}) is proportional to the third
group invariant $T_F$ as distinct from the right-hand side of Eq.\,(\ref{eq:6q}),
proportional to the quark Casimir invariant $C_F$ and the kinetic equation for
$I_{\mu \nu}(p,x)$ with purely gauge sector on the right-hand side \cite{mar2},
proportional to the gluon Casimir invariant $C_A$.

\section{Classification of types of nonlinear wave interactions.
Plasmino and plasmon kinetic equations}
\setcounter{equation}{0}

Now we perform a preliminary analysis of the right-hand sides of the generalized
kinetic equations (\ref{eq:6q}) and (\ref{eq:6e}), the purpose of which
is the identification of the terms with specific physical processes described
by them. To establish this connection it is
convenient to use the Feynman diagrams, defining the contribution of the
leading order in the coupling constant to the damping of soft-fermion
excitations in framework of resummed perturbation theory.
There are known \cite{bra1} to be two diagrams (Figs.\,1(a) and 1(b))
\begin{figure}[hbtp]
\centering
\includegraphics[scale=0.5]{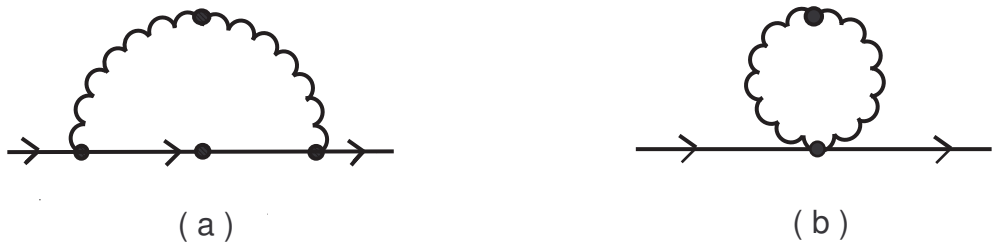}
\caption{\small One-loop diagrams that contribute to the imaginary part of the
self-energy of quarks. The blob stands for HTL resummation.}
\label{f1:pcx}
\end{figure}
with soft loop
momentum that contribute at order $g$ to the effective self-energy for quarks:
the usual self-energy graph at one-loop order (Fig.\,1(a)),
except that all of the vertices
and propagators are effective, and the effective gluon tadpole on the soft-fermion
line (Fig.\,1(b)), which is special to the effective expansion.

Let us consider the second term\footnote{As was mentioned at the end of Sec.\,III,
the first term on the right-hand side of Eq.\,(\ref{eq:6q}), involving the
imaginary part of the HTL self-energy, describes linear Landau damping.
It vanishes, when one studies the decay of on-shell excitations and therefore can
be omitted.}
on the right-hand side of Eq.\,(\ref{eq:6q}).
Physically, this contribution just corresponds to the
stimulated scattering processes of soft-fermion excitations by 
hard-particle QGP through a soft-virtual gluon oscillation,
without the change of statistic type of both soft and hard
excitations, i.e.,
\[
{\rm q} + {\rm Q}\,(\bar{\rm Q}) \rightarrow
{\rm q} + {\rm Q}\,(\bar{\rm Q}), \quad
{\rm q} + {\rm G} \rightarrow {\rm q} + {\rm G} ,
\]
and similarly for $\bar{\rm q}$. Here, ${\rm q}$ represents fermionic collective excitations
(we do not distinguish between normal-particle excitations and plasminos) and
${\rm Q},\,\bar{\rm Q}$, and ${\rm G}$ are quark, antiquark, and gluon excitations with
characteristic
momenta of order $T$. It can be easily inferred if cutting the usual self-energy
graph before spelling out the effective gluon propagator, i.e., inserting a hard
bubble along the gluon line, as depicted in Fig.\,2.

The third term on the right-hand side of Eq.\,(\ref{eq:6q}) is associated
with the so-called decay processes depicted in Fig.\,3
following from the cutting of the usual self-energy graph
$$
{\rm q} \rightleftharpoons {\rm q} + {\rm g},\quad
{\rm q} + \bar{\rm q} \rightarrow {\rm g},
$$
where g is the gluonic collective excitations (here, we do not distinguish between
longitudinal and transverse excitations).
Similar decay processes have been studied in detail,
since they are immediately connected with the processes of
\begin{figure}[hbtp]
\centering
\includegraphics[scale=0.5]{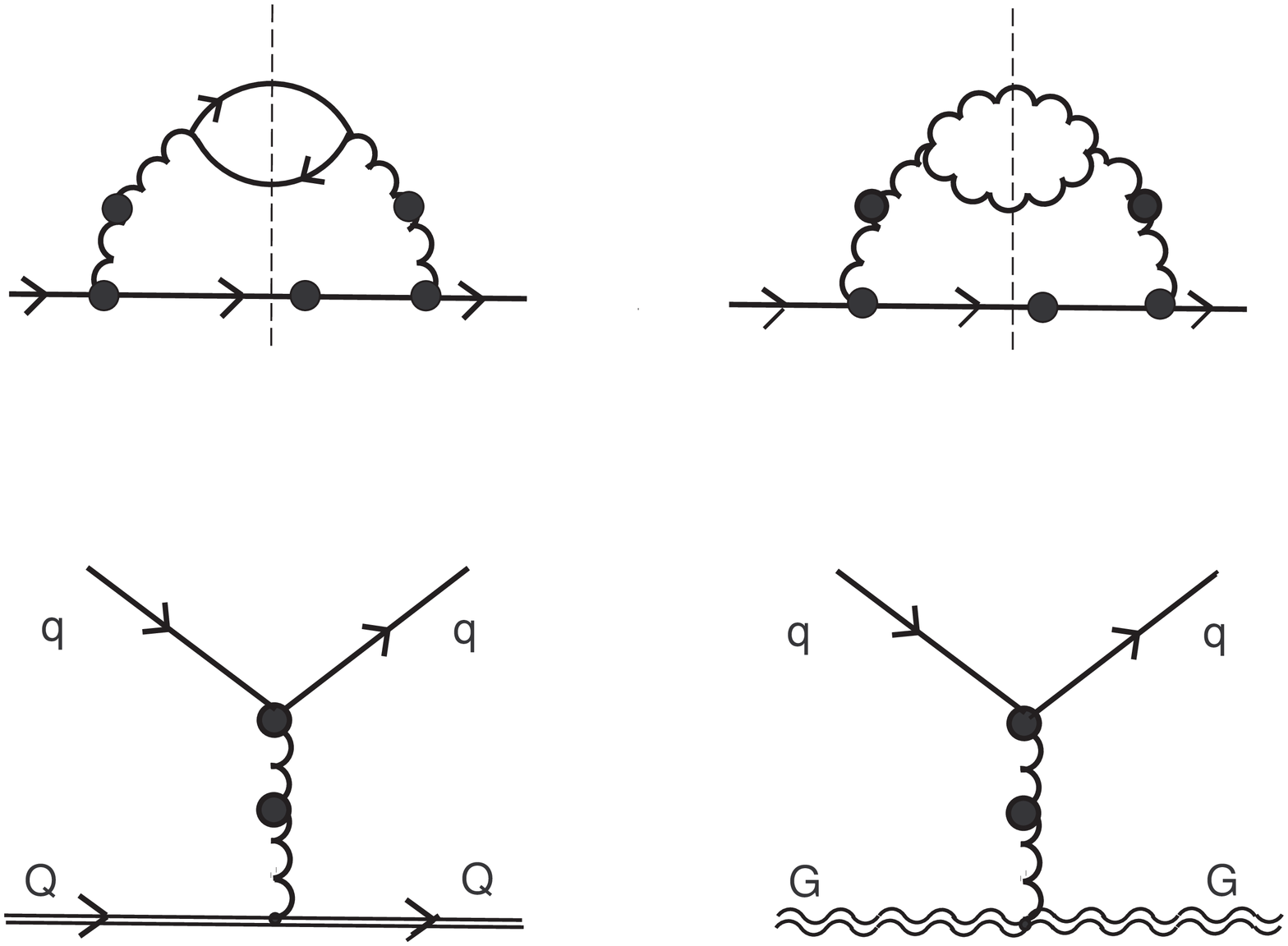}
\caption{\small The process of the stimulated scattering of soft fermionic
excitations by hard QGP particles through a resummed gluon
propagator\,$\,^{\ast}{\cal D}$, where a vertex of a
three-soft-wave interaction
is induced by\,$\,^{\ast}\Gamma^{(G)}_{\mu}$.
The double lines denote hard particles.}
\label{f2:eps}
\end{figure}
\begin{figure}[t]
\centering
\includegraphics[scale=0.5]{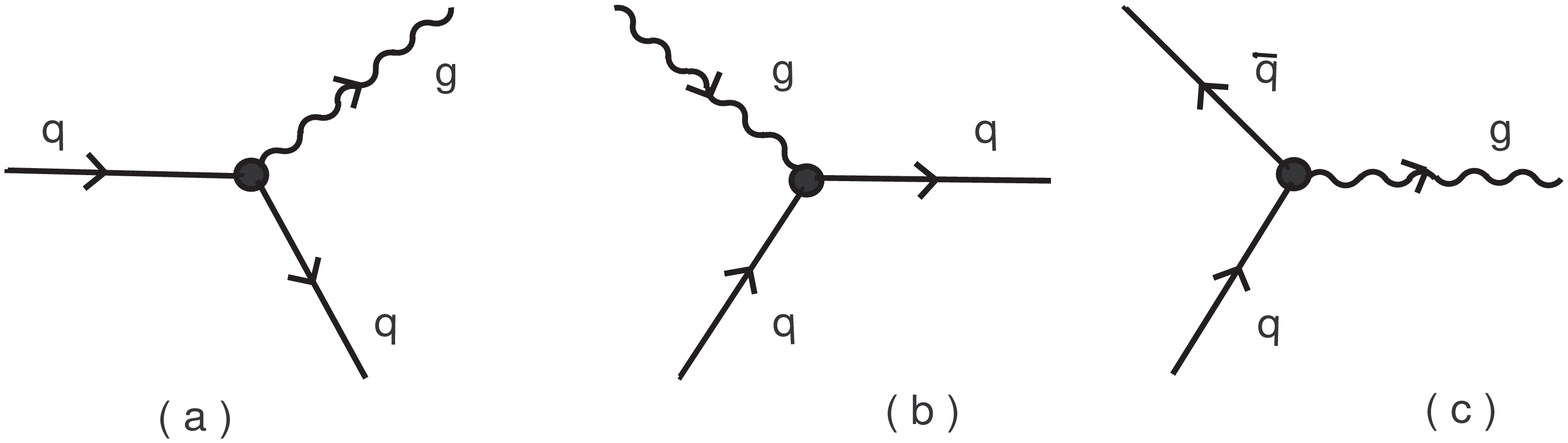}
\caption{\small (a), (b) Radiation (absorption) of the soft-boson
excitations by soft-fermion ones; (c) annihilation of the 
soft-fermion--antifermion excitations into soft-boson excitations.}
\label{f3:eps}
\end{figure}
soft-dilepton productions in hot QCD plasma. In the high-temperature
QCD plasma in the approach based on using dispersion relation methods \cite{bra4}, these
processes are connected with the so-called pole-pole terms.

Finally, the remaining terms grouped on the right-hand side of Eq.\,(\ref{eq:6q})
are connected
with the more interesting process of the nonlinear wave interaction with
stimulated scattering of soft-fermion excitations by hard-particle QGP's 
varying with the change of statistics of excitations:
\[
{\rm q}\,(\bar{\rm q}) + {\rm G} \rightarrow
{\rm g} + {\rm Q}\,(\bar{\rm Q}), \quad
{\rm q}\,(\bar{\rm q}) + \bar{\rm Q}\,({\rm Q}) \rightarrow
{\rm g} + {\rm G}.
\]
It is defined by two different
physical processes. The first represents the Compton
scattering type process and is connected with the effective two-quark--two-gluon vertex
function (more exactly, with its imaginary part)
on the right-hand side of Eq.\,(\ref{eq:6q}). This corresponds diagrammatically
to the appropriate cutting of the gluon ``tadpole'' graph, before illustrating the
effective two-quark--two-gluon vertex function, as depicted in Fig.\,4.
\begin{figure}[htbp]
\centering
\includegraphics[scale=0.5]{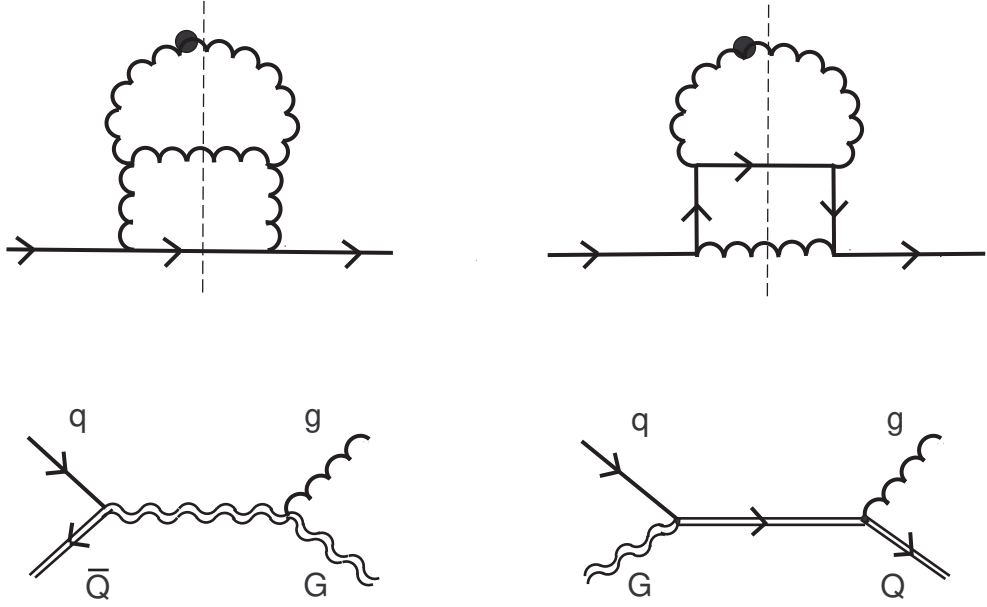}
\caption{\small The Compton-like scattering of soft-fermion excitations
by a hard QGP particle with a change in statistics of soft excitations (s channel).}
\label{f4:eps}
\end{figure}
The second process connected with the remaining terms 
defines the scattering of a fermionic quantum oscillation through a virtual
soft-fermion oscillation. Diagrammatically this process of scattering is defined by
the corresponding cutting of the usual self-energy graph before illustrating the
effective quark propagator as in Fig.\,5.
\begin{figure}[t]
\centering
\includegraphics[scale=0.5]{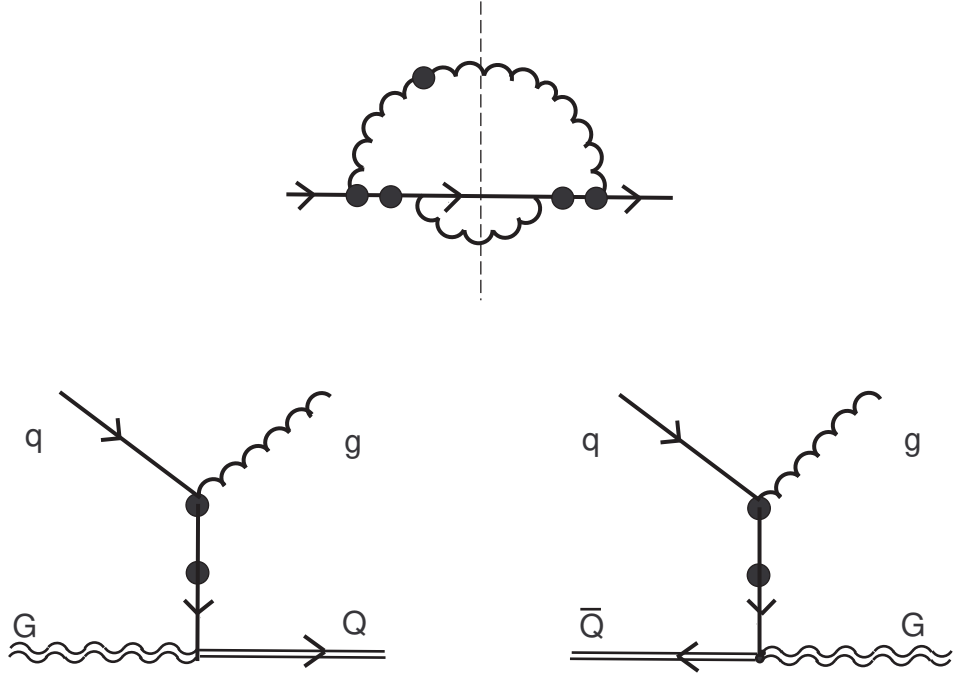}
\caption{\small The process of the stimulated scattering process of the 
soft-fermion excitations by hard QGP particles through a resummed quark
propagator\,$\,^{\ast}\!S$, where
a vertex of a three-soft-wave interaction is induced
by\,$\,^{\ast}\Gamma^{(Q)}_{\mu}$ (t channel).}
\label{f5:eps}
\end{figure}
As it will be shown
in the next section, the terms corresponding to the processes drawn in
Figs.\,4 and 5
form a gauge-invariant function. The process, described by these terms,
will be called further {\it the nonlinear Landau damping of the soft-fermion
excitations} by virtue of the great similarity of the expression defining this
process and the corresponding expression in a purely gauge case \cite{mar2}.

Here, it is necessary to note that the terms on the right-hand side of kinetic
equation (\ref{eq:6q}) corresponding to the process diagrammatically
shown in Fig.\,5 are very close in structure to the terms connected with stimulated
scattering processes, without changing the statistic type
and which are diagrammatically given in Fig.\,2.
However, in the last case the Compton
scattering type contribution is absent, which, in turn, is associated with the absence
of the effective quark tadpole on the soft-quark line. For this reason, we not
assign this process to the one of the nonlinear Landau damping.

Now we turn to the right-hand side of generalized kinetic equation (\ref{eq:6e}). The
second term on the right-hand side of Eq.\,(\ref{eq:6e}) is defined by
time reversal of the decay processes that are drawn in Figs.\,3(a) and 3(c).
The remaining terms
on the right-hand side of Eq.\,(\ref{eq:6e}) are associated with the processes defined by
the time reversal of the processes depicted in Figs.\,4 and 5,
i.e., with the process of the nonlinear Landau damping of the soft-gluon
excitations.

In the remainder of this work, we restrict our attention to the detailed
study of the process of nonlinear Landau damping, and moreover
we restrict ourselves only to the process of the nonlinear interaction between
plasminos and plasmons, i.e., purely collective excitations in hot QCD plasmas.
The exception is only Sec.\,XII, where we consider the difficulties connected with
the calculation of the nonlinear Landau damping rate for normal quark excitations in
the vicinity of the light cone.

As it was mentioned in the Introduction in global equilibrium QGP's for the case of
massless quarks and for zero chemical potential there are two branches of fermion
excitations with a positive energy. In this connection we define the Wigner
function $\Upsilon(q, x)$ in the form of an expansion
\begin{equation}
\Upsilon(q,x) = h_{+}(\hat{\bf q}) \tilde{\Upsilon}_{+}(q,x) +
h_{-}(\hat{\bf q}) \tilde{\Upsilon}_{-}(q,x),
\label{eq:7e}
\end{equation}
where $\tilde{\Upsilon}_{\pm}$ are certain scalar functions and
$h_{\pm}(\hat{\bf q}) = (\gamma^0 \mp \hat{{\bf q}}\cdot {\bf \gamma})/2$ with
$\hat{{\bf q}} \equiv {\bf q}/\vert {\bf q} \vert.$  By the condition (\ref{eq:6w})
the functions $\tilde{\Upsilon}_{\pm}$ are real. Let us define dependence on
variable $q^0$.

For this purpose we omit nonlinear terms and the anti-Hermitian part of the
quark self-energy in Eq.\,(\ref{eq:3r}). Substituting the function
$\delta^{ji}(h_{\pm}(\hat{{\mathbf q}}))_{\beta \alpha} \tilde{\Upsilon}_{\pm}(q,x)
\,\delta(q-q^{\prime})$ instead of $\langle \bar{\psi}^i_{\alpha}(-q)\psi^j_{\beta}
(q^{\prime})\rangle$, we find
$$
-\,\bigl[\,q^0 \mp \bigl(\vert {\bf q} \vert + {\rm Re}\,\delta \Sigma_{\pm}(q)\bigr)
\bigr]\, \tilde{\Upsilon}_{\pm}(q,x) = 0,
$$
where
\begin{equation}
\delta \Sigma_{\pm}(q) = \pm\,\frac{1}{2}\,{\rm tr} \Bigl\{h_{\pm}(\hat{\mathbf q})
\,\delta \Sigma(q) \Bigr\} =
\frac{\omega_0^2}{\vert {\mathbf q} \vert}\biggl[\,1 - \biggl(1 \mp \frac{\vert {\mathbf q}
\vert }{q^0} \biggr) F\biggl(\frac{q^0}{\vert {\mathbf q} \vert} \biggr) \biggr]
\label{eq:7r}
\end{equation}
with
$$
F(z) = \frac{z}{2} \biggl[\,\ln \bigg{\vert} \frac{1 + z}{1 - z} \bigg{\vert}
- i \pi \theta(1 - \vert z \vert) \biggr].
$$
The solutions of these equations have the structure
\begin{equation}
\tilde{\Upsilon}_{\pm}(q,x) =
\Upsilon_{\pm}({\mathbf q},x)\,\delta(q^0\!- \omega_{\pm}({\mathbf q})) +
\Upsilon_{\mp}(-{\mathbf q},x)\,\delta(q^0\!+ \omega_{\mp}({\mathbf q})).
\label{eq:7t}
\end{equation}
Here, $\Upsilon_{\pm}({\mathbf q},x) \equiv \Upsilon^{\pm}_{\mathbf q}$ are certain
functions of a wave vector ${\mathbf q}$ and $\omega_{\pm}({\mathbf q})\equiv
\omega^{\pm}_{{\mathbf q}}$
are frequencies of the normal-particle excitations and plasminos, respectively.
The solutions (\ref{eq:7t}) describe what is called a
{\it quasiparticle approximation}.
In this approximation we assume that the off-equilibrium two-point functions
$\tilde{\Upsilon}_{\pm}(q,x)$
have the same mass-shell conditions as in equilibrium, with an off-equilibrium
deviation in the spectral densities $\Upsilon^{\pm}_{\mathbf q}$.

Furthermore, let us represent the function $I_{\mu \nu}(p,x)
=I_{\mu \nu}$ in the form of an expansion
\begin{equation}
I_{\mu \nu}= Q_{\mu \nu}(p)I_{p}^{l} + P_{\mu \nu}(p)I_{p}^{t}, \quad
I_{p}^{l,t} \equiv I^{l,t}(p,x) ,
\label{eq:7tt}
\end{equation}
where
$$
Q_{\mu \nu}(p) = \frac{\bar{u}_{\mu} (p) \bar{u}_{\nu} (p)}{\bar{u}^2(p)}, \quad
P_{\mu \nu} (p) = g_{\mu \nu} - D_{\mu \nu}(p) - Q_{\mu \nu}(p),
$$
$$
D_{\mu \nu} = p_{\mu} p_{\nu}/p^2, \quad
\bar{u}_{\mu} (p) = p^2 u_{\mu} - p_{\mu} (p u), \quad u_{\mu}=(1,0,0,0)
$$
are longitudinal and transverse projections in the Lorentz covariant form,
respectively, and we take the functions $I_p^{l,t}$ in the form of the quasiparticle
approximation
\begin{equation}
I_{p}^{l}=I_{\mathbf p}^{l} \delta ( p^0 - {\omega}_{\mathbf p}^{l}) +
I_{- \mathbf p}^{l} \delta( p^0 + {\omega}_{\mathbf p}^{l}), \;
I_{p}^{t}=I_{\mathbf p}^{t} \delta ( p^0 - {\omega}_{\mathbf p}^{t}) +
I_{- \mathbf p}^{t} \delta( p^0 + {\omega}_{\mathbf p}^{t}).
\label{eq:7ttt}
\end{equation}
In Eq.\,(\ref{eq:7ttt}) $I_{\mathbf p}^{l,t}$ are certain functions of a
wave vector ${\mathbf p}$,
and ${\omega}_{\mathbf p}^{l,t} \equiv {\omega}^{l,t}({\mathbf p})$ are frequencies
of the gluonic longitudinal and transverse eigenwaves in a QGP.

The equations describing the variation of the spectral densities
$\Upsilon^{-}_{\mathbf q}$ and $I^l_{\mathbf p}$ are obtained from
Eqs.\,(\ref{eq:6q}) and (\ref{eq:6e}) by the replacements
\begin{equation}
\Upsilon(q,x) \rightarrow
h_{-}(\hat{\mathbf q})\Upsilon^{-}_{\mathbf q}
\delta(q^0 - \omega^{-}_{\mathbf q}), \;
I_{\mu \nu}(p,x) \rightarrow
Q_{\mu \nu}(p)\bigl[\,I_{\mathbf p}^{l} \delta ( p^0 -
{\omega}_{\mathbf p}^{l}) + I_{- \mathbf p}^{l} \delta( p^0 +
{\omega}_{\bf p}^{l})\bigl]
\label{eq:7tttt}
\end{equation}
for Eq.\,(\ref{eq:6q}), and
$$
I_{\mu \nu}(p,x) \rightarrow
Q_{\mu \nu}(p)I_{\mathbf p}^{l} \delta ( p^0 - {\omega}_{\mathbf p}^{l}), \;
\Upsilon(q,x) \rightarrow
h_{-}(\hat{\mathbf q})\Upsilon^{-}_{\mathbf q} \delta(q^0 - \omega^{-}_{\mathbf q}) +
h_{+}(\hat{\mathbf q}) \Upsilon^{-}_{-{\mathbf q}} \delta(q^0 + \omega^{-}_{\mathbf q})
$$
for Eq.\,(\ref{eq:6e}). $\delta$ functions enable us to remove integration over
$q^0$ and $p^0$. Retaining on the right-hand side the terms responsible for the
process of nonlinear Landau damping only, we have instead of
Eqs.\,(\ref{eq:6q}) and (\ref{eq:6e}) 
\begin{equation}
2\,\bigg( \frac{\partial}{\partial q^{\mu}}\, [\,q^0\! + \vert {\bf q} \vert +
\,{\rm Re}\, \delta \Sigma_{-}(q)] \bigg)_{q^0=\omega^{-}_{\mathbf q}}
\frac{\partial \Upsilon^{-}_{\mathbf q}}{\partial x_{\mu}}
= 2\,g^2C_F \Upsilon^{-}_{\mathbf q} \int\!{\rm d}{\mathbf p}\,
I_{\mathbf p}^l\,{\rm Im}
\,T^{(Q)}({\mathbf q},{\mathbf p}),
\label{eq:7y}
\end{equation}
where
$$
T^{(Q)}({\mathbf q},{\mathbf p}) \equiv
{\rm tr} \Big\{ \Big( \delta \Gamma^{(Q)\, \mu \nu}(p,-p;q,-q) -\!
\,^{\ast}\Gamma^{(Q)\,\mu}(p;q-p,-q) \,^{\ast}\!S(q-p)
\,^{\ast}\Gamma^{(Q)\,\nu}(-p;q,-q+p)
$$
\begin{equation}
- \,^{\ast}\Gamma^{(Q)\,\nu}(-p;q+p,-q) \,^{\ast}\!S(q+p)
\,^{\ast}\Gamma^{(Q)\,\mu}(p;q,-q-p) \Big) h_{-}(\hat{\bf q})\Big\}Q_{\mu \nu}(p)
\Big\vert_{q^0=\omega^{-}_{\mathbf q},\;p^0=\omega^l_{\mathbf p}} ,
\label{eq:7u}
\end{equation}
and
\begin{equation}
\bigg( \frac{\partial}{\partial p^{\mu}}\,[\,p^2\! +
{\rm Re}\,\delta \Pi^l(p)] \bigg)_{p^0=\omega^{l}_{\mathbf p}}
\frac{\partial I^{l}_{\mathbf p}}{\partial x_{\mu}}
= -2\,g^2T_F I^l_{\mathbf p} \int\!{\rm d}{\mathbf q}\,\Upsilon^{-}_{\mathbf q}
\,{\rm Im}\,T^{(G)}({\mathbf q},{\mathbf p}),
\label{eq:7i}
\end{equation}
where
$$
T^{(G)}({\mathbf q},{\mathbf p}) \equiv
{\rm tr} \Big\{\!\Big( \delta \Gamma^{(G)\, \mu \nu}(p,-p;q,-q) +\!
\,^{\ast}\Gamma^{(Q)\,\nu}(-p;p-q,q)\!\,^{\ast}\!S(p-q)
\!\,^{\ast}\Gamma^{(G)\,\mu}(p;-p+q,\!-q)
$$
\begin{equation}
- \,^{\ast}\Gamma^{(G)\,\mu}(p;q,-p-q) \,^{\ast}\!S(p+q)
\,^{\ast}\Gamma^{(Q)\,\nu}(-p;p+q,-q) \Big) h_{-}(\hat{\bf q})\Big\}
Q_{\mu \nu}(p)\Big\vert_{q^0=\omega^{-}_{\bf q},\;p^0=\omega^l_{\mathbf p}}.
\label{eq:7o}
\end{equation}
The expressions $T^{(Q)}({\mathbf q},{\mathbf p})$ and
$T^{(G)}({\mathbf q},{\mathbf p})$ contain the factors
$$
\frac{1}{v\cdot q + i \epsilon}\,, \quad \frac{1}{v\cdot q - i \epsilon}\,,\quad
\frac{1}{v\cdot (q-p)  + i \epsilon}\,,\ldots\,.
$$
The imaginary parts of the first two factors should be set equal to zero
because they are connected with the linear Landau damping of a plasmino (which is
absent in a QGP) and therefore, the imaginary part of the expressions
(\ref{eq:7u}) and (\ref{eq:7o}) will be defined as
$$
{\rm Im}\,\frac{1}{v\cdot (q-p) + i \epsilon}\,
\bigg\vert_{q^0=\omega^{-}_{\mathbf q},\,
p^0=\omega^l_{\mathbf p}} = - \pi\delta(\omega^{-}_{\mathbf q} -
\omega^l_{\mathbf p} -
{\mathbf v}\cdot ({\mathbf q}-{\mathbf p})), \ldots\,.
$$
Because of the absence of the linear Landau damping for the plasmino from the definitions of
effective two-quark--two-gluon vertex functions (\ref{eq:4yy}) and (\ref{eq:6r}) and
two-quark--one-gluon vertex functions (\ref{eq:4y}) and (\ref{eq:4a}), the next
equalities are as follows
\begin{equation}
\delta\Gamma^{(Q)\,\mu \nu}(p,-p;q,-q)=\delta\Gamma^{(G)\,\mu \nu}(p,-p;q,-q),
 \,^{\ast}\Gamma^{(Q)\,\mu}(p;q-p,-q)=\!\,^{\ast}\Gamma^{(G)\,\mu}(p;q-p,-q),
\ldots ,
\label{eq:7p}
\end{equation}
and as consequence, we have
\begin{equation}
T^{(Q)}({\mathbf q},{\mathbf p})=T^{(G)}({\mathbf q},{\mathbf p}).
\label{eq:7a}
\end{equation}
Therefore, in the subsequent discussion the superscripts $(Q)$ and $(G)$
will be dropped. Let us emphasize that the equalities
(\ref{eq:7p}) and (\ref{eq:7a}) hold only for on-shell plasma excitations.

\section{Gauge invariance of
I\lowercase{m}\,$T({\bf \lowercase{q}},{\bf \lowercase{p}})$}
\setcounter{equation}{0}

The issue of the gauge dependence of the plasmino nonlinear Landau damping rate
is closely associated with the issue of the gauge dependence of ${\rm Im}\,
T({\mathbf q},{\mathbf p})$. In the next section the expression
${\rm Im}\,T({\mathbf q},{\mathbf p})$ is shown to define 
the plasmino--hard-particle
scattering matrix element squared, i.e., it has direct physical relevance.
To establish the gauge invariance of the matrix element for plasmino--hard-particle
scattering we need to show that the expression
${\rm Im}\,T({\mathbf q},{\mathbf p})$ in the covariant gauge equals a similar
one in the temporal gauge.

We prove the gauge invariance for a more general expression, which in a covariant
gauge has the form
$$
\,^{\ast}\tilde{\Gamma}(p_1,p_2;q_1,q_2) \equiv
\Big\{\delta \Gamma^{\mu \nu}(p_1,p_2;q_1,q_2) -\!
\,^{\ast}\Gamma^{\mu}(p_1;q_1+p_2,q_2) \,^{\ast}\!S(q_1+p_2)
\,^{\ast}\Gamma^{\nu}(p_2;q_1,q_2+p_1)
$$
\begin{equation}
-\! \,^{\ast}\Gamma^{\nu}(p_2;q_1+p_1,q_2) \,^{\ast}\!S(q_1+p_1)
\,^{\ast}\Gamma^{\mu}(p_1;q_1,q_2+p_2)\Big\}\bar{u}_{\mu}(p_1)
\bar{u}_{\nu}(p_2)\Big\vert_{\rm on-shell}.
\label{eq:8q}
\end{equation}
Here, $p_1+p_2+q_1+q_2=0$. Notice that the order of the Lorentz indices
of the last two terms in curly brackets on the right-hand side of
Eq.\,(\ref{eq:8q}) is important. The association of the expression
(\ref{eq:7u}) with (\ref{eq:8q}) is given by
$$
{\rm Im}\,T({\mathbf q},{\mathbf p}) =
\frac{1}{\sqrt{\bar{u}^2(p)\bar{u}^2(p_1)}}
\,{\rm Im}\,{\rm tr}\Bigl\{\!\,^{\ast}\tilde{\Gamma}(p_1,p_2;q_1,q_2)
h_{-}(\hat{\mathbf q}_1)\Bigr\} \Big\vert_{p_1=-p_2=p,\;q_1=-q_2=q}.
$$
A similar expression (\ref{eq:8q}) in the temporal gauge is obtained with
the replacements
$$
\bar{u}_{\mu}(p) \rightarrow \tilde{u}_{\mu}(p)\equiv
p^2(p_{\mu} - u_{\mu}(p\cdot u))/(p\cdot u),\;\; \mbox{etc}.
$$

The gauge-invariance proof is based on using identities analogous to
the effective Ward identities in hot gauge theory \cite{bra1,tay}.
It can be shown that the following equalities hold:
$$
\delta\Gamma^{\mu \nu}(p_1,p_2;q_1,q_2)p_{1 \mu} =
\,^{\ast}\Gamma^{\nu}(p_2;q_1,q_2+p_1) -\!\,^{\ast}\Gamma^{\nu}(p_2;q_1+p_1,q_2),
$$
\begin{equation}
\delta\Gamma^{\mu \nu}(p_1,p_2;q_1,q_2)p_{2 \nu} =
\,^{\ast}\Gamma^{\mu}(p_1;q_1,q_2+p_2) -\!\,^{\ast}\Gamma^{\mu}(p_1;q_1+p_2,q_2),
\label{eq:8w}
\end{equation}
$$
\,^{\ast}\Gamma^{\mu}(p;q_1,q_2)p_{\mu} =
\,^{\ast}\!S^{-1}(q_1) -\!\,^{\ast}\!S^{-1}(q_1+p).
$$

Initially we calculate the convolution with the effective two-quark--two-gluon
vertex function $\delta\Gamma^{\mu \nu}$. Slightly cumbersome but not complicated
computations by using the effective Ward identities (\ref{eq:8w}) and the
mass-shell conditions lead to the following expression
\begin{equation}
\delta \Gamma^{\mu \nu}(p_1,p_2;q_1,q_2)
\bar{u}_{\mu}(p_1)\bar{u}_{\nu}(p_2)\Big\vert_{\rm on shell} =
p_1^2p_2^2\,\delta\Gamma^{00}(p_1,p_2;q_1,q_2) + \Xi(p_1,p_2;q_1,q_2) ,
\label{eq:8e}
\end{equation}
where
$$
\Xi(p_1,p_2;q_1,q_2) = p_1^0p_2^2 \Bigl\{\!\,^{\ast}\Gamma^0(p_2;q_1+p_1,q_2)
-\!\,^{\ast}\Gamma^0(p_2;q_1,q_2+p_1) \Bigr\}
$$
$$
+\,p_2^0p_1^2 \Bigl\{\!\,^{\ast}\Gamma^0(p_1;q_1+p_2,q_2)
-\! \,^{\ast}\Gamma^0(p_1;q_1,q_2+p_2) \Bigr\}
-p_1^0p_2^0 \Bigl\{\!\,^{\ast}\!S^{-1}(q_1+p_2)
+\!\,^{\ast}\!S^{-1}(q_1+p_1)\Bigr\}.
$$

Furthermore, we calculate the convolution with the terms containing the
two-quark--one-gluon effective vertex in Eq.\,(\ref{eq:8q}). Here, we derive the
expression
\begin{equation}
\Bigl\{\!\,^{\ast}\Gamma^{\mu}(p_1;q_1+p_2,q_2)\!\,^{\ast}\!S(q_1+p_2)
\!\,^{\ast}\Gamma^{\nu}(p_2;q_1,q_2+p_1) + (p_1 \leftrightarrow p_2,\,
\mu \leftrightarrow \nu)\Bigr\}\bar{u}_{\mu}(p_1)
\bar{u}_{\nu}(p_2)\Big\vert_{\rm on-shell}
\label{eq:8r}
\end{equation}
$$
= p_1^2p_2^2 \Bigr\{\!\,^{\ast}\Gamma^0(p_1;q_1+p_2,q_2)\!\,^{\ast}\!S(q_1+p_2)
\!\,^{\ast}\Gamma^0(p_2;q_1,q_2+p_1)  + (p_1 \leftrightarrow p_2) \Bigr\} +
\Xi(p_1,p_2;q_1,q_2).
$$
Subtracting Eq.\,(\ref{eq:8r}) from (\ref{eq:8e}), we arrive at the desired expression
$$
\,^{\ast}\tilde{\Gamma}(p_1,p_2;q_1,q_2) =
p_1^2p_2^2\Big\{\delta \Gamma^{00}(p_1,p_2;q_1,q_2)-\,^{\ast}\Gamma^{0}
(p_1;q_1+p_2,q_2)\,^{\ast}S(q_1+p_2)\,^{\ast}\Gamma^{0}(p_2;q_1,q_2+p_1)
$$
\begin{equation}
-\!\,^{\ast}\Gamma^{0}(p_2;q_1+p_1,q_2) \,^{\ast}S(q_1+p_1)
\,^{\ast}\Gamma^{0}(p_1;q_1,q_2+p_2)\Big\}.
\label{eq:8t}
\end{equation}

Now we consider the structure of $\,^{\ast}\tilde{\Gamma}$ in the temporal gauge.
For this purpose we replace $\bar{u}_{\mu}$ by $\tilde{u}_{\mu}$ in Eq.\,(\ref{eq:8q}).
The convolution with an effective two-quark--two-gluon vertex function leads to
\begin{equation}
\delta \Gamma^{\mu \nu}(p_1,p_2;q_1,q_2)
\tilde{u}_{\mu}(p_1)\tilde{u}_{\nu}(p_2)\Big\vert_{\rm on shell} =
p_1^2p_2^2\,\delta\Gamma^{00}(p_1,p_2;q_1,q_2) + \tilde{\Xi}(p_1,p_2;q_1,q_2) ,
\label{eq:8y}
\end{equation}
where
$$
\tilde{\Xi}(p_1,p_2;q_1,q_2) = \frac{p_1^2p_2^2}{p_1^0p_2^0} \Bigl\{ \Bigl(
p_1^0\,^{\ast}\Gamma^0(p_1;q_1+p_2,q_2) -\!\,^{\ast}\!S^{-1}(q_1+p_2) \Bigr)
+ ( p_1 \leftrightarrow p_2 ) \Big\}.
$$
Convolution with the terms containing $\,^{\ast}\Gamma^{\mu}$ yields
\begin{equation}
\Bigl\{\!\,^{\ast}\Gamma^{\mu}(p_1;q_1+p_2,q_2)\!\,^{\ast}\!S(q_1+p_2)
\!\,^{\ast}\Gamma^{\nu}(p_2;q_1,q_2+p_1) + (p_1 \leftrightarrow p_2,\,
\mu \leftrightarrow \nu)\Bigr\}\tilde{u}_{\mu}(p_1)
\tilde{u}_{\nu}(p_2)\Big\vert_{\rm on-shell}
\label{eq:8u}
\end{equation}
$$
= p_1^2p_2^2\, \Bigr\{\!\,^{\ast}\Gamma^0(p_1;q_1+p_2,q_2)\!\,^{\ast}\!S(q_1+p_2)
\!\,^{\ast}\Gamma^0(p_2;q_1,q_2+p_1)  + (p_1 \leftrightarrow p_2) \Bigr\} +
\tilde{\Xi}(p_1,p_2;q_1,q_2).
$$
Subtracting Eq.\,(\ref{eq:8u}) from (\ref{eq:8y}), we obtain a similar expression
(\ref{eq:8t}). Thus, we have shown that at least in the class of covariant and
temporal gauges, the expression ${\rm Im}\,T({\bf q},{\bf p})$ is
gauge invariant.

\section{Nonlinear Landau damping rate of a plasmino}
\setcounter{equation}{0}

In our work \cite{mar2} the transformation of the nonlinear Landau damping rate for
a plasmon in a purely gauge sector of soft QGP excitations was performed. An
expression similar to ${\rm Im}\,T({\mathbf q}, {\mathbf p})$ was transformed
to module squared of a sum of the terms, which are interpreted as
the scattering amplitude of specific physical processes.
In this section we represent a
similar transformation for the function ${\rm Im}\,T({\mathbf p}, {\mathbf q})$.
Since this function is not dependent
on the choice of a gauge, we choose a temporal gauge for simplicity.

The right-hand side of Eq.\,(\ref{eq:7y}) contains the contribution of two
different processes. The first is associated with absorption of a plasmino by
QGP particles with frequency $\omega_{\mathbf q}^{-}$ and wave vector
${\mathbf q}$,
and consequent radiation of a plasmon with frequency $\omega_{\mathbf p}^l$ and
wave vector ${\mathbf p}$. It is defined by the second term on the right-hand side of
Eq.\,(\ref{eq:7u}). The frequencies and wave vectors of an incident plasmino
and a recoil plasmon satisfy the conservation law
\begin{equation}
\omega^{-}_{\mathbf q} - \omega^l_{\mathbf p} - {\mathbf v}\cdot ({\mathbf q}
- {\mathbf p})=0.
\label{eq:9q}
\end{equation}
The second process represents simultaneous radiation (or absorption) of a plasmino
and plasmon with frequencies $\omega_{\mathbf q}^{-}$, $\omega_{\mathbf p}^l$
and wave vectors ${\mathbf q}$, ${\mathbf p}$ satisfying the conservation law
\begin{equation}
\omega^{-}_{\mathbf q} + \omega^l_{\mathbf p} - {\mathbf v}\cdot ({\mathbf q}
+ {\mathbf p})=0 ,
\label{eq:9w}
\end{equation}
and is defined by the third term in Eq.\,(\ref{eq:7u}). The contribution of the
second process to the order of interest is not important. The first term on the
right-hand side of Eq.\,(\ref{eq:7u}), associated with an effective 
two-quark--two-gluon
vertex function, contains both processes. Furthermore, we take into account the
terms in Eq.\,(\ref{eq:9q}) and we drop the terms that contain the $\delta$ function
of Eq.\,(\ref{eq:9w}). Notice that the conservation law (\ref{eq:9q}) defines, in a very
nontrivial manner, accessible kinematic regions of wave vectors ${\mathbf q}$ and
${\mathbf p}$ in the process of nonlinear Landau damping for a plasmino.
In Sec.\,XI
we consider only the simplest limiting case of accessible kinematic region
of plasmon wave vector ${\mathbf p}$ for the case of the ${\mathbf q}=0$ mode.

Now we consider the expression with $\delta \Gamma^{\mu \nu}$. With regard to
the discussion above and to the definition (\ref{eq:6r}), the contribution
of a given term to ${\rm Im}\,T({\mathbf q},{\mathbf p})$ can be represented as
\begin{equation}
2\pi\omega_0^2 \Biggl[\,\frac{p^2}{{\mathbf p}^2(p^0)^2}
\int\!\frac{{\rm d}\Omega}{4\pi}
\,\frac{(1+{\bf v}\cdot \hat{{\mathbf q}}\,)\,({\mathbf v}\cdot {\mathbf p})^2}
{(v\cdot q)^2}\,\delta(\omega^{-}_{\mathbf q} - \omega^l_{\mathbf p}
- {\mathbf v}\cdot ({\mathbf q} - {\mathbf p})) \Biggr]_{\rm on-shell}.
\label{eq:9e}
\end{equation}

Let us consider now a more complicated second term in Eq.\,(\ref{eq:7u}) associated with
the two-quark--one-gluon vertices. By the definition (\ref{eq:4t}) and (\ref{eq:4y}),
the following equality is obeyed:
$$
\,^{\ast}\Gamma^{\mu}(p;l,-q) =\!\,^{\ast}\Gamma^{\mu}(-p;q,-l)
\equiv\!\,^{\ast}\Gamma^{\mu}
$$
(hereafter, $l \equiv q-p$). By using this relation and the decomposition of the
effective quark propagator $\!\,^{\ast}\!S(l)$ onto $h_{\pm}(\hat{\bf l})$,
$$
\,^{\ast}\!S(l) = h_{+}(\hat{\mathbf l}) \,^{\ast}\!\triangle_{+}(l) +
h_{-}(\hat{\mathbf l}) \,^{\ast}\!\triangle_{-}(l),
$$
where
\begin{equation}
\,^{\ast}\!\triangle_{\pm}(l) =
-\,\frac{1}{l^0 \mp [\,\vert{\bf l} \vert + \delta \Sigma_{\pm}(l)]},
\label{eq:9r}
\end{equation}
the contribution to ${\rm Im}\,T({\mathbf q},{\mathbf p})$ of the term
associated with the effective two-quark--one-gluon vertex functions can be
represented as
\begin{equation}
\Biggl[\,\frac{p^2}{{\mathbf p}^2(p^0)^2}\,{\rm Im}\,\Bigl(\!\,^{\ast}\!\triangle_{+}
(l)\,{\rm tr} \Bigl\{
h_{-}(\hat{\mathbf q})(\,^{\ast}\Gamma^i{\rm p}^i)
h_{+}(\hat{\mathbf l})(\,^{\ast}\Gamma^j{\rm p}^j) \Bigr\}
+ \,^{\ast}\!\triangle_{-}(l)\,{\rm tr} \Bigl\{
h_{-}(\hat{\mathbf q})(\,^{\ast}\Gamma^i{\rm p}^i)
h_{-}(\hat{\mathbf l})(\,^{\ast}\Gamma^j{\rm p}^j) \Bigr\} \Bigr) \Biggr]_{\rm on-shell}.
\label{eq:9t}
\end{equation}
Let us compute the traces in the last expression. For this purpose  we research
the matrix structure of the function $\,^{\ast}\Gamma^i{\rm p}^i$ in more detail. From
the analysis of the corresponding expressions for HTL functions, derived by Frenkel
and Taylor (in particular, the expression (3.38) in Ref.\,\cite{fre}), it is easy to see
that $\,^{\ast}\Gamma^i{\rm p}^i$ can be represented in the form of the expansion
\begin{equation}
\,^{\ast}\Gamma^i{\rm p}^i = \gamma^0\delta\!{\it \Gamma}_0 +
({\mathbf l}\cdot {\mathbf \gamma})
\,^{\ast}\!{\mathit \Gamma}_{\parallel} + (({\mathbf n}\times {\mathbf l})\cdot {\mathbf \gamma})
\,^{\ast}\!{\mathit \Gamma}_{\perp},
\label{eq:9y}
\end{equation}
where ${\mathbf n} \equiv {\mathbf q}\times {\mathbf p}$ and the coefficient
functions are defined as
$$
\delta\!{\mathit \Gamma}_0 = \omega_0^2\!\int\!\frac{{\rm d}\Omega}{4\pi}
\,\frac{{\mathbf v}\cdot {\mathbf p}}{(v\cdot l + i\epsilon )(v\cdot q)},
$$
\begin{equation}
\,^{\ast}\!{\it \Gamma}_{\parallel} =
\frac{{\mathbf p}\cdot {\mathbf l}}{{\mathbf l}^2} + \delta\!{\mathit \Gamma}_{\parallel}
\equiv\frac{{\mathbf p}\cdot {\mathbf l}}{{\mathbf l}^2} -
\frac{\omega_0^2}{{\mathbf l}^2}\!\int\!\frac{{\rm d}\Omega}{4\pi}
\,\frac{({\mathbf v}\cdot {\mathbf p})\,({\mathbf v}\cdot {\mathbf l})}
{(v\cdot l + i\epsilon)(v\cdot q)},
\label{eq:9u}
\end{equation}
$$
\,^{\ast}\!{\mathit \Gamma}_{\perp} =
\frac{1}{{\mathbf l}^2} + \delta\!{\mathit \Gamma}_{\perp} \equiv
\frac{1}{{\mathbf l}^2} -
\frac{\omega_0^2}{{\mathbf l}^2\,{\mathbf n}^2}\!\int\!\frac{{\rm d}\Omega}
{4\pi}\,\frac{({\mathbf v}\cdot {\mathbf p})\,({\mathbf v}\cdot ({\mathbf n}
\times {\mathbf l}))}{(v\cdot l + i\epsilon )(v\cdot q)}.
$$
The matrix basis in the expansion (\ref{eq:9y}) is convenient in that it is
``ortogonal'' in trace computing. Substituting expression (\ref{eq:9y}) into
(\ref{eq:9t}) we can compute desired traces in Eq.\,(\ref{eq:9t}) in terms of the
functions (\ref{eq:9u}). However, this direct approach is not quite convenient in
view of its nontransparency and the necessity of cumbersome
calculations. We overdetermine the expansion (\ref{eq:9y}), which enables us to
represent the function $\,^{\ast}\Gamma^i{\rm p}^i$ in a more appropriate form for
trace computing. Instead of expansion (\ref{eq:9y}) we can write (the details
of overdetermination are given in Appendix~A~)
\begin{equation}
\,^{\ast}\Gamma^i{\rm p}^i =
-\,h_{-}(\hat{\mathbf l})\,^{\ast}\!{\mathit \Gamma}_{+}
-h_{+}(\hat{\mathbf l})\,^{\ast}\!{\mathit \Gamma}_{-}
+ 2h_{-}(\hat{\mathbf q})\,{\mathbf l}^2 \vert {\mathbf q}\vert
\,^{\ast}\!{\mathit \Gamma}_{\perp},
\label{eq:9i}
\end{equation}
where
\begin{equation}
\,^{\ast}\!{\mathit \Gamma}_{\pm} \equiv -\,\delta\!{\mathit \Gamma}_0 \mp
\vert {\mathbf l}\vert \,^{\ast}\!{\mathit \Gamma}_{\parallel}
+ \frac{{\mathbf n}^2}{\vert {\mathbf q} \vert}\,\frac{1}{1\mp \hat{\mathbf q}\cdot
\hat{\mathbf l}} \,^{\ast}\!{\mathit \Gamma}_{\perp}.
\label{eq:9o}
\end{equation}
The explicit selectivity of the matrix $h_{-}(\hat{\mathbf q})$ in the
expansion (\ref{eq:9i})
is connected with the fact that we have restricted our consideration to the study
of a plasmino branch of fermion excitations only. In the case of the branch
describing normal-particle excitations, it is necessary
to use the following expansion instead of Eq.\,(\ref{eq:9i}):
\begin{equation}
\,^{\ast}\Gamma^i{\rm p}^i =
-\,h_{-}(\hat{\mathbf l})\,^{\ast}\!\acute{{\mathit \Gamma}}_{+}
-h_{+}(\hat{\mathbf l})\,^{\ast}\!\acute{{\mathit \Gamma}}_{-}
- 2h_{+}(\hat{\mathbf q})\,{\mathbf l}^2 \vert {\mathbf q} \vert
\,^{\ast}\!{\mathit \Gamma}_{\perp},
\label{eq:9p}
\end{equation}
where now
\begin{equation}
\,^{\ast}\!\acute{{\mathit \Gamma}}_{\pm} \equiv -\,\delta\!{\mathit \Gamma}_0 \mp
\vert {\mathbf l}\vert \,^{\ast}\!{\mathit \Gamma}_{\parallel}
- \frac{{\mathbf n}^2}{\vert {\mathbf q} \vert}\,\frac{1}{1\pm \hat{\mathbf q}\cdot
\hat{\mathbf l}} \,^{\ast}\!{\mathit \Gamma}_{\perp}.
\label{eq:9a}
\end{equation}

Substituting expression (\ref{eq:9i}) into (\ref{eq:9t}), and using the
identities for $h_{\pm}(\hat{\mathbf l})$ matrices,
$$
h_{\pm}(\hat{\mathbf l})h_{\pm}(\hat{\mathbf l}) = 0, \quad
h_{\pm}(\hat{\mathbf l})h_{\mp}(\hat{\mathbf l})h_{\pm}(\hat{\mathbf l}) =
h_{\pm}(\hat{\mathbf l}),
$$
after computing the trivial traces, we obtain
\begin{equation}
\Biggl[\, \frac{p^2}{{\mathbf p}^2(p^0)^2} \bigg\{ (1-\hat{\mathbf q}
\cdot\hat{\mathbf l}\,)
\,{\rm Im}\,(\!\,^{\ast}\!\triangle_{+}(l)(\!\,^{\ast}\!{\mathit \Gamma}_{+})^2)
+ (1+\hat{\mathbf q}\cdot\hat{\mathbf l}\,)
\,{\rm Im}\,(\!\,^{\ast}\!\triangle_{-}(l)(\!\,^{\ast}\!{\mathit \Gamma}_{-})^2)
\bigg\}\Biggr]_{\rm on-shell}.
\label{eq:9s}
\end{equation}

Furthermore, we use the relation
$$
{\rm Im}\,(\!\,^{\ast}\!\triangle_{\pm}(l)(\!\,^{\ast}\!{\mathit \Gamma}_{\pm})^2) =
-\,{\rm Im}\,(\!\,^{\ast}\!\triangle_{\pm}^{-1}(l))\,\vert\!\,^{\ast}\!\triangle_{\pm}(l)
\,^{\ast}\!{\mathit \Gamma}_{\pm}\vert^2 +
2\,{\rm Im}\,(\!\,^{\ast}\!{\mathit \Gamma}_{\pm})\,{\rm Re}\,
(\!\,^{\ast}\!\triangle_{\pm}(l)\,^{\ast}\!{\mathit \Gamma}_{\pm}).
$$

Taking into consideration the last relation and the equalities
$$
{\rm Im}\,^{\ast}\!\triangle^{-1}_{\pm}(l) =
-\,\pi\omega_0^2\!\int\!\frac{{\rm d}\Omega}
{4\pi}\,(1\mp {\mathbf v}\cdot \hat{\mathbf l}\,)\,\delta(v\cdot l),
$$
$$
{\rm Im}\,^{\ast}\!{\mathit \Gamma}_{\pm} =
\pi\omega_0^2\!\int\!\frac{{\rm d}\Omega}{4\pi}\,\frac{{\mathbf v}
\cdot {\mathbf p}}{v\cdot q}\,\biggl(\!
1\mp{\mathbf v}\cdot \hat{\mathbf l} + \frac{{\mathbf v}\cdot ({\mathbf n}
\times {\mathbf l})}
{\vert{\mathbf q}\vert\,{\mathbf l}^2(1\mp \hat{\mathbf q}\cdot \hat{\mathbf l}\,)}\,
\biggr)\,\delta(v\cdot l),
$$
subtracting Eq.\,(\ref{eq:9s}) from (\ref{eq:9e}), we define the desired expression
for ${\rm Im}\,T$:
$$
{\rm Im}\,T({\mathbf q},{\mathbf p}) =
\pi\omega_0^2\,\Biggl[\,\frac{p^2}{{\mathbf p}^2(p^0)^2}
\int\!\frac{{\rm d}\Omega}{4\pi}\,\biggl\{
2\,\frac{({\mathbf v}\cdot {\mathbf p})^2}{(v\cdot q)^2}
\,(1+{\mathbf v}\cdot \hat{{\mathbf q}}\,)
$$
\begin{equation}
+\,(1-\hat{\mathbf q}\cdot \hat{\mathbf l}\,)(1-{\mathbf v}\cdot \hat{\mathbf l}\,)
\,\vert\!\,^{\ast}\!\triangle_{+}(l)
\,^{\ast}\!{\mathit \Gamma}_{+}\vert^2
+ 2\,\frac{{\mathbf v}\cdot {\mathbf p}}{v\cdot q}\,\varrho_{+}({\mathbf v};
\hat{\mathbf q},\hat{\mathbf p})\,{\rm Re}\,
(\!\,^{\ast}\!\triangle_{+}(l)\,^{\ast}\!{\mathit \Gamma}_{+})
\label{eq:9d}
\end{equation}
$$
+\,(1+\hat{\mathbf q}\cdot \hat{\mathbf l}\,)(1+{\mathbf v}\cdot \hat{\mathbf l}\,)
\,\vert\!\,^{\ast}\!\triangle_{-}(l)
\,^{\ast}\!{\mathit \Gamma}_{-}\vert^2
+ 2\,\frac{{\mathbf v}\cdot {\mathbf p}}{v\cdot q}\,\varrho_{-}({\mathbf v};
\hat{\mathbf q},\hat{\mathbf p})\,{\rm Re}\,
(\!\,^{\ast}\!\triangle_{-}(l)\,^{\ast}\!{\mathit \Gamma}_{-})
\biggr\}\,\delta(v\cdot l)\Biggr]_{\rm on-shell}.
$$
Here,
\begin{equation}
\varrho_{\pm}({\mathbf v};\hat{\mathbf q},\hat{\mathbf p}) \equiv
(1\mp\hat{\mathbf q}\cdot \hat{\mathbf l}\,)(1\mp{\mathbf v}\cdot
\hat{\mathbf l}\,)
+ \frac{{\mathbf v}\cdot ({\mathbf n}\times {\mathbf l})}{\vert{\mathbf q}\vert
\,{\mathbf l}^2}.
\label{eq:9f}
\end{equation}
The expression (\ref{eq:9d}) can be led to a more descriptive form. For this
purpose we add to the expression in the curly brackets on the right-hand
side of Eq.\,(\ref{eq:9d}) the term
$$
\frac{{\mathbf v}\cdot ({\mathbf n}\times {\mathbf l})}{\vert{\mathbf q}\vert
\,{\mathbf l}^2}
\,\bigl(\vert\!\,^{\ast}\!\triangle_{+}(l) \,^{\ast}\!{\mathit \Gamma}_{+}\vert^2
+ \vert\!\,^{\ast}\!\triangle_{-}(l) \,^{\ast}\!{\mathit \Gamma}_{-}\vert^2 \bigr),
$$
equals zero over solid integration. In addition, we rewrite the first term in 
curly brackets by using the relation
$$
2\,(1 + {\mathbf v}\cdot \hat{\mathbf q}\,) =
\varrho_{+}({\mathbf v};\hat{\mathbf q},\hat{\mathbf p}) +
\varrho_{-}({\mathbf v};\hat{\mathbf q},\hat{\mathbf p})  ,
$$
which is a consequence of the definitions (\ref{eq:9f}). Taking into account
the discussion above, we rewrite expression (\ref{eq:9d}) in the following
form:
\begin{equation}
{\rm Im}\,T({\mathbf q},{\mathbf p}) =
\pi \omega_0^2
\frac{(\omega^l_{\mathbf p})^2 - {\mathbf p}^2}{(\omega_{\mathbf p}^l)^2
{\mathbf p}^2}\!\int\!\frac{{\rm d}\Omega}{4\pi}\,\delta(\omega_{\mathbf q}^{-} -
\omega_{\mathbf p}^l - {\mathbf v}\cdot ({\mathbf q} - {\mathbf p})) \{
\varrho_{+}({\mathbf v};\hat{\mathbf q},\hat{\mathbf p})
{\mathit w}_{\mathbf v}^{+}({\mathbf q},{\mathbf p})
+ \varrho_{-}({\mathbf v};\hat{\mathbf q},\hat{\mathbf p})
{\mathit w}_{\mathbf v}^{-}({\mathbf q},{\mathbf p})\},
\label{eq:9g}
\end{equation}
where
\begin{equation}
{\mathit w}_{\mathbf v}^{\pm}({\mathbf q},{\mathbf p}) =
\vert{\mathcal M}_{\pm}({\mathbf q},{\mathbf p})\vert^2 \equiv
\left\vert\,\frac{{\mathbf v}\cdot {\mathbf p}}{v\cdot q}
+\!\,^{\ast}\!\triangle_{\pm}(l)\,^{\ast}\!{\mathit \Gamma}_{\pm}
\,\right\vert^2_{\rm on-shell}.
\label{eq:9h}
\end{equation}
The denominator $v\cdot q$ in Eq.\,(\ref{eq:9h}) is eikonal, which was expected
in an approximation to the small-angle scattering of a high-energy particle.
It defines the effective (here, effectiveness is not in terms of HTL resummation)
propagator in the Compton type scattering process depicted in Fig.~4.
The factor ${\mathbf v}\cdot {\mathbf p}$ is connected with the effective
vertex with one soft external leg and two hard ones.
Accordingly, the second term in the amplitudes ${\mathcal M}_{\pm}({\mathbf q},
{\mathbf p})$ represents the second type of scattering process depicted
in Fig.~5. Here, the quark propagator transfers
the soft momentum and is defined by the scalar functions
$\,^{\ast}\!\triangle_{+}$ or $\,^{\ast}\!\triangle_{-}$, and a vertex with
soft external momenta is defined by the scalar functions
$\,^{\ast}\!{\mathit \Gamma}_{+}$ or $\,^{\ast}\!{\mathit \Gamma}_{-}$ for
corresponding propagators.

Furthermore, we are relating the Wigner functions $\Upsilon_{\mathbf q}^{-}$
and $I_{\mathbf p}^l$ to the plasmino and plasmon number densities, setting
accordingly
\begin{equation}
n_{\mathbf q}^{-} = 2\,{\rm Z}_{-}^{-1}({\mathbf q}) \Upsilon_{\mathbf q}^{-},
\quad
{\mathit N}_{\bf p}^l = -\,2\,\omega_{\mathbf p}^l {\rm Z}_l^{-1}
({\mathbf p})I_{\mathbf p}^l.
\label{eq:9j}
\end{equation}
Here,
$$
{\rm Z}_{-}^{-1}({\mathbf q}) = 1 + \biggl(\frac{\partial{\rm Re}\,\delta
\Sigma_{-}(q)}{\partial q^0} \biggr)_{q^0=\omega_{\mathbf q}^{-}}, \qquad
{\rm Z}_{l}^{-1}({\mathbf p}) = 1 + \biggl(\frac{\partial{\rm Re}\,\delta
\Pi^l(p)}{\partial (p^0)^2} \biggr)_{p^0=\omega_{\mathbf p}^{l}}
$$
are the residues of the effective quark and gluon propagators at the appropriate
poles, respectively. The factor 2 in front of ${\rm Z}_{-}^{-1}({\mathbf q})$
takes into account the presence in QGP's of antiplasminos. By using the definition
(\ref{eq:9j}),
Eqs.\,(\ref{eq:7y}) and (\ref{eq:7i}) can be rewritten in the usual form of
kinetic equations (containing on the left-hand side drift terms, and on the
right-hand side the terms responsible for collisions), where the role
distribution functions of quasiparticles--plasminos and plasmons--fulfill the
functions (\ref{eq:9j})
\begin{equation}
\frac{\partial n_{\mathbf q}^{-}}{\partial t} +
{\mathbf V}_{\mathbf q}^{-}\cdot \frac{\partial n_{\mathbf q}^{-}}
{\partial {\mathbf x}} = - \,\gamma^{-}({\mathbf q})n_{\mathbf q}^{-} \equiv
-\,g^2C_F\,n_{\mathbf q}^{-}\!\int\!{\rm d}{\mathbf p}\,
{\mathcal Q}({\mathbf q},{\mathbf p}) N_{\mathbf p}^l,
\label{eq:9k}
\end{equation}
\begin{equation}
\frac{\partial N_{\mathbf p}^{l}}{\partial t} +
{\mathbf V}_{\mathbf p}^{l}\cdot \frac{\partial N_{\mathbf p}^{l}}
{\partial {\mathbf x}} = +\,\gamma^{l}({\mathbf p})N_{\mathbf p}^{l} \equiv
+\,g^2T_FN_{\mathbf p}^{l}\!\int\!{\rm d}{\mathbf q}\,
{\mathcal Q}({\mathbf q},{\mathbf p})n_{\mathbf q}^{-},
\label{eq:9l}
\end{equation}
where
${\mathbf V}_{\mathbf q}^{-} \equiv \partial\omega_{\mathbf q}^{-}/
\partial {\mathbf q}$ and
${\mathbf V}_{\mathbf p}^{l} \equiv \partial\omega_{\mathbf p}^{l}/
\partial {\mathbf p}$ are
group velocities of plasminos and plasmons respectively, and the kernel
${\mathcal Q}({\mathbf q},{\mathbf p})$ is defined by
\begin{equation}
{\mathcal Q}({\mathbf q},{\mathbf p}) =
\frac{1}{2\omega_{\mathbf p}^l}\, {\rm Z}_{-}({\mathbf q}) {\rm Z}_{l}({\mathbf p})\,
{\rm Im}\,T({\mathbf q},{\mathbf p}).
\label{eq:9z}
\end{equation}
The functions $\gamma^{-}({\mathbf q})$ and $\gamma^{l}({\mathbf p})$ on the
right-hand side of Eqs.\,(\ref{eq:9k}) and (\ref{eq:9l}) represent the nonlinear
Landau damping rates for plasminos with momentum ${\mathbf q}$ and for plasmons
with momentum ${\mathbf p}$, respectively.

The structure of kernel (\ref{eq:9z}) is rather unexpected. As we see from
the expression (\ref{eq:9g}), this kernel is not reduced to the squared modulus of
one scalar function, as this occurs in a purely gauge case \cite{mar2}. Here,
${\mathcal Q}({\mathbf q},{\mathbf p})$ is defined by a sum of the squared moduli of
two independent scalar functions\footnote{As will be shown in the next section,
the coefficient functions $\varrho_{\pm}$ in Eq.\,(\ref{eq:9g}) are not, in general,
case definite; therefore in principle it is impossible to express
the expression ${\rm Im}\,T$ as the squared modulus of one scalar
function--the total scattering amplitude.}: amplitudes
${\mathcal M}_{+}({\mathbf q},{\mathbf p})$ and
${\mathcal M}_{-}({\mathbf q},{\mathbf p})$. This is a
point that we find difficult to interpret and therefore additional analysis of
this problem is required. The only remark that may be made is that
the interference between the scattering processes depicted in Fig.~5
proceeding through intermediate quark virtual oscillations with propagators
$\,^{\ast}\!\triangle_{+}(l)$ and $\,^{\ast}\!\triangle_{-}(l)$ accordingly,
vanishes by the relation
$$
h_{-}(\hat{\mathbf l}) \biggl[\, \int\!\frac{{\rm d}\Omega}{4\pi}\,
\not\!v\,\delta(\omega_{\mathbf q}^{-} - \omega_{\mathbf p}^l -{\mathbf v}
\cdot({\mathbf q}-{\mathbf p})) \biggr] h_{+}(\hat{\mathbf l}) = 0.
$$
This relation is the analog of the relation in a purely gauge case \cite{mar2}
$$
({\mathbf p}_2)^i \biggl[\,\int\!\frac{{\rm d}\Omega}{4\pi}\,v^iv^j\,
\delta(\omega_{\mathbf p}^{l} - \omega_{{\mathbf p}_1}^l -{\mathbf v}
\cdot({\mathbf p}-
{\mathbf p}_1)) \biggr] (\check{\mathbf n}\times {\mathbf p}_2)^j = 0,\quad
\check{\mathbf n} \equiv {\mathbf p}\times {\mathbf p}_1,\;{\mathbf p}_2
\equiv {\mathbf p} - {\mathbf p}_1,
$$
that is responsible for the absence of interference between the scattering of
plasmon by a QGP thermal particle through the longitudinal and transverse
virtual gluon oscillations.

\section{Decomposition of kernel
${\mathcal Q}({\bf \lowercase{q}},{\bf \lowercase{p}})$ into positive
and negative parts}
\setcounter{equation}{0}

In this section we consider the problem of the direction
of the effective pumping over of plasma excitation energy in the process
of the nonlinear interaction of plasminos 
and plasmons. For this purpose, first, we study in more detail the
structure of the 
$\varrho_{\pm}({\mathbf v};\hat{\mathbf q},\hat{\mathbf p})$ function.

By using the expansion (A4), we represent the functions $\varrho_{\pm}$
(\ref{eq:9f}) in the form
\begin{equation}
\varrho_{\pm}({\mathbf v};\hat{\mathbf q},\hat{\mathbf p}) =
1 + {\mathbf v}\cdot \hat{\mathbf q} \mp (\hat{\mathbf q}\cdot \hat{\mathbf l}
+ {\mathbf v}\cdot \hat{\mathbf l}\,).
\label{eq:10q}
\end{equation}
Let us introduce the coordinate system in which axis $0Z$
is aligned with vector $\hat{\bf l}$; then the coordinates of vectors
$\hat{\mathbf q}$ and ${\mathbf v}$ are equal to
$\hat{\mathbf q} = ( 1, \alpha, \beta)$ and ${\mathbf v} = (1, \theta, \varphi),$
respectively. By $\Phi$ we denote the angle between
${\mathbf v}$ and $\hat{\mathbf q}$: ${\mathbf v}\cdot \hat{\mathbf q} = \cos \Phi$.
The angle $\Phi$ can be expressed as
\begin{equation}
\cos\Phi = \sin\theta \sin\alpha \cos( \varphi - \beta) +
\cos\theta \cos\alpha.
\label{eq:10w}
\end{equation}
In the fixed coordinate system expression (\ref{eq:10q}) reads
\begin{equation}
\varrho_{\pm}(\alpha,\beta;\theta,\varphi) =
1 + \sin \theta \sin \alpha \cos( \varphi - \beta) +
\cos \theta \cos \alpha \mp (\cos \alpha + \cos \theta).
\label{eq:10e}
\end{equation}
By using trigonometric formulas, it is easily to show that the last expression can
be represented in the homogeneous quadratic form with respect to
the variables
$\cos\frac{1}{2}(\theta - \alpha)$ and $\cos\frac{1}{2}(\theta + \alpha)$
\begin{equation}
\varrho_{\pm}(\alpha,\beta;\theta,\varphi) =
{\mathbf y}^{\dagger} {\mathcal A}_{\pm} {\mathbf y},
\label{eq:10r}
\end{equation}
where
$$
{\mathbf y} \equiv {\bf y}(\alpha,\theta) \equiv
\left(
\begin{array}{c}
\cos\frac{1}{2}(\theta - \alpha) \\
\cos\frac{1}{2}(\theta + \alpha)
\end{array}
\right), \;
{\mathcal A}_{\pm} \equiv {\mathcal A}_{\pm}(\beta,\varphi) \equiv
\left(
\begin{array}{cc}
2\cos^2\frac{1}{2}(\varphi -\beta) & \mp1 \\
\mp1 & 2\sin^2\frac{1}{2}(\varphi -\beta)
\end{array}
\right).
$$
The eigenvalues of matrices ${\mathcal A}_{\pm}$ are equal to
\begin{equation}
\lambda^{(+)} = 1 + \sqrt{\,1 + \cos^2(\varphi - \beta)} > 0, \quad
\lambda^{(-)} = 1 - \sqrt{\,1 + \cos^2(\varphi - \beta)} \leq 0.
\label{eq:10t}
\end{equation}
From the theory of matrices \cite{gan} it is known that the real symmetric quadratic form
is indefinite if and only if the eigenvalues $\lambda$ of
matrix ${\mathcal A}$ have the different signs. Thus, by virtue of Eq.\,(\ref{eq:10t})
we have proved that the coefficient functions $\varrho_{\pm}$ are indefinite.
By the linear transformation of the variables
$\cos\frac{1}{2}(\theta - \alpha)$ and $\cos\frac{1}{2}(\theta + \alpha)$,
the quadratic form (\ref{eq:10r}) can be expressed as the canonical one, where
the positive and negative parts of $\varrho_{\pm}$ are explicitly displayed
\begin{equation}
\varrho_{\pm}(\alpha,\beta;\theta,\varphi) =
\lambda^{(+)} (\chi_{\pm}^{(+)})^2 +\lambda^{(-)} (\chi_{\pm}^{(-)})^2.
\label{eq:10y}
\end{equation}
Here, $\chi_{\pm}^{(+)}$ and $\chi_{\pm}^{(-)}$ are certain linear functions
of $\cos\frac{1}{2}(\theta - \alpha)$ and $\cos\frac{1}{2}(\theta + \alpha)$.

The explicit expressions for $\chi_{\pm}^{(+)}$ and $\chi_{\pm}^{(-)}$
are defined with the help of the transformations
\begin{equation}
\left(
\begin{array}{c}
\chi_{\pm}^{(+)} \\
\chi_{\pm}^{(-)}
\end{array}
\right) = {\mathcal U}_{\pm}
\left(
\begin{array}{c}
\cos\frac{1}{2}(\theta - \alpha)\\
\cos\frac{1}{2}(\theta + \alpha)
\end{array}
\right),
\label{eq:10u}
\end{equation}
where, in turn, the transformation matrices ${\mathcal U}_{\pm}$ are defined by
the solutions of the corresponding matrix equations
$$
{\mathcal U}_{\pm}{\mathcal A}_{\pm} = {\rm diag}(\lambda^{(+)},\lambda^{(-)})\,
{\mathcal U}_{\pm}.
$$
The solutions of the last equations have the following structure
$$
{\mathcal U}_{\pm} = \left(
\begin{array}{cc}
\cos\vartheta & \pm \sin\vartheta \\
\mp \sin\vartheta & \cos\vartheta
\end{array}
\right), \qquad \tan\vartheta = -\,\frac{1}{\cos(\varphi - \beta) +
\sqrt{\,1+\cos^2(\varphi - \beta)}}.
$$
By using a given expression and Eq.\,(\ref{eq:10u}), we derive the desired
expressions for $\chi_{\pm}^{(+)}$ and $\chi_{\pm}^{(-)}$. The functions
$\chi_{\pm}^{(+)}$ and $\chi_{\pm}^{(-)}$ can be easily rewritten in the terms of
initial vectors ${\mathbf l}$, ${\bf q}$ and ${\mathbf v}$. For example, from
Eq.\,(\ref{eq:10w}) it follows that
$$
\cos(\varphi - \beta) = -\,\frac{{\mathbf n}\cdot ({\mathbf v}\times {\mathbf l})}
{\vert{\mathbf n}\vert \vert{\mathbf v}\times {\mathbf l}\vert}\quad \mbox{etc}.
$$

By substituting the expansion (\ref{eq:10y}) into (\ref{eq:9g}), we can represent
the kernel ${\mathcal Q}({\mathbf q},{\mathbf p})$ in the form of a sum of
positive and negative parts
\begin{equation}
{\mathcal Q}({\mathbf q},{\mathbf p}) = {\mathcal Q}^{(+)}
({\mathbf q},{\mathbf p})-
{\mathcal Q}^{(-)}({\mathbf q},{\mathbf p}),
\quad{\mathcal Q}^{(\pm)}({\mathbf q},{\mathbf p})\geq 0,
\label{eq:10i}
\end{equation}
where
\begin{equation}
{\mathcal Q}^{(\pm)}({\mathbf q},{\mathbf p})
= \pm\,\pi\omega_0^2 \left[
\frac{(\omega^l_{\mathbf p})^2 -
{\mathbf p}^2}{2(\omega_{\mathbf p}^l)^3{\mathbf p}^2}
{\rm Z}_{-}({\mathbf q}){\rm Z}_l({\mathbf p}) \right]
\label{eq:10o}
\end{equation}
$$
\times\int\!\frac{{\rm d}\Omega}{4\pi}\,\delta(\omega_{\mathbf q}^{-} -
\omega_{\mathbf p}^l - {\mathbf v}\cdot ({\mathbf q} - {\mathbf p})) \Bigr\{
\lambda^{(\pm)}[(\chi_{+}^{(\pm)})^2 {\mathit w}_{\mathbf v}^{+}({\mathbf q},{\mathbf p})
+ (\chi_{-}^{(\pm)})^2 {\mathit w}_{\mathbf v}^{-}({\mathbf q},{\mathbf p})]\Big\}.
$$
We use the decomposition (\ref{eq:10i}) in general analysis of the problem of a
direction of the pumping over of excitation energy in the study of the process of
nonlinear interaction between plasminos and plasmons.

To consider the model problem of interaction of two infinitely narrow packets
with typical wave vectors ${\mathbf q}_0$ and ${\mathbf p}_0$, let us introduce
the number densities $n_{\mathbf q}^{-}$ and $N_{\mathbf p}^l$ as follows:
$$
n_{\mathbf q}^{-}(t) = n^{-}(t)\,\delta({\mathbf q}-{\mathbf q}_0),\qquad
N_{\mathbf p}^l(t)=N^l(t)\,\delta({\mathbf p}-{\mathbf p}_0).
$$
We have restricted ourselves to the spatially homogeneous case. Substituting
the last expressions into Eqs.\,(\ref{eq:9k}) and (\ref{eq:9l}), we obtain the
coupled nonlinear equations:
\begin{equation}
\frac{\partial n^{-}}{\partial t} =
-\,g^2C_F{\mathcal Q}\,n^{-}N^l, \quad n^{-}(t_0) = n_0^{-},
\label{eq:10p}
\end{equation}
$$
\frac{\partial N^{l}}{\partial t} =
+\,g^2T_F{\mathcal Q}\,n^{-}N^l, \quad N^{l}(t_0) = N_0^{l}.
$$
Here, ${\mathcal Q}\equiv {\mathcal Q}({\mathbf q}_0,{\mathbf p}_0).$ The system of the
equations (\ref{eq:10p}) possesses the integral of motion
$$
{\mathcal C} \equiv T_F\,n^{-}(t) + C_FN^l(t) = T_F\,n^{-}_0 + C_FN^l_0.
$$
The general solution of this system, with regard to decomposition of the kernel
(\ref{eq:10i}), has the form
$$
n^{-}(t) = n_0^{-}\,{\mathcal C}\,
\displaystyle\frac{{\rm e}^{-g^2{\mathcal C}{\mathcal Q}^{(+)}(t - t_0)}}
{C_FN_0^l\,{\rm e}^{-g^2{\mathcal C}{\mathcal Q}^{(-)}(t - t_0)} +
T_F\,n_0^{-}{\rm e}^{-g^2{\mathcal C}{\mathcal Q}^{(+)}(t - t_0)}},
$$
$$
N^{l}(t) = N_0^{l}\,{\mathcal C}\,
\displaystyle\frac{{\rm e}^{-g^2{\mathcal C}{\mathcal Q}^{(-)}(t - t_0)}}
{C_FN_0^l\,{\rm e}^{-g^2{\cal C}{\mathcal Q}^{(-)}(t - t_0)} +
T_F\,n_0^{-}{\rm e}^{-g^2{\mathcal C}{\mathcal Q}^{(+)}(t - t_0)}}.
$$
Let us analyze the behavior of these solutions in the limit for $t \rightarrow
\infty$.
\begin{enumerate}
\item Let us assume that the values of the wave vectors ${\mathbf q}_0$
and ${\mathbf p}_0$ are such that the following inequality
\begin{equation}
{\mathcal Q}^{(+)}({\mathbf q}_0,{\mathbf p}_0) >
{\mathcal Q}^{(-)}({\mathbf q}_0,{\mathbf p}_0)
\label{eq:10a}
\end{equation}
is true. Then in the limit for $t \rightarrow \infty$ we have
\begin{equation}
n^{-}(t)\rightarrow 0, \qquad N^l(t) \rightarrow N_0^l + (T_F/C_F)\,n_0^{-}.
\label{eq:10s}
\end{equation}
Thus we see that as a result of the nonlinear interaction of two infinitely
narrow packets with fermion and boson quantum numbers, the effective
pumping over of energy from the first packet to the second one takes place.
\item If the inverse inequality holds,
\begin{equation}
{\mathcal Q}^{(+)}({\mathbf q}_0,{\mathbf p}_0) <
{\mathcal Q}^{(-)}({\mathbf q}_0,{\mathbf p}_0),
\label{eq:10d}
\end{equation}
then in the limit for $t \rightarrow \infty$ we have
$$
n^{-}(t)\rightarrow n^{-}_0 + (C_F/T_F)N_0^{l}, \quad
N^l(t) \rightarrow 0.
$$
Here, we have an inverse case--the damping of bosonic excitations and the growth
of fermionic excitations.
\item Finally, in the limiting case
${\mathcal Q}^{(+)}({\mathbf q}_0,{\mathbf p}_0) =
{\mathcal Q}^{(-)}({\mathbf q}_0,{\mathbf p}_0)$,
the interaction between wave packets is absent.
\end{enumerate}

The inequalities (\ref{eq:10a}) and (\ref{eq:10d}) define the kinematic relations
between wave vectors of excitations with different statistics, such that one or
another process of pumping over of energy occurs. However, the general analysis of
these inequalities is a complicated problem by virtue of the complexity of the
expressions for positive and negative parts of the kernel
${\mathcal Q}$. In the next section we shall study in detail only the simplest
limiting case of the interaction of ``standing'' plasmino $({\mathbf q}=0)$ with
a plasmon. Here, the direction of the effective pumping over energy is easily
defined.

\section{The damping rate of a plasmino at rest}
\setcounter{equation}{0}

The solution of the kinetic equation (\ref{eq:9k}), defining a change of the
plasmino number density, can be formally represented in the form
(for a spatially homogeneous case)
$$
n_{\mathbf q}^{(-)}(t) = n_{\mathbf q}^{(-)}(t_0)
\exp\Bigl\{-\,g^2C_F\!\!\int\limits_{t_0}^t\! {\rm d}t^{\prime}\!\int\!{\rm d}
{\mathbf p}
\,{\mathcal Q}({\mathbf q},{\mathbf p}) N_{\mathbf p}^l(t^{\prime})\Bigr\}.
$$
Let us approximate the plasmon number density $N_{\mathbf p}^l$ by its equilibrium
value or Planck distribution
$$
N_{\mathbf p}^l \simeq N_{eq}^l(\vert {\mathbf p}\vert) =
\frac{1}{(2\pi)^3}\,\displaystyle\frac{1}{{\rm e}^{\omega_{\mathbf p}^l/T} - 1}
$$
and define the quasiparticle {\it damping rate} of the standing plasmino by
means of the relation
\begin{equation}
\gamma^{-}(0) = \frac{1}{2}\,g^2C_F
\lim\limits_{\vert{\bf q}\vert \to 0}\!\int\!{\rm d}{\mathbf p}
\,{\mathcal Q}({\mathbf q},{\mathbf p}) N_{eq}^l(\vert{\mathbf p}\vert).
\label{eq:11q}
\end{equation}

Now we represent a complete calculation of $\gamma^{-}(0)$. We start with a
representation of a kernel ${\mathcal Q}({\mathbf q},{\mathbf p})$
in the form (\ref{eq:9z}),
where the function ${\rm Im}\,T({\mathbf q},{\mathbf p})$ is determined through
Eqs.\,(\ref{eq:9g}) and (\ref{eq:9h}). Here, it is convenient for us to introduce a new
coordinate system in which axis $0Z$ is aligned with the vector $\hat{\mathbf p}$;
then the coordinates of vectors $\hat{\mathbf q}$ and ${\mathbf v}$ are equal to
$\hat{\mathbf q} = (1,\alpha^{\prime},\beta^{\prime})$ and
${\mathbf v}=(1,\theta^{\prime},
\varphi^{\prime})$, respectively. By $\Phi^{\prime}$ we denote the angle between
${\mathbf v}$, and $\hat{\mathbf q}$. The angle $\Phi^{\prime}$ is connected with
$\alpha^{\prime}$, $\beta^{\prime}$, $\theta^{\prime}$ and $\varphi^{\prime}$
via the relation (\ref{eq:10w}) with a corresponding change $\alpha \rightarrow
\alpha^{\prime}$, etc.

In the limit for $\vert{\mathbf q}\vert \rightarrow 0$, the kernel
${\mathcal Q}({\mathbf q},{\mathbf p})$ is reduced to
\begin{equation}
{\mathcal Q}(0,{\mathbf p})
= \pi \omega_0^2\,
\frac{(\omega^l_{\mathbf p})^2 -
{\mathbf p}^2}{8(\omega_{\mathbf p}^l)^3\vert{\mathbf p}
\vert^3}\,{\rm Z}_l({\mathbf p})
\int\!\frac{d \Omega}{4\pi}\,
\delta(\cos\theta^{\prime} - \rho^l_{\mathbf p}) \{
\varrho_{+}({\mathbf v};\hat{\mathbf q},\hat{\mathbf p})
\vert{\mathcal M}_{+}(0,{\mathbf p})\vert^2
+ \varrho_{-}({\mathbf v};\hat{\mathbf q},\hat{\mathbf p})
\vert{\mathcal M}_{-}(0,{\mathbf p})\vert^2\},
\label{eq:11w}
\end{equation}
where $\rho_{\mathbf p}^l\equiv(\omega_{\mathbf p}^l
- \omega_0)/\vert {\mathbf p}\vert$;
$d \Omega = \sin\theta^{\prime}\,{\rm d}\theta^{\prime}\,
{\rm d}\varphi^{\prime}.$
The expressions for $\varrho_{\pm}({\bf v};\hat{\mathbf q},\hat{\mathbf p})$
are defined
by Eq.\,(\ref{eq:10q}). In a new coordinate system, instead of Eq.\,(\ref{eq:10e}),
we have the function representation $\varrho_{\pm}$ in terms of angles
\begin{equation}
\varrho_{\pm}(\alpha^{\prime},\beta^{\prime};\theta^{\prime},\varphi^{\prime}) =
1 + \sin\theta^{\prime} \sin\alpha^{\prime} \cos(\varphi^{\prime} - \beta^{\prime}) +
\cos\theta^{\prime} \cos\alpha^{\prime} \pm (\cos\alpha^{\prime} +
\cos\theta^{\prime}).
\label{eq:11e}
\end{equation}
The limits of the scattering amplitudes ${\mathcal M}_{\pm}$, by its definitions
(\ref{eq:9h}) equals
\begin{equation}
{\mathcal M}_{\pm}(0,{\mathbf p}) =
\frac{\vert {\mathbf p}\vert}{\omega_0}\cos\theta^{\prime}
+ \lim\limits_{\vert {\mathbf q}\vert \to 0}(\!\,^{\ast}\!\triangle_{\pm}(l)
\,^{\ast}\!{\mathit \Gamma}_{\pm}).
\label{eq:11r}
\end{equation}
Thus, the problem of computing of $\gamma^{-}(0)$ is reduced to the calculation
of limits of scalar functions:
the effective propagators $\,^{\ast}\!\triangle_{\pm}$
and the effective vertices $\,^{\ast}\!{\mathit \Gamma}_{\pm}$.

By using the definitions of the 
$\,^{\ast}\!{\mathit \Gamma}_{\pm}$ effective scalar vertex functions 
(\ref{eq:9o}) and (\ref{eq:9u}), after slightly cumbersome computations, we define
\begin{equation}
\lim\limits_{\vert{\mathbf q}\vert \to 0}\!\,^{\ast}\!{\mathit \Gamma}_{\pm} =
-\,\frac{\vert{\mathbf p}\vert}{\omega_0}\,\rho_{\mathbf p}^l
\lim\limits_{\vert{\bf q}\vert\to 0}\!\,^{\ast}\!\triangle^{-1}_{\pm}(l) -
[\,\omega_0 \mp \vert{\mathbf p}\vert \mp \frac{{\mathbf p}^2}{\omega_0}\,
\rho_{\mathbf p}^l (1\pm\rho_{\mathbf p}^l)\,].
\label{eq:11t}
\end{equation}
In the last equality we use the $\,^{\ast}\!\triangle^{-1}_{\pm}(q)$ 
definition (\ref{eq:9r}) as the $F(q_0/\vert{\mathbf q}\vert)$ function (\ref{eq:7r})
(see Braaten and Pisarski \cite{bra2}).
Inserting Eq.\,(\ref{eq:11t}) into (\ref{eq:11r}), we reduce the scattering amplitudes to
$$
{\mathcal M}_{\pm}(0,{\mathbf p}) = \frac{\vert {\mathbf p}\vert}
{\omega_0}\cos\theta^{\prime}
- \frac{\vert{\mathbf p}\vert}{\omega_0}\,\rho_{\mathbf p}^l-
[\,\omega_0 \mp \vert{\mathbf p}\vert \mp \frac{{\mathbf p}^2}{\omega_0}\,
\rho_{\mathbf p}^l (1\pm\rho_{\mathbf p}^l)\,]
\lim\limits_{\vert{\mathbf q}\vert\to 0}\!\!
\,^{\ast}\!\triangle_{\pm}(l).
$$
By the $\delta$ function in Eq.\,(\ref{eq:11w}), all terms in
${\mathcal M}_{\pm}(0,{\mathbf p})$ not containing the propagators
$\,^{\ast}\!\triangle_{\pm}$, are relatively reduced
in the limit of the ${\mathbf q}=0$ mode.

The remaining terms, after substitution into Eq.\,(\ref{eq:11w}) and integration
over the solid angle with regard to Eq.\,(\ref{eq:11e}), yield
$$
{\mathcal Q}(0,{\mathbf p})
= \pi \omega_0^2
\,\frac{(\omega^l_{\mathbf p})^2 -
{\mathbf p}^2}{16(\omega_{\bf p}^l)^3\vert{\mathbf p}\vert^3}
\,{\rm Z}_l({\mathbf p})\theta(1-\vert\rho_{\mathbf p}^l\vert)
\biggr\{(1+\cos\alpha^{\prime})(1+\rho_{\mathbf p}^l)\Bigr[\,
\omega_0 - \vert{\mathbf p}\vert - \frac{{\mathbf p}^2}{\omega_0}\,
\rho_{\mathbf p}^l (1+\rho_{\mathbf p}^l)\Bigl]^2
\,\Bigl\vert\lim\limits_{\vert{\mathbf q}\vert\to 0}\!\!\,^{\ast}\!\triangle_{+}
(l)\Bigr\vert^2
$$
\begin{equation}
+\,(1-\cos\alpha^{\prime})(1-\rho_{\mathbf p}^l)\Bigr[\,
\omega_0 + \vert{\mathbf p}\vert + \frac{{\mathbf p}^2}{\omega_0}\,
\rho_{\mathbf p}^l (1-\rho_{\mathbf p}^l)\Bigl]^2
\,\Bigr\vert\!\lim\limits_{\vert{\mathbf q}\vert\to 0}\!\!\,^{\ast}\!\triangle_{-}(l)
\Bigr\vert^2\biggl\}.
\label{eq:11y}
\end{equation}
We note that this expression is not dependent on the angle $\beta^{\prime}$. This
enables us to represent the integration measure on the right-hand side of 
Eq.\,(\ref{eq:11q}) in the form
$$
\int\!{\rm d}{\mathbf p} = 2\pi\!\int\limits_{0}^{\infty}\!{\mathbf p}^2
{\rm d}\vert{\mathbf p}\vert
\int\limits_{1}^{- 1}\!{\rm d}(\cos\alpha^{\prime}).
$$
Substituting Eq.\,(\ref{eq:11y}) into (\ref{eq:11q}) and performing an angular
integration over $\alpha^{\prime}$, we finally obtain $\gamma^{-}(0)$
\begin{equation}
\gamma^{-}(0) = g^2C_F\!\int\limits_{0}^{\infty}\!{\rm d}\vert{\mathbf p}\vert
\,\theta(1-\vert\rho_{\mathbf p}^l\vert)\,{\mathcal Q}^{-}
(\vert{\mathbf p}\vert)N_{eq}^l(\vert{\mathbf p}\vert),
\label{eq:11u}
\end{equation}
where the kernel ${\mathcal Q}^{-}(\vert{\mathbf p}\vert)$ has the form
\begin{equation}
{\mathcal Q}^{-}(\vert{\mathbf p}\vert)
= \pi^2 \omega_0^2
\,\frac{(\omega^l_{\mathbf p})^2 - {\mathbf p}^2}
{4(\omega_{\mathbf p}^l)^3\vert{\mathbf p}\vert}\,{\rm Z}_l({\mathbf p})
\biggr\{(1+\rho_{\mathbf p}^l)\Bigr[\,
\omega_0 - \vert{\mathbf p}\vert - \frac{{\mathbf p}^2}{\omega_0}\,
\rho_{\mathbf p}^l (1+\rho_{\mathbf p}^l)\,\Bigl]^2
\Bigl\vert\lim\limits_{\vert{\mathbf q}\vert\to 0}\!\!\,^{\ast}\!\triangle_{+}
(l)\Bigr\vert^2
\label{eq:11i}
\end{equation}
$$
+\,(1-\rho_{\mathbf p}^l)\Bigr[\,
\omega_0 + \vert{\mathbf p}\vert + \frac{{\mathbf p}^2}{\omega_0}\,
\rho_{\mathbf p}^l (1-\rho_{\mathbf p}^l)\,\Bigl]^2
\,\Bigr\vert\!\lim\limits_{\vert{\mathbf q}\vert\to 0}\!\!\,^{\ast}\!\triangle_{-}(l)
\Bigr\vert^2\biggl\}.
$$
By virtue of the $\theta$ function in integrand (\ref{eq:11u}), the kernel
${\mathcal Q}^{-}(\vert{\mathbf p}\vert)$ is positive; therefore
$$
\gamma^{-}(0) > 0,
$$
i.e., the standing plasmino is damped. Thus, in the process of nonlinear
Landau damping, the pumping over of the excitation energy of standing plasminos
into plasmon branch of plasma excitations occurs, and therefore the first
case (\ref{eq:10s}) is true.

The function $\rho_{\mathbf p}^l$, entering
the argument of $\theta$ function and in a kernel (\ref{eq:11i}), decreases
with momentum $\vert {\mathbf p} \vert$ from $+\infty$ reaching a minimum, and
then monotonically increases at large $\vert {\mathbf p} \vert$, asymptotically
tending to +1 from below. The
equation $\rho_{\mathbf p}^l=1$ defines a lower limit of integration over the momentum
of the recoil plasmon. The numerical solution of a given equation yields
$$
\vert{\mathbf p}^{\ast}\vert \simeq 0.447\,\omega_{\rm pl}\; \; \mbox{for}\;\;
N_f=2,\qquad
\vert{\mathbf p}^{\ast}\vert \simeq 0.495\,\omega_{\rm pl}\; \; \mbox{for}\;\;
N_f=3
$$
(hereafter, the numerical estimates are presented for the ${\rm SU}(3)$ color group).
The function $\rho_{\mathbf p}^l$ reaches an absolute minimum at
$$
\vert{\mathbf p}^{\ast\ast}\vert \simeq 1.236\,\omega_{\rm pl},\quad
\rho_{{\mathbf p}^{\ast\ast}}^l\simeq 0.664 \;\; \mbox{for}\;\;N_f=2,
\qquad
\vert{\mathbf p}^{\ast\ast}\vert \simeq 1.305\,\omega_{\rm pl},\quad
\rho_{{\mathbf p}^{\ast\ast}}^l\simeq 0.692 \;\; \mbox{for}\;\;N_f=3.
$$
Notice that the value of the function $\rho_{\mathbf p}^l$ at a minimum point
coincides with the value of the plasmon group velocity ${\rm d}\omega_{\mathbf p}^l/
{\rm d}\vert{\mathbf p}\vert$. Thus, if the incident plasmino is initially at rest,
both the recoil plasmon and the effective quark that propagates below the
light cone carry nonzero energies and momenta of order $gT$.

At the end of this section we compare derived expressions for the
damping rate (\ref{eq:11u}) and (\ref{eq:11i}) of standing plasmino 
with the similar
expression obtained in the framework of the HTL approximation \cite{bra2,kob}.
For this purpose in Eqs.\,(\ref{eq:11u}) and (\ref{eq:11i}) we rescale
$\omega_{\mathbf p}^l \rightarrow \omega_0\, \widetilde{\omega}_{\mathbf p}^l$ and
$\vert {\mathbf p} \vert \rightarrow \omega_0\vert\widetilde{\mathbf p}\vert$.
The Planck distribution $N_{\rm eq}^l(\vert {\mathbf p}\vert)$ in the integrand
in Eq.\,(\ref{eq:11u}) should be set equal
$T/((2\pi)^3\omega_0\,\widetilde{\omega}_{\mathbf p}^l)$,
since the energy is soft. Further rewriting the kernel (\ref{eq:11i}) in terms
of the functions accepted in the paper \cite{kob} and defining the damping
constant as
\[
\gamma^{-}(0) = \check{a}(N_c,N_f)\,\frac{g^2TC_F}{4\pi},
\]
we derive the expression for the coefficient $\check{a}(N_c,N_f)$ required for
comparison
\[
\check{a}(N_c,N_f) = \frac{1}{2}\int\limits_{0}^{\infty}
\frac{\vert\widetilde{\bf p}\vert^2 {\rm d}\vert\widetilde{\mathbf p}\vert}
{\widetilde{\omega}^l_{\mathbf p}}\,
\widetilde{R}_l(\widetilde{\omega}^l_{\mathbf p},\vert\widetilde{\mathbf p}\vert)
\,\theta(\vert\widetilde{\mathbf p}\vert
- \vert\widetilde{\omega}_{\mathbf p}^l-1\vert)
\]
\[
\times\{[\,\vert\widetilde{\mathbf p}\vert
+ \widetilde{\omega}_{\mathbf p}^l -2\,]^2
\tilde{\beta}_{+}(1-\widetilde{\omega}^l_{\mathbf p},
\vert\widetilde{\mathbf p}\vert)
+[\,\vert\widetilde{\mathbf p}\vert - \widetilde{\omega}_{\bf p}^l +2\,]^2
\tilde{\beta}_{-}(1-\widetilde{\omega}^l_{\mathbf p},
\vert\widetilde{\mathbf p}\vert)\},
\]
where
\[
\widetilde{R}_l(\widetilde{\rm p}_0,\vert\widetilde{\mathbf p}\vert)=
-\frac{\widetilde{\rm p}_0}{\widetilde{\mathbf p}^2}
\frac{\widetilde{\rm p}_0^2 - \widetilde{\mathbf p}^2}
{3r - \widetilde{\rm p}_0^2 + \widetilde{\mathbf p}^2},\quad
r=\frac{\omega_{pl}^2}{\omega_0^2}.
\]
For the definitions of more cumbersome functions
$\tilde{\beta}_{\pm}(\widetilde{\rm p}_0,\vert\widetilde{\mathbf p}\vert)$ see
Ref.\,\cite{kob}. The coefficient $\check{a}(N_c,N_f)$
exactly coincides with the corresponding part in a similar
coefficient $a(N_c,N_f)$ in the expression of the damping rate of the standing
plasmino, responsible for the scattering process of plasminos by hard-particle
QGP's varying with the change of excitation statistic, derived by Kobes, Kunstatter,
and Mak (\cite{kob}, Eq.\,(5.11)).
We have shown in doing so, by really remaining in the context of
the Blaizot-Iancu equations (\ref{eq:2q})--(\ref{eq:2yyyy}), we are able to
compute not only a spectrum of fermionic excitations in QGP's, but also their
(gauge-invariant) damping rate at the leading order in the coupling constant,
which corresponds to the damping rate from the resummed perturbation theory
\cite{bra2,kob}.

\section{Light-cone singularity: Improved Blaizot-Iancu equations}
\setcounter{equation}{0}

In this section we discuss the behavior of the nonlinear Landau damping rate
for normal-particle excitations, when the quark spectrum $\omega^{+}_{\mathbf q}$
approaches the light cone. Here, the initial expression is
\begin{equation}
\gamma^{+}({\mathbf q}) =
\frac{1}{2}\,g^2C_F\!\int\!{\rm d}{\mathbf p}\,
\acute{\mathcal Q}({\mathbf q},{\mathbf p}) N_{eq}^l({\mathbf p}),\quad
\acute{\mathcal Q}({\mathbf q},{\mathbf p}) =
\frac{1}{2\omega_{\mathbf p}^l}\,{\rm Z}_{+}({\mathbf q}) {\rm Z}_{l}
({\mathbf p})\,{\rm Im}\,\acute{T}({\mathbf q},{\mathbf p}),
\label{eq:12q}
\end{equation}
where
\begin{equation}
{\rm Im}\,\acute{T}({\mathbf q},{\mathbf p}) =
\pi \omega_0^2
\frac{(\omega^l_{\mathbf p})^2 - {\mathbf p}^2}
{(\omega_{\mathbf p}^l)^2{\mathbf p}^2}
\!\int\!\frac{{\rm d}\Omega}{4\pi}\,\delta(\omega_{\mathbf q}^{+} -
\omega_{\mathbf p}^l - {\mathbf v}\cdot ({\mathbf q} - {\mathbf p})) \{
\acute{\varrho}_{+}({\mathbf v};\hat{\mathbf q},\hat{\mathbf p})
\acute{\mathit w}_{\mathbf v}^{+}({\mathbf q},{\mathbf p})
+ \acute{\varrho}_{-}({\mathbf v};\hat{\mathbf q},\hat{\mathbf p})
\acute{\mathit w}_{\mathbf v}^{-}({\mathbf q},{\mathbf p})\},
\label{eq:12w}
\end{equation}
and
\begin{equation}
\acute{\varrho}_{\pm}({\mathbf v};\hat{\mathbf q},\hat{\mathbf p}) \equiv
(1\pm\hat{\mathbf q}\cdot \hat{\mathbf l}\,)(1\mp{\mathbf v}\cdot \hat{\mathbf l}\,)
-\,\frac{{\mathbf v}\cdot ({\mathbf n}\times {\mathbf l})}
{\vert{\mathbf q}\vert\,{\mathbf l}^2},
\label{eq:12e}
\end{equation}
\begin{equation}
\acute{\mathit w}_{\mathbf v}^{\pm}({\mathbf q},{\mathbf p}) =
\vert\acute{\mathcal M}_{\pm}({\mathbf q},{\mathbf p})\vert^2 \equiv
\left\vert\,\frac{{\mathbf v}\cdot {\mathbf p}}{v\cdot q}
+\!\,^{\ast}\!\triangle_{\pm}(l)\,^{\ast}\!\acute{\mathit \Gamma}_{\pm}
\,\right\vert^2_{\rm on shell}.
\label{eq:12r}
\end{equation}
The scalar vertex functions $\,^{\ast}\!\acute{\mathit \Gamma}_{\pm}$ are defined
by Eq.\,(\ref{eq:9a}). On the right-hand side of Eq.\,(\ref{eq:12q}) we have taken
into account
for simplicity only the contribution from the nonlinear interaction with longitudinal
bosonic excitations, but all subsequent derivations easily extend to the case of
the nonlinear interaction with transverse excitations.

Near the light cone, the HTL-vertex pieces
$\delta\!{\mathit \Gamma}_0$, $\delta\!{\mathit \Gamma}_{\parallel}$, and
$\delta\!{\mathit \Gamma}_{\perp}$, the explicit forms of which are given by
Eq.\,(\ref{eq:9u}), are the origin of the strongest singularity of order
$1/{\varepsilon}$, where $\varepsilon^2\!\equiv\!((\omega^{+}_{\mathbf q})^2
- {\mathbf q}^2)/{\mathbf q}^2$. The terms linear in $\delta\!{\mathit \Gamma}$
in the $\acute{\mathit w}^{\pm}_{\mathbf v}$ probabilities (\ref{eq:12r}) can
lead to the logarithm of $\varepsilon$ only
\cite{fle2} (the first term in the amplitudes
$\acute{\mathcal M}_{\pm}$, defining Compton-like scattering processes, also results
only in logarithmic divergence). Hence we restrict our consideration to the
terms quadratic in $\delta\!{\mathit \Gamma}$.

Now let us single out in the HTL-vertex pieces the terms generating 
near-light-cone $1/\varepsilon$-contributions to $\gamma^{+}({\mathbf q})$.
As an example we consider the scalar function $\delta\!{\mathit \Gamma}_0$.
By using the explicit expression for HTL amplitudes derived by Frenkel and
Taylor \cite{fre}, we find
$$
\delta\!{\mathit \Gamma}_0 = \omega_0^2\!\int\!\frac{d\Omega}{4\pi}
\,\frac{{\mathbf v}\cdot {\mathbf p}}{(v\cdot l + i\epsilon )(v\cdot q)}\simeq
\omega_0^2\!\int\!\frac{d\Omega}{4\pi}
\,\frac{{\mathcal P}}{v\cdot l}\,\frac{{\mathbf v}\cdot{\mathbf p}}{v\cdot q}
$$
$$
\simeq\frac{1}{\,{\mathbf n}^2}\,[\,l^0({\mathbf n}\times{\mathbf q})\cdot{\mathbf p}
- q^0({\mathbf n}\times{\bf l})\cdot{\mathbf p}\,]\,{\rm M}(l,q) = -\,\omega_0^2\,p^0
{\rm M}(l,q).
$$
In the above expression, ${\mathcal P}$ indicates the principal-value prescription;
here and in the following $\simeq$ indicates that we have dropped less-divergent
terms. The Lorentz-invariant function ${\rm M}(l,q)$ is defined by
\begin{equation}
{\rm M}(l,q) = \left\{
\begin{array}{ll}
\displaystyle\frac{1}{2\sqrt{-\Delta(l,q)}}
\ln\!\left(\displaystyle\frac{l\cdot q + \sqrt{-\Delta(l,q)}}
{l\cdot q - \sqrt{-\Delta(l,q)}}\right), & \Delta(l,q)
\equiv l^2q^2-(l\cdot q)^2 < 0 \\
\displaystyle\frac{1}{\sqrt{\Delta(l,q)}}
\arctan\!\left(\displaystyle\frac{\sqrt{\Delta(l,q)}}{l\cdot q}\right), &
\Delta(l,q)>0.
\end{array}
\right.
\label{eq:12t}
\end{equation}
Similarly, the terms containing the function ${\rm M}(l,q)$ in the HTL-vertex
pieces $\delta\!{\mathit \Gamma}_{\parallel}$ and $\delta\!{\it \Gamma}_{\perp}$
are singled out. Substituting next thus derived expressions into 
Eq.\,(\ref{eq:9a}), we obtain
$$
\,^{\ast}\!\acute{\mathit \Gamma}_{\pm} \simeq \omega_0^2\!\left[
(q^0 - l^0)\pm (q^0 + l^0)\,\frac
{q^0\vert{\bf q}\vert \mp l^0\vert{\mathbf l}\vert}
{\vert{\mathbf q}\vert\vert{\mathbf l}\vert \pm {\mathbf q}\cdot{\mathbf l}}
\,\right]\!{\rm M}(l,q)
\equiv\omega_0^2\,f_{\pm}(\vert{\mathbf l}\vert,\vert{\mathbf q}\vert,
\hat{\mathbf l}\cdot\hat{\mathbf q})\,{\rm M}(l,q).
$$

Furthermore, substituting the last expression into the
$\acute{\mathit w}_{\mathbf v}^{\pm}$ probabilities (\ref{eq:12r}) and keeping only the terms
quadratic in ${\rm M}(l,q)$, we have
$$
\acute{\mathit w}_{\bf v}^{\pm}({\mathbf q},{\mathbf p})\simeq\omega_0^4\,\vert\!
\,^{\ast}\!\triangle_{\pm}(l)\vert^2 f_{\pm}^2\,\vert{\rm M}(l,q)\vert^2,
\quad f_{\pm} \equiv f_{\pm}(\vert{\mathbf l}\vert,\vert{\mathbf q}\vert,
\hat{\mathbf l}\cdot\hat{\mathbf q}).
$$
The dependence on the unit vector ${\mathbf v}$ in the integrand of the expression
(\ref{eq:12w}) is defined only by the $\delta$ function and $\acute{\varrho}_{\pm}$
function (\ref{eq:12e}), that enables easily to perform the angular integration in
Eq.\,(\ref{eq:12w}) and thus to derive instead of Eq.\,(\ref{eq:12w})
\begin{equation}
{\rm Im}\,\acute{T}({\mathbf q},{\mathbf p}) \simeq
\pi\omega_0^6\,
\frac{(\omega^l_{\mathbf p})^2 - {\mathbf p}^2}{(\omega_{\mathbf p}^l)^2
{\mathbf p}^2}\frac{1}{2\vert{\mathbf l}\vert}\,\theta(-l^2)
\label{eq:12y}
\end{equation}
$$
\times\Bigl\{
(1-\frac{l^0}{\vert{\mathbf l}\vert})(1 + \hat{\mathbf q}\cdot\hat{\mathbf l})
f_{+}^2\,\vert\!\,^{\ast}\!\triangle_{+}(l)\vert^2 +
(1+\frac{l^0}{\vert{\mathbf l}\vert})(1-\hat{\mathbf q}\cdot\hat{\mathbf l})
f_{-}^2\,\vert\!\,^{\ast}\!\triangle_{-}(l)\vert^2 \Bigr\}\,
\vert{\rm M}(l,q)\vert^2.
$$
The function ${\rm Im}\,\acute{T}({\mathbf q},{\bf p})$ is different from zero
for
$l^2=(q-p)^2\!<\!0$. It is not difficult to show that the condition $l^2\!<\!0$
on the mass shell of plasma excitations leads to $\Delta(l,q)\!<\!0$, and therefore
in Eq.\,(\ref{eq:12t}) it is necessary to take the first
expression.

With $q^2\rightarrow 0$, the square of ${\rm M}(l,q)$ becomes $1/\varepsilon$
times a representation of the $\delta$ function
\begin{equation}
\vert{\rm M}(l,q)\vert^2 \stackrel{q^2\rightarrow 0}{\longrightarrow}
\frac{1}{\varepsilon}
\,\frac{\pi^3}{\vert{\mathbf q}\vert\sqrt{-l^2}}\,\delta(q\cdot l).
\label{eq:12u}
\end{equation}
Substituting further the expression (\ref{eq:12y}) into Eq.\,(\ref{eq:12q}) and
taking into account Eq.\,(\ref{eq:12u}) and ${\rm Z}_{+}({\mathbf q})
\stackrel{q^2\rightarrow 0}{\longrightarrow}1$, we find the most singular
contribution to the nonlinear Landau damping rate of the normal quark excitations
near the light cone:
\begin{equation}
\gamma^{+}({\mathbf q}) \simeq \frac{1}{\varepsilon}\,
\frac{\pi^3\omega_0^6}{2\,\vert{\mathbf q}\vert}\,g^2C_F\!\int\!
\frac{{\rm d}\Omega_{\hat{\bf p}}}{4\pi}\!\int\limits_{0}^{\infty}\!{\rm d}
\vert{\mathbf p}\vert\,
\frac{(\omega_{\mathbf p}^l)^2 - {\mathbf p}^2}{(\omega_{\mathbf p}^l)^3
\vert{\mathbf l}\vert\sqrt{-l^2}}\,{\rm Z}_l({\mathbf p}) N_{eq}^l({\mathbf p})
\label{eq:12i}
\end{equation}
$$
\times\Bigl\{
(1-\frac{l^0}{\vert{\mathbf l}\vert})(1 + \hat{\mathbf q}\cdot\hat{\mathbf l})
f_{+}^2\,\vert\!\,^{\ast}\!\triangle_{+}(l)\vert^2 +
(1+\frac{l^0}{\vert{\mathbf l}\vert})(1-\hat{\mathbf q}\cdot\hat{\mathbf l})
f_{-}^2\,\vert\!\,^{\ast}\!\triangle_{-}(l)\vert^2 \Bigr\}\,
\theta(-l^2)\,\delta(q\cdot l).
$$
Here, the solid integral is over the directions of the unit vector
$\hat{\mathbf p}$.
The $\delta$ function in integrand (\ref{eq:12i}) enables us in principle to
perform the angular integration. Because of the fact that the functions
${\rm Z}_l({\mathbf p})$ and $N_{\rm eq}^l({\mathbf p})$ vanish exponentially at
large
$\vert{\mathbf p}\vert$, the above expression is ultraviolet convergent. Notice
also
that by virtue of the $\theta$ function in the integrand, the damping rate (\ref{eq:12i})
is positive. Thus we have shown that really the nonlinear Landau damping rate of
the normal quark mode near the light cone diverges as $1/\sqrt{q^2}$, 
and thus signals
the need to have further improvement of the Blaizot-Iancu equations, beyond
the HTL approximation.

As mentioned in the Introduction, the light-cone singularities are associated with
the massless basic constituents of a hot plasma, in our case, with massless
hard quarks and hard transverse gluons. The inclusion of the asymptotic thermal
masses for the basic constituents removes this type of singularity.
The effective way of entering thermal
masses into the resummed perturbation theory without spoiling gauge invariance
was suggested by Flechsig and Rebhan \cite{fle3}. Below we use this approach,
reformulating it into a ``kinetic'' language.

The light-cone singularity is generated by the HTL piece of the effective
two-quark--one-gluon vertex function (\ref{eq:4y}) (or (\ref{eq:4a})). In turn,
the expression for $\delta\Gamma(p;q_1,q_2)$ is defined with the help of
the Blaizot-Iancu equation (\ref{eq:2yyy}) for the function $\delta n_{\pm}^{\Psi}$
or by application of Eq.\,({\ref{eq:2yyyy}) for function $\not\!\!\Lambda$.
To derive the expression for $\delta\Gamma(p;q_1,q_2)$, which is free from the
light-cone singularity, we use improved Blaizot-Iancu equations instead of
Eqs.\,(\ref{eq:2yyy}) and (\ref{eq:2yyyy}) (the details of deriving these equations
are given in Appendix B)
$$
[v\cdot D_X,\delta n_{\pm}^{\Psi} ({\mathbf k}, X)] =
\pm\,\frac{ig}{2 \epsilon_k}
\,t^a\,(\bar{\psi}(X) t^a\!\not\!\!\Lambda^{\pm}({\mathbf k}, X)
-\overline{{\not\!\!\Lambda}^{\pm}}({\mathbf k},X) t^a\psi (X)),
\eqno{(2.8^{\,\prime})}
$$
$$
(v\cdot D_X)\!\not\!\!\Lambda^{\pm}({\mathbf k},X)
\pm\frac{\triangle m_{\infty}^2}
{2i\epsilon_k}\not\!\!\Lambda^{\pm}({\mathbf k},X)
= -\,ig\,C_F\,[N(\epsilon_k)
+ n(\epsilon_k)]\!\not\!v\,\psi (X),
\eqno{(2.9^{\,\prime})}
$$
where $\overline{{\not\!\!\Lambda}^{\pm}}({\mathbf k},X) =
(\not\!\!\Lambda^{\pm}({\mathbf k},X))^{\dagger}\gamma^0$,
$\triangle m^2_{\infty}=m_{q,\,\infty}^2-m_{g,\,\infty}^2$, and
$m_{q,\,\infty}$ and $m_{g,\,\infty}$ are asymptotic thermal masses for hard
quarks and gluons, respectively. Instead of the expression (\ref{eq:2e})
for the induced source now we have
$$
\eta(X) = g\!\int\!\!\frac{{\rm d}{\mathbf k}}{(2 \pi)^3} \,\frac{1}{2\epsilon_k}
\not\!\!\Lambda^{+}({\mathbf k},X) +
g\!\int\!\!\frac{{\rm d}{\mathbf k}}{(2 \pi)^3} \,\frac{1}{2\epsilon_k}
\not\!\!\Lambda^{-}({\mathbf k},X).
\eqno{(2.3^{\,\prime})}
$$
The expression for the color current induced by the soft-quark field
(\ref{eq:2t}), is not changed.

Employing the improved Blaizot-Iancu equations $(2.8^{\,\prime})$ and
$(2.9^{\,\prime})$, and expressions for induced source $(2.3^{\,\prime})$ and
the color current (\ref{eq:2t}), it is not difficult to obtain the improved
soft-quark self-energy and the effective two-quark--one-gluon vertex function
used above. These expressions coincide with similar ones derived
in the context of thermal field theory \cite{fle3}. Thus, for example, instead
of Eq.\,(\ref{eq:4y}) now we have
$$
\delta \Gamma^{(Q)}_{\mu}(p;q_1,q_2) =
-\,2\omega^2_0\,\Biggl\langle
\int\!\frac{{\rm d} \Omega}{4 \pi}
\,\Biggl\{
\frac{v_{\mu} \not\!v}
{\Bigl(v\cdot q_1 + \displaystyle\frac{\triangle m_{\infty}^2}{2T\alpha}
+i\epsilon\Bigr)\Bigl(v\cdot q_2 - \displaystyle\frac{\triangle m_{\infty}^2}
{2T\alpha}-i\epsilon\Bigr)}
$$
\begin{equation}
+\frac{v_{\mu} \not\!v}
{\Bigl(v\cdot q_1 - \displaystyle\frac{\triangle m_{\infty}^2}{2T\alpha}
+i\epsilon\Bigr)\Bigl(v\cdot q_2 + \displaystyle\frac{\triangle m_{\infty}^2}
{2T\alpha}-i\epsilon\Bigr)}\Biggr\}\Biggr\rangle_{\alpha}.
\label{eq:12o}
\end{equation}
Here,
$$
\langle{\cal O}(\alpha)\rangle_{\alpha} \equiv
\frac{2}{\pi^2}\!\int\!d{\alpha}\,\frac{\alpha {\rm e}^{\alpha}}
{{\rm e}^{2\alpha} - 1}\,{\cal O}(\alpha),\quad \alpha\equiv\epsilon_k/T.
$$
The use of expression (\ref{eq:12o}) gives a finite nonlinear Landau damping
rate $\gamma^{+}({\mathbf q})$ near the light cone. The additional terms in
denominators of Eq.\,(\ref{eq:12o}) lead to the replacement of the divergence factor
$1/\varepsilon$ on the right-hand side of Eq.\,(\ref{eq:12i}):
$$
\frac{1}{\varepsilon}\rightarrow \langle \alpha^{1/2}\rangle_{\alpha}
\,\frac{1}{2}\!\left(\frac{T\vert{\bf q}\vert}{\triangle m_{\infty}^2}
\right)^{1/2},
$$
where $\langle\alpha^{1/2}\rangle_{\alpha}=(3/4\pi^{3/2})(1-2^{-2/3})
\,\zeta(5/2)$ and $\zeta$ is a Riemann zeta function.

At the end of this section notice that we run into similar divergence in
research of a behavior near the light cone of the nonlinear Landau damping rate
for the transverse bosonic excitations in purely gauge sector \cite{mar2}.
The light cone
singularity here is generated by the HTL piece of the effective three-gluon vertex
function. It is clear that in this case for elimination of this singularity
it is necessary to use the improved Blaizot-Iancu equations for functions
$\delta n_{\pm}^{A}$ and $\delta N^{A}$ instead of Eqs.\,(\ref{eq:2y}) and
(\ref{eq:2yy}). Here, we have not presented their explicit forms, noting
only that their derivation is rather different from deriving improved equations
$(2.8^{\,\prime})$ and $(2.9^{\,\prime})$.

\section{Conclusion}\setcounter{equation}{0}
          
In Sec.\,X the nonlinear interaction of plasminos with
plasmons is shown to lead to the effective pumping over of plasma excitation energy from
the fermionic branch of plasma oscillations to the bosonic branch and vice versa,
and the kinematic relations between wave vectors of excitations are defined,
such that one or another process of pumping over of energy occurs. However,
it is clear that conclusions made in Sec.\,X are somewhat restricted
since they were made with allowance only for one process--the
process of nonlinear Landau damping. For precise study of nonlinear dynamics
of soft-fermion and soft-boson excitations in hot QCD plasma it is
necessary also to take into account in the plasmino and plasmon kinetic equations
the remaining terms on the right-hand sides of generalized kinetic equations
(\ref{eq:6q}) and (\ref{eq:6e}), responsible for the decay processes etc,
and consider the existence of soft normal quark excitations
and soft transverse gluon ones (the kinetic equations for them are defined
from Eqs.\,(\ref{eq:6q}) and (\ref{eq:6e}) by the replacements similar to
Eq.\,(\ref{eq:7tttt})). In addition to the right-hand side of the generalized kinetic equation
(\ref{eq:6e}) should be supplemented with the terms from purely gauge sector
\cite{mar2}.

Thus obtained self-consistent nonlinear system of four kinetic equations
will contain the maximum comprehensive information on soft excitations dynamics
in hot QCD plasma, which may be obtained in the context of the initial
Blaizot-Iancu equations (\ref{eq:2q})--(\ref{eq:2yyyy}) (or its improved
variant) in the first nonvanishing approximation\footnote{The following
term in the expansion over this small parameter leads to the processes of
Boltzmann type:
$$
{\rm q} + {\rm q}(\bar{\rm q}) \rightleftharpoons
{\rm q} + {\rm q}(\bar{\rm q}),\quad
{\rm q} + \bar{\rm q} \rightleftharpoons {\rm g} + {\rm g},\quad
{\rm g} + {\rm g} \rightleftharpoons {\rm g} + {\rm g}, \quad \mbox{etc.},
$$
going without exchange of energy between hard thermal particles and soft
plasma waves. The example of the construction of the Boltzmann equation, describing
the elastic scattering process of colorless plasmons between themselves
can be found in Ref.\,\cite{mar3}.} in the expansion over the small parameter
$gRA_{\mu}$, where $R\sim (gT)^{-1}$ and $\vert A_{\mu} \vert \ll T$.
Certain nonlinear processes will be
kinematically forbidden and therefore relevant contributions on the
right-hand side of kinetic equations will drop out. However, even with
allowance for the last case, this system of equations is especially complicated
for any analytical research (similar to studies carried out in Sec.\,X of the present
work and in Sec.\,10 in Ref.\,\cite{mar2}), and therefore here
invoking numerical methods is required.

\section*{\bf Acknowledgment}\setcounter{equation}{0}
This work was supported by an INTAS grant (No.\,2000-15) and the
Grant of the 6th Competition for Young Scientist of RAS (No.\,1999-80).

\section*{Appendix A}\setcounter{equation}{0}

By using the relations $\gamma^0 = h_{-}(\hat{\mathbf l})
+ h_{+}(\hat{\mathbf l})$
and ${\mathbf l}\cdot {\mathbf \gamma} = h_{-}(\hat{\mathbf l})
- h_{+}(\hat{\mathbf l})$,
we rewrite the expansion (\ref{eq:9y}) in the form
$$
\hspace{1.5cm}
\,^{\ast}\Gamma^i{\rm p}^i = h_{-}(\hat{\mathbf l})\,(
\delta\!{\mathit \Gamma}_0
+ \vert {\mathbf l}\vert\,^{\ast}\!{\mathit \Gamma}_{\parallel})
+ h_{+}(\hat{\mathbf l})\,(\delta\!{\mathit \Gamma}_0 - \vert {\mathbf l}\vert
\,^{\ast}\!{\mathit \Gamma}_{\parallel})
+ (({\mathbf n}\times {\mathbf l})\cdot
{\mathbf \gamma})\,^{\ast}\!{\mathit \Gamma}_{\perp}.
\eqno{({\rm A}1)}
$$
Let us add to the expression in the first parentheses on the right-hand side
of Eq.\,(A1) and then subtract from it the function
${\mathbf n}^2/(\vert{\mathbf q}\vert
(1-\hat{\mathbf q}\cdot\hat{\mathbf l}))$ and
in the second parentheses add and subtract the function
${\mathbf n}^2/(\vert{\mathbf q}\vert (1+\hat{\mathbf q}\cdot\hat{\mathbf l}))$
correspondingly. Let us introduce the following functions:
$$
\,^{\ast}\!{\mathit \Gamma}_{\pm} \equiv -\,\delta\!{\mathit \Gamma}_0 \mp
\vert {\mathbf l}\vert \,^{\ast}\!{\mathit \Gamma}_{\parallel}
+ \frac{{\mathbf n}^2}{\vert {\mathbf q} \vert}\,\frac{1}{1\mp
\hat{\mathbf q}\cdot
\hat{\mathbf l}} \,^{\ast}\!{\mathit \Gamma}_{\perp}.
$$
Instead of Eq.\,(A1), now we have
$$
\,^{\ast}\Gamma^i{\rm p}^i =
-\,h_{-}(\hat{\mathbf l})\,^{\ast}\!{\mathit \Gamma}_{+}
-h_{+}(\hat{\mathbf l})\,^{\ast}\!{\mathit \Gamma}_{-}
+\,\biggl[\,\frac{{\mathbf n}^2}{\vert {\mathbf q} \vert} \biggl(
\frac{1}{1 - \hat{\mathbf q}\cdot \hat{\mathbf l}}\,h_{-}(\hat{\mathbf l}) +
\frac{1}{1 + \hat{\mathbf q}\cdot \hat{\mathbf l}}\,h_{+}(\hat{\mathbf l})\!\biggr)
+ (({\mathbf n}\times {\mathbf l})\cdot {\mathbf \gamma})
\biggr]\!\,^{\ast}\!{\mathit \Gamma}_{\perp}.
\eqno{({\rm A}2)}
$$

Furthermore, we transform the expression in the parentheses on the right-hand
side of Eq.\,(A2). By using the definitions of matrices
$h_{\pm}(\hat{\mathbf l})$, this expression can be rewritten in the form
$$
\frac{1}{2}\biggl[\,\gamma^0
\biggl(\frac{1}{1-\hat{\mathbf q}\cdot\hat{\mathbf l}} +
\frac{1}{1+\hat{\mathbf q}\cdot\hat{\mathbf l}}\biggr) +
(\hat{\mathbf l}\cdot\gamma)
\biggl(\frac{1}{1-\hat{\mathbf q}\cdot\hat{\mathbf l}} -
\frac{1}{1+\hat{\mathbf q}\cdot\hat{\mathbf l}}\biggr)\biggr]
$$
$$
= \frac{{\mathbf q}^2 {\mathbf l}^2}{{\mathbf n}^2}(\gamma^0
+ (\hat{\mathbf l}\cdot{\mathbf \gamma})(\hat{\mathbf q}\cdot \hat{\mathbf l})) =
2\,\frac{{\mathbf q}^2 {\mathbf l}^2}{{\mathbf n}^2}\,h_{-}(\hat{\mathbf q}) -
\frac{\vert {\mathbf q}\vert}{{\mathbf n}^2}\,(({\mathbf n}\times {\mathbf l})
\cdot {\mathbf \gamma}).
\eqno{({\rm A}3)}
$$
In deriving the last equality we have used the expansion of vector ${\mathbf q}$
into two mutually ortogonal vectors, ${\mathbf l}$ and ${\mathbf n}\times{\mathbf l}$:
$$
{\mathbf q} = \frac{{\mathbf q}\cdot{\mathbf l}}{{\mathbf l}^2}\,{\mathbf l} +
\frac{1}{{\mathbf l}^2}\,({\mathbf n}\times {\mathbf l}).
\eqno{({\rm A}4)}
$$
Substituting Eq.\,(A3) into (A2) and reducing the terms with vector
product, finally  we derive the expression (\ref{eq:9i}) for convolution
$\,^{\ast}\Gamma^i{\rm p}^i$, instead of Eq.\,(\ref{eq:9y}).\\

\section*{Appendix B}
\setcounter{equation}{0}

Here we show how one can obtain the improved Blaizot-Iancu equations, used
in Sec.\,XII for deriving the finite nonlinear Landau damping rate for 
normal-particle excitations near the light cone. As the initial equations for obtaining
an improved equation for a one-body density matrix mixing fermion and boson degrees
of freedom, we take Eqs.\,(3.33) and (3.34) in the original paper of Blaizot and
Iancu \cite{bla2} (see the accepted notations and definitions therein)
$$
[\,\partial_{s}^2 + \partial_{s}\cdot\partial_X + 2igA(X)\cdot\partial_s]
\,K_a^{\mu\,<}(s,X) =ig\,(\not\!\partial_s \Delta)\gamma^{\mu}t^a\psi(X),
\eqno{({\rm B}1)}
$$
$$
[\,\delta^{ab}g_{\mu \nu}(\partial_s^2 - \partial_s\cdot\partial_X)
-gf^{abc}\Gamma_{\mu\nu\rho\lambda}A_c^{\rho}(X)\,\partial_s^{\lambda}]
\,K_b^{\nu\,<}(s,X) = -\,gS_0^{<}(s)\gamma_{\mu}t^a\psi(X).
\eqno{({\rm B}2)}
$$
Furthermore, we expand $K^< = K^{(0)} + K^{(1)} + \ldots\,$, with
$K^{(1)} \sim gK^{(0)}$, etc. The dominant terms in Eqs.\,(B1) and (B2)
lead to the consistency condition
$$
\partial_s^2 K^{(0)}(s,X) = 0.
\eqno{({\rm B}3)}
$$
Equation (B3) can be considered as the Klein-Gordon equation for hard, massless, free
particles. However, due to interaction with the hot medium, the basic
constituents of the plasma are known to acquire the dynamical (asymptotic) temperature-induced
masses. To account for this fact we modify the condition (B3), replacing it in the
case of Eq.\,(B1) by
$$
\partial_s^2 K^{(0)}(s,X) \approx - \, m_{q,\,\infty}^2 K^{(0)}(s,X),
\eqno{({\rm B}4)}
$$
where $m_{q,\,\infty}^2 = g^2T^2\,C_F/4$ is the asymptotic thermal mass for
hard quark, and in the case of Eq.\,(B2) by
$$
\partial_s^2 K^{(0)}(s,X) \approx - \, m_{g,\,\infty}^2 K^{(0)}(s,X),
\eqno{({\rm B}5)}
$$
where $m_{g,\,\infty}^2 = g^2T^2(N_f+2N_c)/12$ is the asymptotic thermal mass for
the hard transverse gluon. It is necessary to keep the terms $\partial_s^2$ inside
the brackets on the left-hand sides of Eqs.\,(B1) and (B2) setting them equals
the right-hand side of Eqs.\,(B4) and (B5), accordingly. Further, subtracting
Eq.\,(B2) from (B1) and performing transformations similar to those in Ref.\,\cite{bla2},
we come to improved equation for a density $\not\!\!{\mathcal K}(k,X) = t^a
\gamma^{\mu}K^a_{\mu}(k,X)$:
$$
(v \cdot D_X)\!\not\!\!{\mathcal K}(k,X) -\frac{i}{2}\triangle m_{\infty}^2
\not\!\!{\mathcal K}(k,X)
= -\,ig \, C_F\rho_0(k) \,[N (k_0) + n(k_0)]\!\not\!k\,\psi (X) .
$$
Here, $\triangle m_{\infty}^2 \equiv m_{q,\,\infty}^2 - m_{g,\,\infty}^2.$
Substituting into the last equation the expansion
$$
\not\!\!{\mathcal K}(k,X) = 2\pi \delta(k^2)\{
\theta(k^0)\!\not\!\!\Lambda^{+}({\mathbf k},X) +
\theta(-k^0)\!\not\!\!\Lambda^{-}(-{\mathbf k},X)\},
\eqno{({\rm B}6)}
$$
replacing ${\mathbf k}\rightarrow -\,{\mathbf k}$, where it is necessary,
and dividing by
$\epsilon_k$, we obtain the improved Eq.\,$(2.9^{\,\prime})$. By virtue of the fact
that $\triangle m_{\infty}^2 \neq 0$, the discrepancy
$\not\!\!\Lambda^{+}({\mathbf k},X)\!\neq\,\not\!\!\Lambda^{-}({\mathbf k},X)$
will hold.
The last case is an actual reflection at the kinetic description level of the
discrepancy of the two terms in the integrands of the improved hard thermal loops
with external fermions when exchanging ${\mathbf k}\rightarrow -\,{\mathbf k}$,
found by Flechsig and Rebhan \cite{fle3}.

Now we turn to deriving the improved kinetic equations for
$\delta n_{\pm}^{\Psi}$.
The equations for functions $\delta n_{\pm}^{\Psi}$ are defined from the equation
for the current density $J_{\mu}^{\Psi}(k,X)$ \cite{bla2}:
$$
[v \cdot D_X, J_{\mu}^{\Psi} (k,X)] = ig\,k_{\mu}
t^a\{\bar{\psi}(X) t^a\!\not\!\!{\mathcal K}(k,X)
- \not\!\!{\mathcal H}(k,X) t^a \psi (X)\}
\eqno{({\rm B}7)}
$$
by substituting to it the expansion
$J_{\mu}^{\Psi}(k,X) = k_{\mu}4\pi\delta (k^2)
\{\theta (k_0)\delta n_{+}^{\Psi}({\mathbf k},X) +
\theta (-k_0)\delta n_{-}^{\Psi}(-{\mathbf k},X)\}$.
The relevant modification
of the equation for $\delta n_{\pm}^{\Psi}$ here is achieved by the simple
requirement of using the expansion (B6) on the right-hand side of Eq.\,(B7),
where the functions $\not\!\!\Lambda^{\pm}$ obey the improved equations
$(2.8^{\,\prime})$.

\end{document}